\documentclass[10pt,preprint]{aastex}

\newcommand{\hii}{H{\scriptsize II} }
\newcommand{\Msol}{~M$_{\odot}$}
\newcommand{\tnm}[1]{\tablenotemark{#1}} 

\shorttitle{{\em Chandra} study of M17} 

\shortauthors{Broos et al.}

\slugcomment{Re-submitted November, 2006 for \apj}

\begin{document}

\title{The young stellar population in M17 revealed by {\em Chandra}}

\author{
Patrick S. Broos\altaffilmark{1}\altaffilmark{*}, 
Eric D. Feigelson\altaffilmark{1}, 
Leisa K. Townsley\altaffilmark{1}, 
Konstantin V. Getman\altaffilmark{1}, 
Junfeng Wang\altaffilmark{1}, 
Gordon P. Garmire \altaffilmark{1}, 
Zhibo Jiang\altaffilmark{2}, 
Yohko Tsuboi \altaffilmark{3} 
}

\altaffiltext{*}{Send requests to P. Broos at patb@astro.psu.edu.}
\altaffiltext{1}{Department of Astronomy \& Astrophysics, 525 Davey Laboratory, Pennsylvania State University, University Park, PA 16802} 

\altaffiltext{2} {Purple Mountain Observatory, National Astronomical 
Observatories, Chinese Academy of Sciences, 2 Beijing Xi Lu, Nanjing, 
Jiangsu 210008, China}
               
\altaffiltext{3} {Department of Physics, Faculty of Science and Engineering,
Chuo University, 1-13-27 Kasuga, Bunkyo-ku, Tokyo 112-8551, Japan}

\begin{abstract}

We report here results from a {\em Chandra} ACIS observation of the 
stellar populations in and around the Messier~17 \hii region.  
The field reveals 886 sources with observed X-ray luminosities (uncorrected for absorption) between $\sim29.3 
< \log L_x < 32.8$~ergs~s$^{-1}$, 771 of which have stellar 
counterparts in infrared images.  
In addition to comprehensive 
tables of X-ray source properties, several results are presented: 
\begin{enumerate}

\item The X-ray Luminosity Function is calibrated to that of the Orion 
Nebula Cluster population to infer a total population of roughly 8000--10,000 
stars in M17, one-third lying in the central NGC~6618 
cluster.  

\item About 40\% 
of the ACIS sources are heavily obscured with $A_V > 
10$~mag.  Some are concentrated around well-studied star-forming
regions---IRS~5/UC1, the Kleinmann-Wright Object, and M17-North---but most are 
distributed across the field. As previously shown, 
star formation appears to be widely distributed in the molecular clouds.
X-ray emission is detected from 64 of the hundreds of Class I protostar candidates that can be identified by near- and mid-infrared colors.  These constitute the most likely protostar candidates known in M17.  

\item The spatial distribution of X-ray stars is complex: in addition 
to the central NGC~6618 cluster and well-known embedded groups, we 
find a new embedded cluster (designated M17-X), a 2 pc-long arc of 
young stars along the southwest edge of the M17 \hii region, and 0.1~pc 
substructure within various populations.  These structures may indicate that the 
populations are dynamically young.

\item All (14/14) of the known O stars but only about half (19/34) of the known B0--B3 stars in the M17 field are 
detected.  These stars exhibit the long-reported correlation 
between X-ray and bolometric luminosities of $L_x \sim 10^{-7}L_{bol}$.
While many O and early B stars show the soft X-ray emission expected from microshocks in their winds or moderately hard emission that could be caused by magnetically channeled wind shocks, six of these stars exhibit very hard thermal plasma components ($kT>4$~keV) that may be due to colliding wind binaries.
More than 100 candidate new OB stars are found, including 28 X-ray detected intermediate- and high-mass protostar candidates with infrared excesses.                                                           

\item Only a small fraction (perhaps 10\%) of X-ray selected high- and 
intermediate-mass stars exhibit $K$-band emitting protoplanetary 
disks, providing further evidence that inner disks evolve very 
rapidly around more massive stars.

\end{enumerate}

\end{abstract}


\section{Introduction \label{sec:intro}}

In the last 30 years, most observational studies of the stellar 
denizens of massive star-forming regions (MSFRs) were based on 
broadband visual photometry, supplemented by occasional visual 
spectroscopy of the most massive cluster members, and near-infrared (NIR) 
photometry of small regions ($<10\arcmin$) situated on the cluster 
cores. 
In a few visually obscured regions this was augmented by NIR spectroscopy of the brightest OB cluster members.
The surrounding \hii regions and molecular clouds were often 
well-studied in radio/millimeter wavelengths, including 
arcminute-scale maps of molecular line emission and arcsecond-scale 
maps of continuum structures.  More recently, heated dust structures have been 
mapped in the mid-infrared (MIR) with space-based instruments on the {\em Midcourse Space Experiment (MSX)} \citep[e.g.][]{Kraemer03} and the 
{\em Spitzer Space Telescope} \citep[e.g.][]{Churchwell04,Allen05}.  

In the high-energy regime, lower-mass young stars in nearby 
star-forming regions were studied extensively with X-ray missions 
during the 1980-90s \citep{Feigelson99}, but studies of the more 
distant MSFRs were hampered by the limited spatial resolution of 
these telescopes.  The launch of the {\em Chandra X-ray Observatory} 
\citep{Weisskopf02} in 1999 makes X-ray studies of MSFRs much more 
feasible, due to the sub-arcsecond spatial resolution of its mirrors, 
their high reflectivity at shorter wavelengths that penetrate 
interstellar absorption, and the excellent noise characteristics of 
its primary imaging camera, the Advanced CCD Imaging Spectrometer 
(ACIS). High-resolution X-ray images of MSFRs can reveal a complex 
menagerie of hundreds of stellar sources (protostars, pre-main sequence stars, 
and OB/Wolf-Rayet stars) and diffuse structures such as wind-swept bubbles, 
superbubbles, chimneys, and supernova remnants \citep[see review 
by][]{Feigelson06}.  X-ray selected samples are surprisingly 
advantageous for uncovering the stellar populations of young clusters. 
By studying a number of these fields with varying ages and stellar 
content with {\em Chandra}, we hope to characterize these stellar 
populations and to learn more about the origins of any diffuse X-ray 
emission that may be present.  Some of the most recent {\em Chandra} 
MSFR studies include  NGC~6193 \citep{Skinner05}, NGC~6334 
\citep{Ezoe06}, 30~Doradus \citep{Townsley06a,Townsley06b}, 
Cepheus B/OB3b \citep{Getman06a}, RCW~38 \citep{Wolk06}, Westerlund 1 
\citep{Skinner06,Muno06b}, W~49A \citep{Tsujimoto06}, and NGC~6357 \citep{Wang06}.
These studies all show rich stellar populations, 
many uniquely revealed by the high-resolution {\em Chandra} observations.

These X-ray observations are complementary to
high-resolution IR imaging data from 2MASS, {\em Spitzer}, and modern
ground-based facilities.  We have shown that {\em Chandra} observations of 
Galactic MSFRs are not strongly contaminated by X-ray 
sources that are not cluster members; in a typical {\em Chandra}  
observation of a cluster around $1-2$ kpc distant, simulations and 
data show that only a handful of foreground stars and background active galactic nuclei (AGN) are 
present ($\sim$5\% of the X-ray sources), while hundreds of cluster members are revealed 
\citep{Getman05a,Getman06a,Wang06}.  These X-ray detections can be used 
to identify cluster members for further IR study without the usual 
selection by IR excess.  This provides a disk-unbiased 
sample for further IR study, at least for lower-mass T~Tauri (pre-main sequence) stars.  X-ray observations also are not 
hampered by the confusion due to bright dust and gas emission that 
plagues IR and visual studies of \hii regions.  X-rays are often the 
only way to detect fainter cluster members along ionization fronts or in
the vicinity of bright OB stars.

\begin{figure}
\centering 
\mbox{
\includegraphics[height=4in]{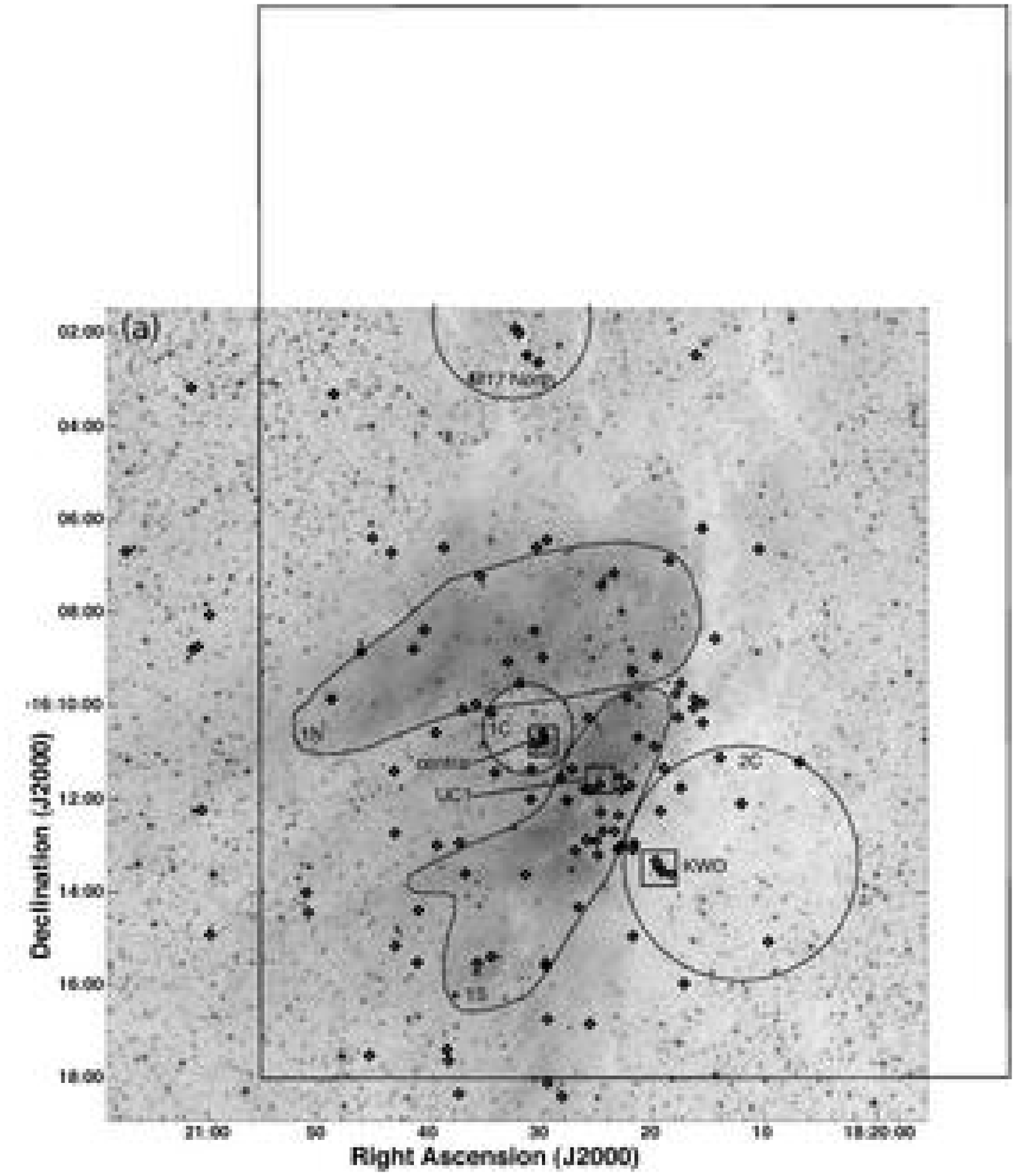} 
\includegraphics[height=4in]{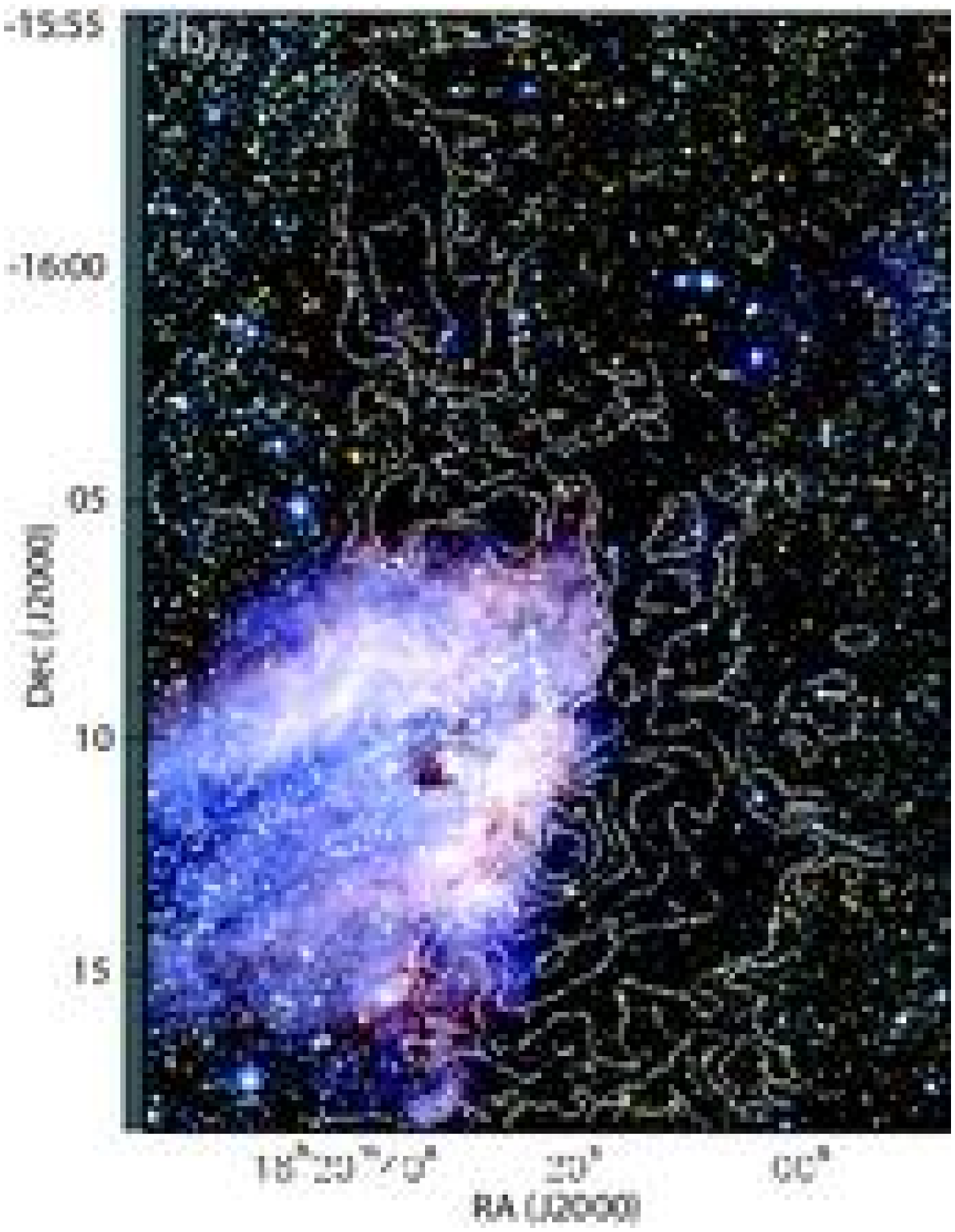} 
}
\caption{{\bf (a)} 2MASS $K$-band mosaic of M17 covering the $17\arcmin 
\times 17\arcmin\/$ {\em Chandra} field. The 115 small diamonds mark 
ACIS sources without identified counterparts (\S \ref{sec:isolated}). 
Two polygons, reproduced with their labels from Figure 2c in \citet{Jiang02}, outline the North Bar
and South Bar in the \hii region.  Two circles, reproduced with their labels from the same figure, outline the 
central NGC~6618 stellar cluster core and the embedded M17-SW region.  
A third circle outlines the embedded M17-North region described by \citet{Henning98}.
Three small boxes indicate the  
expanded views in Figures \ref{fig:UC1_K-X} and \ref{central_img.fig}.  
{\bf (b)} C$^{18}$O contours (-5, -2, 2, 4, 10, 20, 30, 40, 50, and 60~K~km~s$^{-1}$) shown on an image from the 2MASS survey,
taken from the right-hand panel in Figure~1 of \citet{Wilson03} and reproduced here with permission of the authors.
This field of view is marked by a large rectangle in panel {\it a}.
\label{fig:K_17x17}
}
\end{figure}

Here we present the first high-spatial-resolution X-ray point source 
study of \object{M17} ($\Leftrightarrow$ W~38 $\Leftrightarrow$ S~45 $\Leftrightarrow$ RCW~160 $\Leftrightarrow$ the Omega Nebula), a famous and spectacular                                        
nearby MSFR (Figure~\ref{fig:K_17x17}). M17 is a blister \hii region at the northeast edge of 
one of the largest giant molecular clouds in the Galaxy, with an extent of 
4$^\circ$ \citep[$\sim$110~pc,][]{Elmegreen79}. The local geometry 
resembles the Orion Nebula \hii region except that it is viewed 
edge-on rather than face-on \citep{Meixner92}.  Molecular material is 
concentrated in a clumpy structure known as M17-SW.  Active star 
formation is concentrated in three molecular cores (Figure~\ref{fig:K_17x17}{\it a}): around the 
ultracompact \hii region M17-UC1 and its associated masers, the double B star 
known as the Kleinmann-Wright Object (KWO) and its associated cluster 
\citep{Chini04a}, and the protostars in M17-North.  Historically, their 
proximity to the \hii region was attributed to sequential triggering 
of star formation \citep{Elmegreen77}, although this link has been 
challenged for M17-SW by \citet{Wilson03}, who note that the KWO is too far from the OB cluster and too evolved for its formation to have been caused by the cluster. The clumpy molecular structure of the 
cloud (Figure~\ref{fig:K_17x17}{\it b}) and infrared protostars indicate that star formation is still 
in progress all around the \hii region
\citep{Jiang02, Wilson03, Chini05, Reid06}.  

NGC~6618, the central cluster that illuminates M17, has 100 stars 
earlier than B9 \citep{Lada91}; for comparison, Orion has a dozen.  Several 
obscured O4-O5 stars form a central $1\arcmin$ ring and are principally 
responsible for ionizing the nebula \citep{Hanson97}. NGC~6618 is 
estimated to be no more than 1~Myr old; this extreme youth and the 
absence of evolved stars in the region give good circumstantial 
evidence that we are seeing this cluster before its first supernova, 
making M17 an ideal place to study the interactions of massive stars 
with their environment and stellar disks before they are affected by 
nearby supernovae.  We give a more extensive description of past 
observations of M17 and NGC~6618 in our earlier analysis of this {\em Chandra} observation,  
which focused on the diffuse X-ray emission produced by the 
cluster OB winds \citep[][henceforth TFM03]{TFM03}.  

Distance estimates for M17 range from 1.3~kpc \citep{Hanson97} to 
2.2~kpc \citep{Chini80}.  A thorough study by 
\citet{Nielbock01} gives a distance of $1.6 \pm 0.3$~kpc; we adopt 
this value here.  
This is consistent with TFM03 and with the NIR study of \citet{Jiang02}, whose data are used in this paper.

\begin{figure}
\centering 
\includegraphics[width=0.9\textwidth]{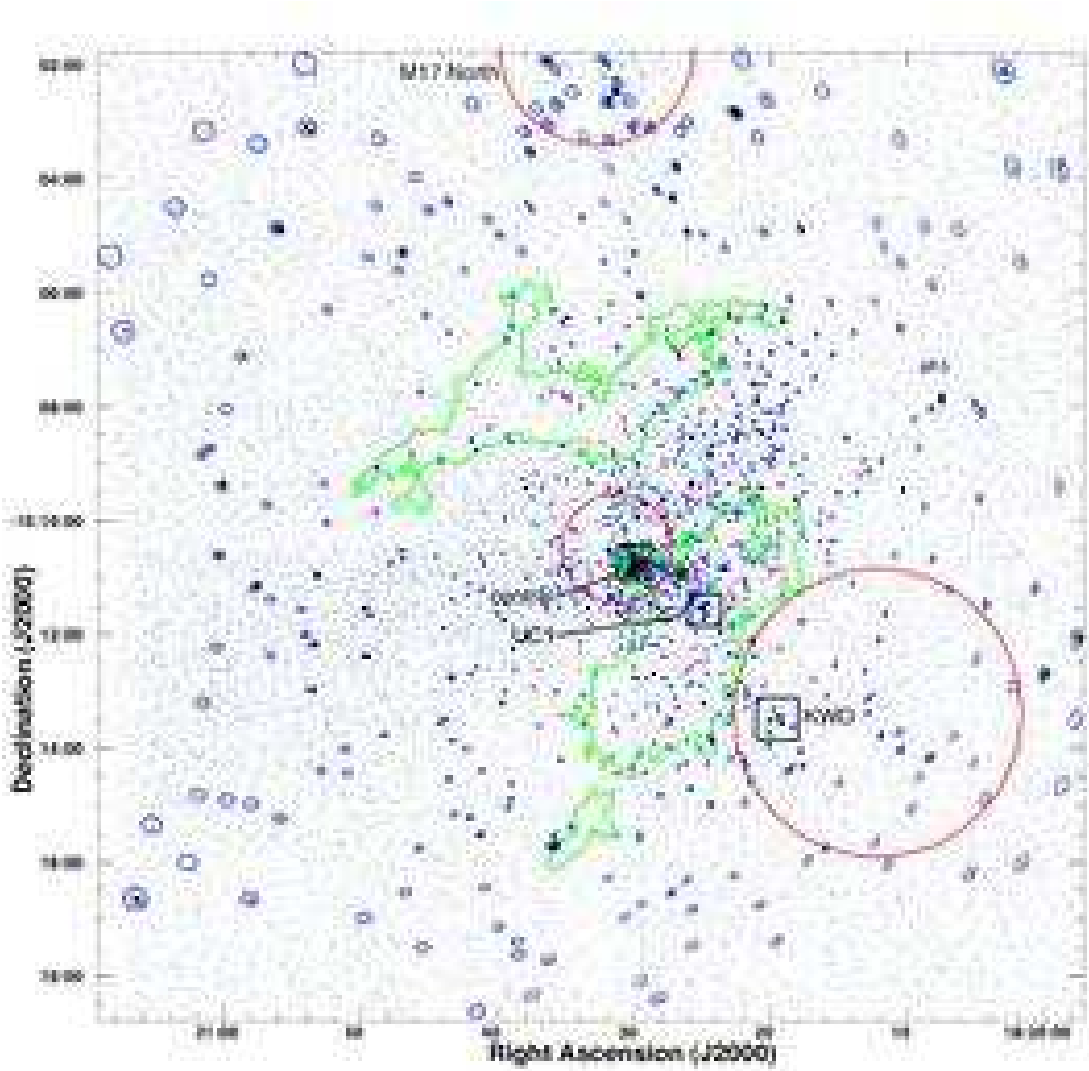} 
\caption{A {\em Chandra} ACIS-I binned image of M17 with 886 source extraction regions designed to match the {\em Chandra} Point Spread Function (PSF) shown in blue.  
The PSF is a complex function of position and photon energy; for example the 90\% encircled energy radius at 1.5~keV varies from $\sim$0.9$\arcsec$ on axis to $\sim$6$\arcsec$ at the field edge (see \S 4 of 
\anchor{http://asc.harvard.edu/proposer/}{The {\em Chandra} Proposers' Observatory Guide}  
at {\it http://asc.harvard.edu/proposer/}).
Contours (green) of bright dust emission from {\em Spitzer} 3.6\micron\ data outline the North Bar and South Bar in the \hii region.  
The circles and squares are reproduced from Figure~\ref{fig:K_17x17}.
\label{fig:ACIS_17x17}
}                                              
\end{figure}

The present {\em Chandra} study is based on a 40~ks pointing with 
the ACIS camera in March 2002; this is the same dataset analyzed by 
TFM03.  This image locates over 800 stellar X-ray sources in the 
NGC~6618 cluster and cloud population with sub-arcsecond accuracy 
(Figure~\ref{fig:ACIS_17x17}).  
{\em Chandra} detects all of the known O stars and about half the 
cataloged early B stars in NGC~6618, and adds dozens of likely new massive, 
intermediate-mass, and protostellar members as well as hundreds of T Tauri stars. 
Sparse clusters around M17-UC1, the KWO, and M17-North are 
found. Soft diffuse X-ray emission arising from thermalized O-star 
winds pervades the \hii region, as discussed by TFM03.

This paper begins with a description of the X-ray sources and their 
stellar counterparts (\S \ref{sec:observation}). We quantify the 
stellar population and its spatial structure in \S 
\ref{sec:stellarpop}. Two important subpopulations, the embedded 
stars and the massive OB stars, are discussed in detail in the next 
two sections (\S \ref{sec:embedded}-\ref{sec:OB}).  
We summarize our findings in \S \ref{sec:summary}.

\section{X-ray Observation, Data Analysis, and Initial Results \label{sec:observation}}

M17 was observed with the Imaging Array of the {\em Chandra}/ACIS 
camera (ACIS-I, $17\arcmin \times 17\arcmin$ field of view) for roughly 40~ks in March 2002. The aim point was centered on the source with the 
earliest spectral type in NGC 6618's central ring of O stars, the 
O4+O4 visual binary (\S \ref{sec:Kleinmann}) called ``Kleinmann's Anonymous Star'' 
\citep{Kleinmann73a}. 

Our data reduction methods are described in detail in Appendix B of 
TFM03; the same reduced dataset used in that study was also used for 
this one, so we refer the reader to that description for details of 
the dataset and our data analysis.  Briefly, the data were corrected 
for charge transfer inefficiency (CTI) problems in the CCDs using 
the Penn State CTI corrector \citep{Townsley02}; the $0.5\arcsec$ 
event position randomization added in the {\em Chandra} X-ray Center's 
standard data processing was removed; a sub-pixel positioning scheme 
\citep{Tsunemi01} was applied to improve source positions.  We then 
created twelve different images of the ACIS field:  soft (0.5--2~keV), 
hard (2--8~keV), and full (0.5--8~keV) X-ray wavebands with four 
different pixel binning scales (4$\times$, 2$\times$, 1$\times$, and 
0.5$\times$ the nominal $0.5\arcsec$ ACIS sky pixel) to cover the full ACIS field, the ACIS 
Imaging Array, the central $8\arcmin$, and the central $4\arcmin$, 
respectively.

\subsection{Source Detection and Extraction \label{sec:extraction}}

A list of candidate point sources was then obtained in each of these 
band-limited images using the {\em wavdetect} wavelet source 
detection algorithm \citep{Freeman02}.  The algorithm was purposefully 
run with a low threshold ($P = 10^{-5}$) that is sensitive to very 
faint sources but also identifies some spurious noise features.  
The $10^{-5}$ {\em wavdetect} threshold is used because experience has shown that more stringent thresholds (e.g.\ the more typical $10^{-6}$) result in many missed sources in regions with complex backgrounds, such as M17's diffuse X-ray emission.
These twelve source lists were merged (keeping the source position from the 
highest-resolution image) to generate the  master list of candidate sources for the 
full observation. The field was visually examined and a few possible 
faint sources missed by the algorithm were added, resulting in 933 candidate source positions.

Photons were extracted using the 
\anchor{http://www.astro.psu.edu/xray/docs/TARA/ae_users_guide.html}{ACIS Extract}\footnote{The ACIS 
Extract software package and User's Guide have been available online at 
\url{http://www.astro.psu.edu/xray/docs/TARA/ae\_users\_guide.html} since 
February 2003. } (AE) software package \citep{Broos02}.  The procedures used in AE 
are described in TFM03 and \citet{Getman05b}.  The source extraction 
regions were usually defined to be $\sim$90\% contours of the local 
Point Spread Function (PSF) at 1~keV; crowded sources were assigned 
smaller extraction regions following visual review.  AE applies an 
energy-dependent correction to each source's calibration, via the Ancillary Response Function (ARF) file, 
to account for its finite extraction region.  

The background event data set from which local background spectra were 
constructed was obtained by masking regions around all the sources in the catalog 
via a process that involved two passes through AE\@.  First, 
conservative circular masks covering $>99$\% of each source PSF were 
constructed, local backgrounds were extracted, and source fluxes were 
estimated.  Then, each source mask was redefined using the generally 
less restrictive criteria that: (a) the mask shall include the source 
extraction region (polygon) itself; and (b) the mask shall also 
include the region where the expected surface brightness from the 
point source is larger than one half the observed smoothed local 
background. A circular background region was then defined 
independently for each source such that the region enclosed $\sim$100 
events in the source-free masked data set. 

For each candidate source AE computed the  
quantity $P_B$, the Poisson 
probability associated with the ``null hypothesis'' that no source 
exists with respect to the local background level.  
We chose two thresholds on $P_B$ to define the final source list: $P_B < 
0.003$ for 845 ``primary'' sources and $0.003 < P_B < 0.01$ for 41 
``tentative'' sources.
Counterparts were identified for 31 of these tentative sources (\S \ref{sec:counterparts}), adding confidence that these sources are real.  
The remaining 47 candidate sources had $P_B > 0.01$ and were rejected as likely background fluctuations.

These 886 sources are shown superposed on 
the ACIS-I field in Figure~\ref{fig:ACIS_17x17}.  
AE and the {\em Chandra} software package CIAO (version 3.2) were used to compute source and local background 
spectra, calibration files\footnote{CIAO calibrations include the effects of contamination on the ACIS-I optical blocking filter.}, background-corrected median energies, light 
curves, variability estimates, broad-band photometry, and other 
quantitative information for each source. Basic properties of the primary and tentative 
sources are presented in Tables~\ref{tbl:src_properties_main} and 
\ref{tbl:src_properties_tentative}.  Sources are listed by increasing 
right ascension and can be identified by their sequence number (e.g.\ 
\#358) or their IAU designation (e.g.\ CXOU J182025.70$-$161649.9).
Please see the table notes for definitions of the columns and 
see \citet{Getman05b} and \citet{Townsley06b} for further details on the derivations of the properties.
The positions reported are estimated by AE;  the {\em Chandra} coordinate system was aligned to 
the {\em Hipparcos} reference frame using 51 matches between strong on-axis ACIS sources and reliable 2MASS stars 
(\S \ref{sec:counterparts}).
The formal uncertainty in the remaining systematic reference frame offset between {\em Chandra} and 2MASS is $\sim$0.04\arcsec\/ in each axis.

\subsection{Source Variability \label{sec:variability}}

\begin{figure} 
\centering                   
\includegraphics[width=0.85\textwidth]{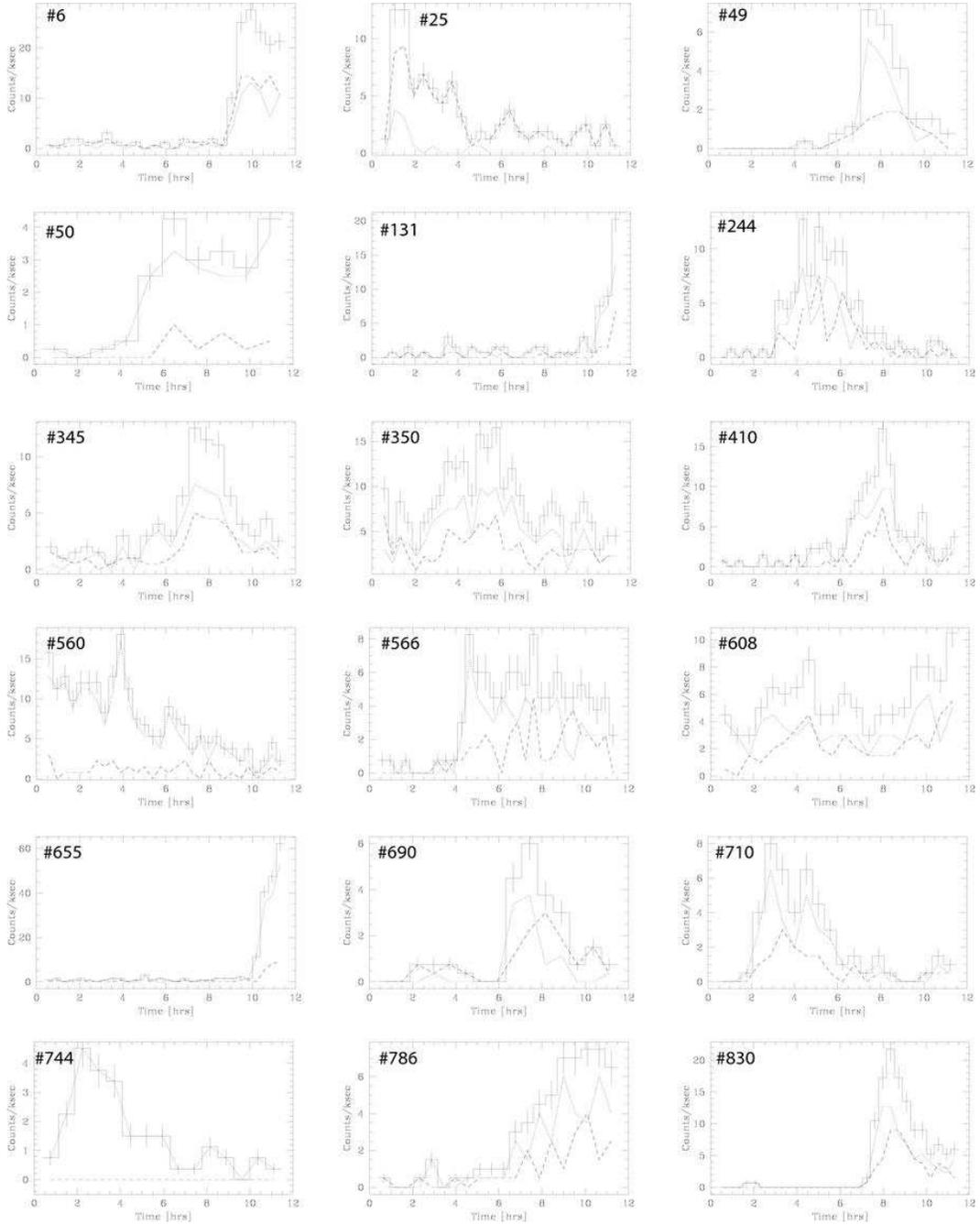} 

\caption{Light curves for some sources exhibiting strong variability 
during the {\em Chandra} observation, labeled by sequence number. 
Histograms show the full ($0.5-8.0$~keV) energy band with $\sqrt{N}$ error bars; 
dashed lines show the soft ($0.5-2.0$~keV) band; dotted lines show the hard ($2.0-8.0$~keV) band.  
Note that the ordinates vary between panels.
\label{fig:lightcurve}}
\end{figure}

Thirty-nine of the M17 ACIS sources show significant X-ray 
variability ($P_{KS} < 0.005$ in column 15 of 
Table~\ref{tbl:src_properties_main}).  We display some of these 
variable light curves in Figure~\ref{fig:lightcurve}.  Many show the 
high-amplitude, fast-rise-slow-decay morphology characteristic of 
solar-type magnetic reconnection flares \citep{Reale01}.  Some show secondary flares 
superposed on the decays of primary flares.  Others show a poorly 
understood slow rise over several hours.  This wide range of flare 
morphologies is characteristic of X-ray luminous pre-main sequence 
stars \citep[e.g.][]{Imanishi01, Grosso04, Favata05, Wolk05}.
A few sources show more random variability not characteristic of T~Tauri flares.
One bright variable star (\#396) is not shown here, but is discussed in Section~\ref{sec:new_OB};
light curves for 24 variable but weak sources are not shown.


\subsection{X-ray Fluxes and Luminosities \label{sec:fluxes}}

Characterization of the spectral properties and luminosities of the 
sources in this data set is a challenging endeavor because the 
quantity of data available for most sources is so low: 95\% of sources have fewer than 
100 net counts.  Such data sets poorly constrain spectral model 
parameters, and ``best fit'' models are often nonphysical.  

Investigators have adopted various approaches to analyze weak ACIS source spectra. 
Some use an assumed intrinsic 
source spectrum and absorbing column to derive a scaling factor 
between intrinsic luminosity and observed flux that is applied to 
all weak sources  \citep{Hong05, Muno06a}.   Some investigators apply 
prior knowledge about individual sources, such as $A_V$ estimates 
from optical or NIR photometry, to constrain spectral 
models \citep{Getman06a,Flaccomio06}. 
Others limit their goals 
to obtaining non-unique spline-like fits to the event energy 
distribution, allowing nonphysical models, and emphasize caution when 
interpreting the spectral parameters \citep{Feigelson02}. All of 
these approaches are reasonable.

Another issue concerns the mathematical method of seeking the 
best-fit spectral model, and then estimating confidence intervals on 
the spectral parameters.  A least-squares approach based on the 
$\chi^2$ statistic applied to binned data is traditional, 
although maximum likelihood estimation of unbinned data is becoming 
more common.  With the nonlinear dependence between spectral 
parameters, it is often difficult to establish their confidence 
intervals for weak sources in an authoritative manner.  Bayesian 
analysis provides a likelihood-based approach that incorporates prior 
knowledge expressed through prior distributions \citep{VanDyk01}.  
The uncertainties of the derived parameters (such as 
absorption-corrected luminosities) are uncovered through posterior 
distributions, and one can marginalize over uninteresting parameters 
if one chooses.  However, standard tools for a full Bayesian spectral 
analysis for X-ray data are not yet available.

In this study, we adopt a strategy most similar to that of 
\citet{Feigelson02}.  Our primary goal is to obtain 
wide-band fluxes using regression models as spline functions.  Fluxes 
are then converted to luminosities, both observed and intrinsic (corrected for 
estimated absorption), using a distance of 1.6~kpc.  Calculations are 
performed using the 
\anchor{http://heasarc.gsfc.nasa.gov/docs/software/lheasoft/xanadu/xspec}{XSPEC package} \citep[version 
12.2.1o,][]{Arnaud96}\footnote{                                        
\url{http://heasarc.gsfc.nasa.gov/docs/software/lheasoft/xanadu/xspec}}   
to fit one- and two-temperature {\em apec} thermal plasma models \citep{Smith01} and power law 
models seen through an absorbing column ($N_H$) of interstellar material with 
cosmic elemental abundances \citep{Morrison83}.  The one-temperature thermal plasma 
model is preferentially used and abundances of $Z = 0.3Z_{\odot}$ are assumed, as found for pre-main sequence stars in other, more nearby, star-forming regions \citep{Imanishi01,Feigelson02}. 
A power law model is adopted if the thermal model poorly described the data or required nonphysical parameters, and if the source is not identified with a known stellar counterpart.
When neither best-fit model is 
acceptable, we freeze the parameter $kT = 2$~keV in the thermal 
model, which is typical for young stars \citep{Getman05b,Preibisch05a}, and then fit for the $N_H$ 
and normalization parameters.  Analysis was not attempted on the 288 
weakest sources whose photometric significance (column 12 in 
Tables~\ref{tbl:src_properties_main} and 
\ref{tbl:src_properties_tentative}) is below 2.0.

The best-fit model was found by the maximum likelihood method 
\citep[XSPEC minimizes the C statistic,][]{Cash79,Wachter79}.  Although the 
accuracy of the parameter error estimation algorithm in XSPEC is not 
well understood for weak sources, we report 90\% confidence intervals when they were 
flagged\footnote{See the {\em tclout error} command in the XSPEC manual.} acceptable by XSPEC.  One should regard the reported 
parameter errors with suitable caution.  A simulation of parameter 
uncertainties for weak sources was reported by \citet{Feigelson02}.

Spectral analysis results for the brighter 598 sources are presented 
in Tables~\ref{tbl:thermal_spectroscopy}~and~\ref{tbl:powerlaw_spectroscopy}.  See the table notes for 
descriptions of the columns. Best-fit absorbing column densities 
range from negligible to $\log N_H \sim 23.7$~cm$^{-2}$, equivalent 
to a visual absorption of $A_V \sim$~250--300~mag. Temperatures range from the 
softest ($kT \sim 0.4$~keV) to the hardest (truncated at $kT = 15$ 
keV) accessible to the {\em Chandra} ACIS-I detectors. 
The range of observed total band ($0.5-8$~keV) luminosities, corrected for absorption,
able to be derived from spectral modeling is  
$29.8 \lesssim \log L_{t,c} \lesssim 33.3$~ergs~s$^{-1}$.
Apparent luminosities range from $\log L_t \sim 29.3$~ergs~s$^{-1}$ (estimated for the faintest ``tentative'' sources in 
Table~\ref{tbl:src_properties_tentative}) to $\log L_t = 32.8$~ergs~s$^{-1}$.
Completeness limits are discussed in \S \ref{sec:XLF}.

These tables report a variety of band-limited luminosities, both apparent and corrected for absorption \citep[see][for details]{Getman05b}.  Absorption-corrected soft-band luminosities ($L_{s,c}$ in our nomenclature) are highly uncertain and are not reported.  For highly obscured sources ($\log N_H > 22.5$~cm$^{-2}$), soft-band luminosities are not reported because they may be misleading; the spectral fits are based solely on hard counts for these sources, so the model parameters may significantly underestimate or even miss entirely contributions from soft spectral components.  Thus, although we could report $L_s$ and $L_{s,c}$ derived from the model fit, we choose to omit these quantities because they are not representative of the true soft-band luminosities of these obscured sources.  The reader is cautioned that total-band luminosities may underestimate the true source luminosities due to such missing soft spectral components in our models.

While a power law fit may suggest that the source could be a background AGN, this result is not definitive.  Of the 52 sources in Table~\ref{tbl:powerlaw_spectroscopy}, all but 12 have IR counterparts (see \S \ref{sec:counterparts}); brighter IR sources are less likely to be AGN.  Conversely, some sources fit with very hard thermal plasmas may be equally well fit with power laws; we preferentially chose the thermal fit in such cases.

\subsection{Stellar Counterparts \label{sec:counterparts}}

Due to the combination of high obscuration and confusion by bright 
nebular emission, there are no adequate visual band catalogs of 
stars in the M17 region.  The best stellar survey of the region is 
in the NIR bands, using the SIRIUS instrument on the InfraRed Survey 
Facility (IRSF) 1.4 m telescope at the South African Astronomical Observatory. 
SIRIUS detected over 
29,000 stars with sensitivity limits around $J < 20$~mag and $K < 19$~mag and 
sub-arcsecond resolution over a $14\arcmin \times 14\arcmin$ region 
\citep{Jiang02}.  For outer portions of the ACIS field outside the 
SIRIUS coverage, we use the 2MASS All-Sky Survey which reaches $K 
\sim 15.5$~mag. A survey of the region with the IRAC detector  on board the {\em Spitzer Space Telescope} 
in four IR ($3.6-8$ $\mu$m) bands has 
recently been released by the Galactic Legacy Infrared Mid-Plane 
Survey Extraordinaire (GLIMPSE) Legacy Team 
\citep[e.g.][]{Churchwell04}.  Here we considered the on-line $\sim$
30 million source Highly Reliable Catalog from the 
\anchor{http://www.astro.wisc.edu/sirtf/glimpsedata.html}{GLIMPSE Enhanced Data Products}\footnote{ 
\url{http://www.astro.wisc.edu/sirtf/glimpsedata.html}}.

We associate ACIS X-ray sources with IR sources using positional 
coincidence criteria\footnote{Software implementing the matching algorithm described here is available in the 
\anchor{http://www.astro.psu.edu/xray/docs/TARA/}{TARA package} at \url{http://www.astro.psu.edu/xray/docs/TARA/} .}.  
We first removed reference frame offsets 
between the ACIS field (aligned to 2MASS) and the SIRIUS and GLIMPSE catalogs using high-quality 
ACIS sources (brighter than 10 counts and within 2\arcmin\ of the aim 
point).  The estimated offsets were 0.6\arcsec\ to SIRIUS and 0.3\arcsec\ to GLIMPSE.  
Then, for every pair of catalog entries (e.g.\ an ACIS and a 
SIRIUS source), we test the hypothesis that 
the two physical sources are spatially coincident, i.e.\ that the two observed positions are random samples drawn from Gaussian distributions with identical means and with the reported standard deviations in position.  If the hypothesis is true then the offsets between the two observed positions (in X and Y) will be drawn from a 2-D Gaussian distribution with zero mean and with X and Y variances equal to the sums of the reported variances.  
We reject this hypothesis if the 
observed positional offsets fall outside a 99\% confidence region of 
this two-dimensional Gaussian.  IR positional errors are obtained 
from the catalogs. 
ACIS positional errors consist of a random component (computed by AE), corresponding to the uncertainty inherent in 
averaging a finite number of event positions, added in quadrature with an arbitrary 0.2\arcsec\ systematic component that avoids unreasonably small error estimates for very bright sources.

The pairs of catalog entries for which the match hypothesis is satisfied are accepted as matches in order of their likelihood.  
When a match is accepted, the participating IR source is removed from 
all pending pairs not yet accepted, thus ensuring that no duplicates 
appear among the associations. In crowded locations, multiple IR 
sources satisfy the match hypothesis for a specific ACIS source.  In 
such situations, if the most likely IR match of the group is still 
available, then it is accepted.  Otherwise the ACIS source remains 
unmatched. 

Likely associations between ACIS sources and IR sources are reported 
in Table~\ref{tbl:counterparts}.
Listed there are 727 SIRIUS, 476 2MASS, and 224 GLIMPSE counterparts;
771 of the 886 ACIS sources (87\%) have a counterpart identified.
GLIMPSE counterparts were notably absent from the North Bar and South Bar which suffer from bright nebular emission.
IR photometry is reproduced from the 
catalogs; $JHK$ magnitudes are from SIRIUS photometry when available, except for four sources (ACIS \#366, 701, 720, and 854) where 2MASS values are more reliable due to saturation effects in the SIRIUS images.  Table~\ref{tbl:counterparts} also gives likely associations to optical stars reported 
by \citet{Chini80} (designated CEN~xxx) and to NIR stars from \citet{Hanson97} derived from \citet{Bumgardner92} (designated B~xxx)
\footnote{
A reference frame offset of 1.9\arcsec\ between ACIS and \citet{Hanson97} positions was removed before matching the catalogs.
}.  Association 
ambiguity flags and notes to other published characteristics of 
selected sources can be found in the table.  

The list of matches in Table~\ref{tbl:counterparts} is not 
perfect, as both false negatives (true physical associations that are 
missed) and false positives (unphysical associations that are listed) 
are likely present. 
To the extent that the actual distributions of position errors follow the stated Gaussian distributions, we expect to miss 1\% of true associations (false-negatives) as specified by the 99\% confidence interval adopted.
In Appendix~\ref{sec:matching} we estimate that 2.5\% of the 2MASS associations, 7\% of the SIRIUS associations, and 9\% of the GLIMPSE associations are expected to be incorrect due to uncertainty in positions.
A non-statistical problem in identifying stellar counterparts is that some cataloged IR sources may be 
condensations in the nebular emission (from atomic lines, PAH bands, 
or heated dust continuum) rather than stars.  
Some NIR sources may be unresolved multiple systems, and \citep[particularly when the primary has spectral type A or B,][]{Stelzer05} the brightest component may not the principal X-ray emitter.
In short, our associations should be considered only ``likely'' rather than confirmed stellar counterparts to the ACIS sources. 

\begin{figure}
\centering
\includegraphics[angle=0,width=0.5\textwidth]{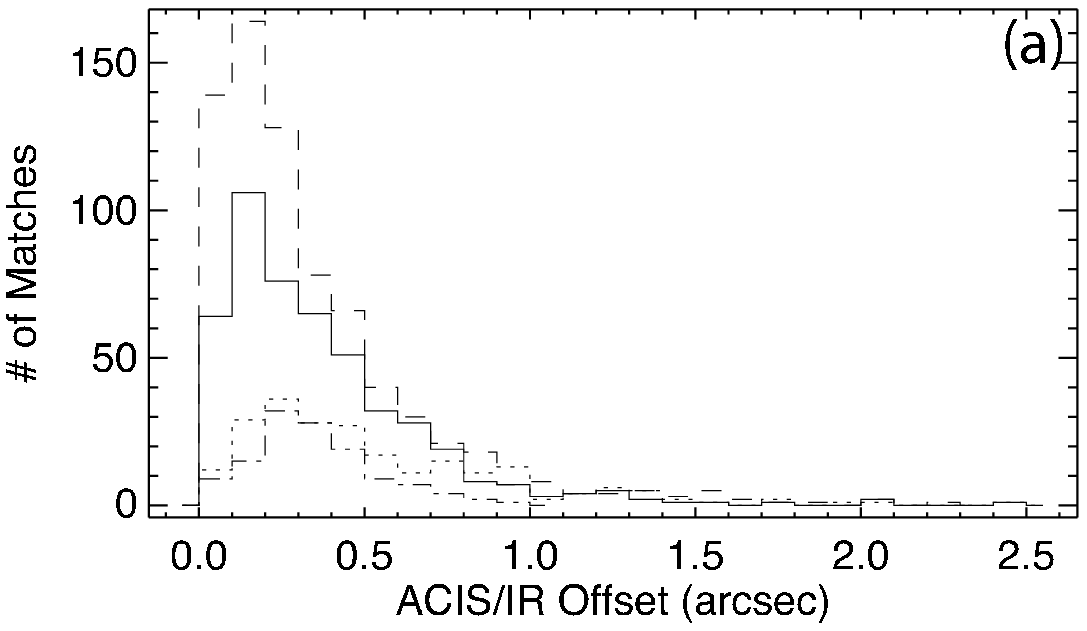} 
\includegraphics[angle=0,width=0.5\textwidth]{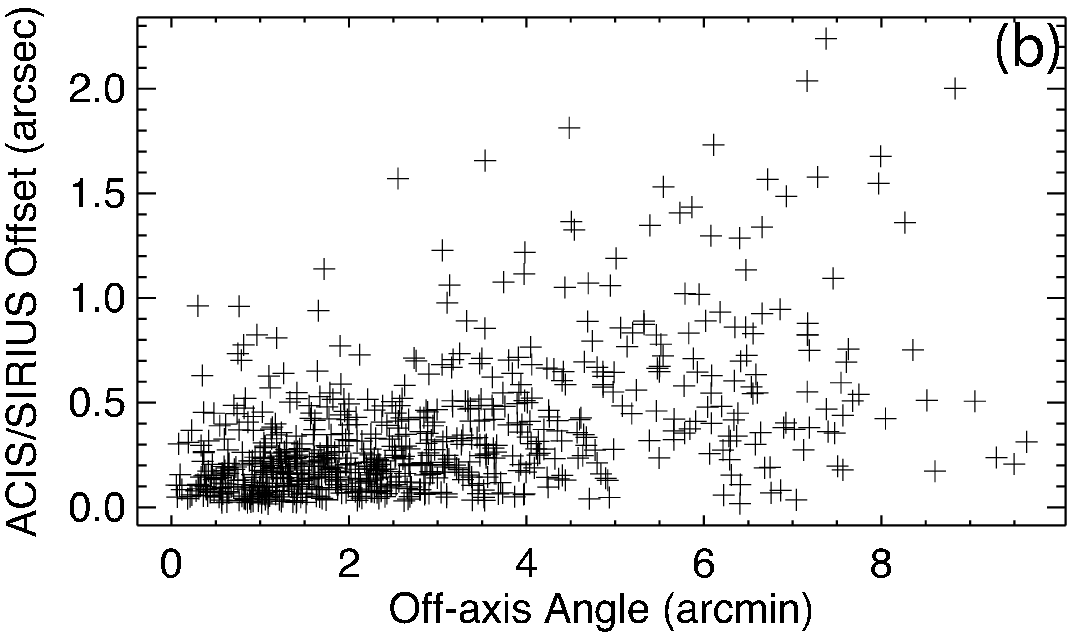} 
\caption{Positional offsets between matching ACIS and IR sources, after removal of reference frame offsets.  Panel {\bf (a)} 
 shows offset distributions for SIRIUS (dashed), 2MASS (solid), 
 GLIMPSE (dotted), and Bumgardner (dash-dot) matches.  The median ACIS/SIRIUS offset is 0.24\arcsec.
Panel {\bf (b)} shows offsets for                                                            
SIRIUS matches vs.\ off-axis angle in the ACIS field. 
\label{fig:offset}} 
\end{figure}

Figure~\ref{fig:offset}{\it a} shows the distributions of offsets between 
the ACIS and IR associations.  
Figure~\ref{fig:offset}{\it b} shows the expected deterioration in positional offsets as the {\em Chandra} PSF broadens 
off-axis.  This can be compared to a similar plot from the {\em Chandra} 
Orion Ultradeep Project (COUP) survey of the Orion Nebula which shows 
median offsets increasing from 0.12\arcsec\/ on-axis to 0.4\arcsec\/ 
off-axis \citep[][Figure~9]{Getman05b}.  The M17 median offsets are 
poorer, ranging from 0.2\arcsec\/ to 1\arcsec\/, due mainly to the 
much weaker ACIS sources in the short M17 observation compared to the long COUP 
observation.


\begin{deluxetable}{rcrrrrrrrrrrrrrccccc}
\centering \rotate \tabletypesize{\tiny} \tablewidth{0pt} 
\tablecolumns{20}

\tablecaption{ Primary {\em Chandra} Catalog:  Basic Source Properties 
\label{tbl:src_properties_main}}

\tablehead{ \multicolumn{2}{c}{Source} &
  &
\multicolumn{4}{c}{Position} &
  &
\multicolumn{5}{c}{Extracted Counts} &
  &
\multicolumn{6}{c}{Characteristics} \\
\cline{1-2} \cline{4-7} \cline{9-13} \cline{15-20}

\colhead{Seq} & \colhead{CXOU J} &
  & 
\colhead{$\alpha_{\rm J2000}$} & \colhead{$\delta_{\rm J2000}$} & 
\colhead{Err} & \colhead{$\theta$} &
  & 
\colhead{Net} & \colhead{$\Delta$Net} & \colhead{Bkgd} & 
\colhead{Net} & \colhead{PSF} &   
  &
\colhead{Signif} & \colhead{$\log P_B$} & \colhead{Anom} & \colhead{Var} &\colhead{EffExp} & \colhead{$E_{median}$}  \\

\colhead{\#} & \colhead{} &
  & 
\colhead{(deg)} & \colhead{(deg)} & \colhead{(\arcsec)} & 
\colhead{(\arcmin)} &
  & 
\colhead{Full} & \colhead{Full} & \colhead{Full} & \colhead{Hard} & 
\colhead{Frac} &
  &
\colhead{} & \colhead{} & \colhead{} & \colhead{} & \colhead{(ks)} & 
\colhead{(keV)}
 \\

\colhead{(1)} & \colhead{(2)} &
  & 
\colhead{(3)} & \colhead{(4)} & \colhead{(5)} & \colhead{(6)} &
  & 
\colhead{(7)} & \colhead{(8)} & \colhead{(9)} & \colhead{(10)} & 
\colhead{(11)} &
  & 
\colhead{(12)} & \colhead{(13)} & \colhead{(14)} & \colhead{(15)} & 
\colhead{(16)} & \colhead{(17)}  }

\startdata
37 & 182009.70$-$161503.7 &  &  275.040439 & -16.251041 &  0.8 &  6.7 &  &     9.7 &   4.0 &   2.3 &     9.4 & 0.90 &  &   2.1 & $<$-5 & . & a &   32.1 & 5.3 \\
51 & 182012.31$-$160447.7 &  &  275.051294 & -16.079918 &  0.8 &  7.2 &  &    16.4 &   5.0 &   3.6 &    11.7 & 0.90 &  &   2.9 & $<$-5 & . & a &   31.6 & 2.3 \\
97 & 182017.84$-$161117.0 &  &  275.074374 & -16.188070 &  0.3 &  3.1 &  &     5.6 &   3.0 &   0.4 &     5.7 & 0.90 &  &   1.6 & $<$-5 & . & a &   36.7 & 3.2 \\
123 & 182019.41$-$161329.1 &  &  275.080890 & -16.224756 &  0.1 &  4.0 &  &    30.8 &   6.1 &   0.2 &    25.9 & 0.60 &  &   4.6 & $<$-5 & . & a &   36.1 & 3.3 \\
128 & 182019.60$-$161038.8 &  &  275.081670 & -16.177456 &  0.3 &  2.6 &  &     8.6 &   3.5 &   0.4 &     8.7 & 0.89 &  &   2.1 & $<$-5 & . & a &   37.4 & 4.8 \\
135 & 182019.88$-$161052.3 &  &  275.082863 & -16.181206 &  0.3 &  2.5 &  &     8.5 &   3.5 &   0.5 &     8.7 & 0.89 &  &   2.1 & $<$-5 & . & b &   37.4 & 5.7 \\
163 & 182020.73$-$160717.9 &  &  275.086416 & -16.121661 &  0.3 &  3.9 &  &     8.2 &   3.5 &   0.8 &     7.5 & 0.89 &  &   2.0 & $<$-5 & . & a &   36.4 & 3.5 \\
177 & 182021.21$-$161222.9 &  &  275.088397 & -16.206382 &  0.2 &  2.9 &  &    16.4 &   4.6 &   0.6 &    12.6 & 0.90 &  &   3.1 & $<$-5 & . & a &   37.2 & 2.8 \\
182 & 182021.32$-$161140.2 &  &  275.088858 & -16.194518 &  0.2 &  2.5 &  &    11.5 &   4.0 &   0.5 &     1.6 & 0.89 &  &   2.5 & $<$-5 & . & b &   37.4 & 1.2 \\
186 & 182021.43$-$160939.1 &  &  275.089300 & -16.160865 &  0.1 &  2.3 &  &    30.3 &   6.1 &   0.7 &    16.6 & 0.89 &  &   4.6 & $<$-5 & . & a &   38.0 & 2.1 \\
187 & 182021.43$-$161206.8 &  &  275.089325 & -16.201912 &  0.0 &  2.7 &  &   363.4 &  19.6 &   0.6 &   288.6 & 0.89 &  &  18.1 & $<$-5 & . & a &   37.3 & 3.1 \\
230 & 182022.58$-$160251.3 &  &  275.094121 & -16.047608 &  0.2 &  7.9 &  &   365.4 &  19.8 &   5.6 &   257.7 & 0.90 &  &  18.0 & $<$-5 & . & a &   31.2 & 2.6 \\
233 & 182022.69$-$160833.9 &  &  275.094578 & -16.142761 &  0.1 &  2.7 &  &    73.1 &   9.1 &   0.9 &    24.6 & 0.89 &  &   7.6 & $<$-5 & . & a &   38.0 & 1.7 \\
236 & 182022.74$-$161123.5 &  &  275.094775 & -16.189866 &  0.1 &  2.0 &  &    60.3 &   8.3 &   0.7 &    45.5 & 0.89 &  &   6.8 & $<$-5 & . & a &   37.9 & 2.6 \\
239 & 182022.87$-$161148.5 &  &  275.095298 & -16.196825 &  0.1 &  2.2 &  &    19.3 &   5.0 &   0.7 &    19.5 & 0.89 &  &   3.5 & $<$-5 & . & a &   37.1 & 4.9 \\
244 & 182022.93$-$161152.7 &  &  275.095573 & -16.197986 &  0.1 &  2.3 &  &   134.4 &  12.1 &   0.6 &    77.6 & 0.89 &  &  10.6 & $<$-5 & g & \nodata &   35.1 & 2.4 \\
246 & 182022.97$-$161131.8 &  &  275.095712 & -16.192185 &  0.2 &  2.1 &  &    12.4 &   4.1 &   0.6 &    12.6 & 0.85 &  &   2.6 & $<$-5 & g & \nodata &   34.1 & 4.8 \\
255 & 182023.16$-$161305.7 &  &  275.096504 & -16.218256 &  0.1 &  3.1 &  &    54.4 &   7.9 &   0.6 &    54.5 & 0.90 &  &   6.4 & $<$-5 & . & a &   35.5 & 4.6 \\
275 & 182023.80$-$161325.7 &  &  275.099198 & -16.223807 &  0.4 &  3.3 &  &     8.5 &   3.5 &   0.5 &     8.6 & 0.90 &  &   2.1 & $<$-5 & . & a &   34.8 & 4.9 \\
281 & 182024.00$-$160818.8 &  &  275.100040 & -16.138573 &  0.1 &  2.7 &  &    29.2 &   6.0 &   0.8 &     1.7 & 0.89 &  &   4.5 & $<$-5 & . & a &   36.2 & 0.9 \\
296 & 182024.39$-$160843.3 &  &  275.101640 & -16.145367 &  0.2 &  2.3 &  &    10.2 &   3.8 &   0.8 &     1.7 & 0.90 &  &   2.3 & $<$-5 & . & a &   38.4 & 1.3 \\
309 & 182024.60$-$161139.2 &  &  275.102519 & -16.194249 &  0.1 &  1.8 &  &    19.0 &   5.0 &   1.0 &    18.3 & 0.90 &  &   3.4 & $<$-5 & . & a &   38.2 & 3.3 \\
324 & 182024.87$-$161127.5 &  &  275.103634 & -16.190973 &  0.1 &  1.6 &  &    49.8 &   7.7 &   1.2 &    31.2 & 0.90 &  &   6.1 & $<$-5 & . & a &   38.2 & 2.7 \\
329 & 182024.94$-$161131.1 &  &  275.103949 & -16.191974 &  0.2 &  1.7 &  &    15.7 &   4.6 &   1.3 &    16.2 & 0.90 &  &   3.0 & $<$-5 & . & c &   38.2 & 4.4 \\
334 & 182025.07$-$161133.9 &  &  275.104499 & -16.192767 &  0.1 &  1.7 &  &    21.8 &   5.3 &   1.2 &    12.2 & 0.90 &  &   3.7 & $<$-5 & . & a &   38.2 & 2.5 \\
350 & 182025.50$-$161053.8 &  &  275.106253 & -16.181624 &  0.0 &  1.2 &  &   308.6 &  18.1 &   1.4 &   188.5 & 0.90 &  &  16.6 & $<$-5 & . & c &   38.6 & 2.5 \\
370 & 182025.93$-$161114.3 &  &  275.108062 & -16.187315 &  0.2 &  1.3 &  &     8.8 &   3.7 &   1.2 &     1.6 & 0.90 &  &   2.1 & $<$-5 & . & a &   38.6 & 1.5 \\
371 & 182025.93$-$160949.4 &  &  275.108070 & -16.163734 &  0.2 &  1.3 &  &     8.1 &   3.5 &   0.9 &     3.6 & 0.90 &  &   2.0 & $<$-5 & . & a &   38.9 & 1.8 \\
372 & 182025.95$-$160938.2 &  &  275.108153 & -16.160626 &  0.3 &  1.4 &  &     5.1 &   3.0 &   0.9 &     3.6 & 0.90 &  &   1.4 &  -3.6 & . & b &   38.8 & 2.5 \\
373 & 182026.00$-$161253.0 &  &  275.108352 & -16.214727 &  0.3 &  2.6 &  &     5.6 &   3.0 &   0.4 &     5.8 & 0.85 &  &   1.6 & $<$-5 & . & a &   37.8 & 4.0 \\
374 & 182026.01$-$161022.4 &  &  275.108377 & -16.172897 &  0.0 &  1.1 &  &   102.3 &  10.7 &   0.7 &    46.7 & 0.90 &  &   9.1 & $<$-5 & g & \nodata &   27.4 & 1.8 \\
375 & 182026.02$-$161240.0 &  &  275.108417 & -16.211132 &  0.2 &  2.4 &  &     8.4 &   3.5 &   0.6 &     4.6 & 0.89 &  &   2.0 & $<$-5 & . & a &   38.0 & 2.3 \\
376 & 182026.04$-$161104.6 &  &  275.108512 & -16.184620 &  0.1 &  1.2 &  &    78.3 &   9.5 &   1.7 &    45.3 & 0.90 &  &   7.8 & $<$-5 & . & a &   38.6 & 2.3 \\
377 & 182026.09$-$161039.9 &  &  275.108747 & -16.177772 &  0.2 &  1.0 &  &    13.9 &   4.4 &   1.1 &     5.5 & 0.90 &  &   2.8 & $<$-5 & . & a &   38.7 & 1.8 \\
378 & 182026.09$-$160301.3 &  &  275.108749 & -16.050381 &  0.7 &  7.5 &  &    11.5 &   4.5 &   4.5 &     0.1 & 0.90 &  &   2.3 &  -4.7 & . & a &   31.8 & 1.1 \\
379 & 182026.11$-$161053.5 &  &  275.108806 & -16.181547 &  0.1 &  1.1 &  &    21.0 &   5.3 &   2.0 &    10.1 & 0.90 &  &   3.6 & $<$-5 & . & a &   38.7 & 1.9 \\
396 & 182026.60$-$161055.7 &  &  275.110852 & -16.182141 &  0.0 &  1.0 &  &  2473.5 &  50.3 &   2.5 &  1622.0 & 0.90 &  &  48.7 & $<$-5 & . & c &   38.7 & 2.7 \\
398 & 182026.62$-$161136.5 &  &  275.110920 & -16.193480 &  0.1 &  1.4 &  &    18.2 &   4.9 &   0.8 &     7.5 & 0.90 &  &   3.3 & $<$-5 & . & a &   38.4 & 1.9 \\
399 & 182026.62$-$160822.9 &  &  275.110937 & -16.139700 &  0.3 &  2.3 &  &     6.2 &   3.2 &   0.8 &     1.7 & 0.90 &  &   1.6 &  -4.8 & . & a &   36.0 & 1.7 \\
433 & 182027.41$-$161331.0 &  &  275.114226 & -16.225280 &  0.0 &  3.1 &  &   263.3 &  16.8 &   0.7 &    18.6 & 0.90 &  &  15.2 & $<$-5 & . & a &   37.5 & 1.3 \\
466 & 182028.15$-$161049.3 &  &  275.117308 & -16.180378 &  0.0 &  0.6 &  &   538.1 &  23.7 &   1.9 &   386.2 & 0.90 &  &  22.2 & $<$-5 & . & a &   38.9 & 2.7 \\
488 & 182028.65$-$161211.6 &  &  275.119414 & -16.203233 &  0.1 &  1.7 &  &    70.6 &   8.9 &   0.4 &    35.8 & 0.89 &  &   7.5 & $<$-5 & . & a &   38.5 & 2.0 \\
495 & 182028.86$-$161042.0 &  &  275.120280 & -16.178337 &  0.2 &  0.4 &  &     8.3 &   3.9 &   2.7 &     0.0 & 0.90 &  &   1.9 &  -4.0 & . & a &   39.0 & 1.6 \\
510 & 182029.21$-$161425.6 &  &  275.121709 & -16.240461 &  0.4 &  3.9 &  &     5.2 &   3.0 &   0.8 &     0.5 & 0.90 &  &   1.4 &  -3.8 & . & a &   36.7 & 1.2 \\
511 & 182029.28$-$161041.4 &  &  275.122029 & -16.178180 &  0.1 &  0.3 &  &    53.0 &   8.2 &   6.0 &    29.7 & 0.90 &  &   6.1 & $<$-5 & . & a &   39.1 & 2.2 \\
512 & 182029.31$-$161047.9 &  &  275.122151 & -16.179989 &  0.1 &  0.4 &  &    77.4 &   9.5 &   3.6 &    50.4 & 0.83 &  &   7.7 & $<$-5 & . & a &   39.1 & 2.5 \\
513 & 182029.32$-$161045.2 &  &  275.122167 & -16.179236 &  0.1 &  0.4 &  &    16.8 &   5.0 &   3.2 &    11.5 & 0.79 &  &   3.0 & $<$-5 & . & a &   39.1 & 2.4 \\
514 & 182029.36$-$160919.2 &  &  275.122353 & -16.155335 &  0.2 &  1.2 &  &     6.1 &   3.2 &   0.9 &     2.6 & 0.91 &  &   1.6 &  -4.4 & . & a &   39.1 & 1.8 \\
515 & 182029.39$-$160943.0 &  &  275.122474 & -16.161946 &  0.0 &  0.8 &  &   155.2 &  13.0 &   0.8 &   102.6 & 0.90 &  &  11.5 & $<$-5 & . & a &   39.3 & 2.5 \\
516 & 182029.39$-$161046.5 &  &  275.122489 & -16.179593 &  0.0 &  0.4 &  &    69.5 &   9.1 &   3.5 &    53.4 & 0.79 &  &   7.2 & $<$-5 & . & a &   39.1 & 3.0 \\
517 & 182029.39$-$161212.9 &  &  275.122489 & -16.203593 &  0.2 &  1.7 &  &     8.5 &   3.5 &   0.5 &     1.8 & 0.89 &  &   2.1 & $<$-5 & . & a &   38.5 & 1.4 \\
518 & 182029.43$-$161050.2 &  &  275.122661 & -16.180616 &  0.1 &  0.4 &  &    47.0 &   7.8 &   6.0 &    29.9 & 0.90 &  &   5.6 & $<$-5 & . & a &   39.1 & 2.7 \\
536 & 182029.81$-$161045.6 &  &  275.124213 & -16.179352 &  0.0 &  0.3 &  &  1912.2 &  44.3 &   6.8 &  1052.5 & 0.80 &  &  42.7 & $<$-5 & . & a &   39.1 & 2.2 \\
543 & 182029.89$-$161044.5 &  &  275.124574 & -16.179031 &  0.0 &  0.3 &  &  3865.9 &  62.7 &   8.1 &  2447.0 & 0.80 &  &  61.1 & $<$-5 & . & a &   39.1 & 2.6 \\
567 & 182030.23$-$161034.9 &  &  275.125982 & -16.176382 &  0.1 &  0.1 &  &    43.6 &   7.5 &   4.4 &     9.4 & 0.90 &  &   5.5 & $<$-5 & . & a &   39.2 & 1.7 \\
574 & 182030.44$-$161053.1 &  &  275.126836 & -16.181427 &  0.1 &  0.4 &  &    89.5 &  10.3 &   5.5 &    22.3 & 0.90 &  &   8.3 & $<$-5 & . & a &   39.2 & 1.7 \\
600 & 182030.95$-$161039.4 &  &  275.128970 & -16.177620 &  0.2 &  0.2 &  &     8.0 &   3.7 &   2.0 &     5.0 & 0.90 &  &   1.9 &  -4.3 & . & a &   39.2 & 2.3 \\
618 & 182031.35$-$160228.4 &  &  275.130660 & -16.041249 &  0.4 &  8.0 &  &    64.4 &   9.3 &  11.6 &    38.7 & 0.75 &  &   6.5 & $<$-5 & . & a &   29.3 & 2.3 \\
647 & 182031.84$-$161138.1 &  &  275.132701 & -16.193943 &  0.3 &  1.2 &  &     3.5 &   2.5 &   0.5 &     0.8 & 0.90 &  &   1.1 &  -2.9 & g & \nodata &   27.5 & 1.6 \\
650 & 182031.89$-$161616.5 &  &  275.132883 & -16.271262 &  0.6 &  5.8 &  &    10.0 &   3.8 &   1.0 &     3.5 & 0.91 &  &   2.3 & $<$-5 & g & \nodata &   18.0 & 1.7 \\
655 & 182031.97$-$161030.5 &  &  275.133240 & -16.175147 &  0.0 &  0.4 &  &   249.3 &  16.3 &   0.7 &   208.6 & 0.90 &  &  14.8 & $<$-5 & g & \nodata &   25.8 & 3.4 \\
706 & 182034.56$-$161523.6 &  &  275.144010 & -16.256565 &  0.5 &  5.0 &  &     9.0 &   3.7 &   1.0 &     9.4 & 0.89 &  &   2.1 & $<$-5 & . & a &   33.6 & 5.0 \\
726 & 182035.63$-$161055.5 &  &  275.148459 & -16.182086 &  0.2 &  1.3 &  &    11.2 &   4.0 &   0.8 &     1.8 & 0.90 &  &   2.5 & $<$-5 & . & a &   39.0 & 1.4 \\
854 & 182053.86$-$160306.5 &  &  275.224439 & -16.051830 &  0.4 &  9.3 &  &   105.5 &  11.3 &   9.5 &     6.0 & 0.91 &  &   8.9 & $<$-5 & . & a &   29.3 & 1.0 \\
\enddata

\tablecomments{Table~\ref{tbl:src_properties_main} is published in 
its entirety in the electronic edition of the {\it Astrophysical 
Journal}.  Interesting sources mentioned in the text are shown here for guidance regarding its form 
and content. }

\tablecomments{{\bf Column 1:} X-ray catalog sequence number, sorted by RA.
{\bf Column 2:} IAU designation.
{\bf Columns 3,4:} Right ascension and declination for epoch J2000.0.
{\bf Column 5:} Estimated standard deviation of the random component of the position error, $\sqrt{\sigma_x^2 + \sigma_y^2}$.  The single-axis position errors, $\sigma_x$ and $\sigma_y$, are estimated from the single-axis standard deviations of the PSF inside the extraction region and the number of counts extracted.
{\bf Column 6:} Off-axis angle.
{\bf Columns 7,8:} Net counts extracted in the total energy band (0.5--8~keV); average of the upper and lower $1\sigma$ errors on column 7.
{\bf Column 9:} Background counts extracted (total band).
{\bf Column 10:} Net counts extracted in the hard energy band (2--8~keV).
{\bf Column 11:} Fraction of the PSF (at 1.497 keV) enclosed within the extraction region.  Note that a reduced PSF fraction (significantly below 90\%) may indicate that the source is in a crowded region. 
{\bf Column 12:} Photometric significance computed as $\frac{\mbox{net counts}}{\mbox{upper error on net counts}}$. 
{\bf Column 13:} Log probability that extracted counts (total band) are solely from background.  Some sources have $P_B$ values above the 1\% threshold that defines the catalog because local background estimates can rise during the final extraction iteration after sources are removed from the catalog.
{\bf Column 14:}  Source anomalies:  g = fractional time that source was on a detector (FRACEXPO from {\em mkarf}) is $<0.9$ ; e = source on field edge; p = source piled up; s = source on readout streak.
{\bf Column 15:} Variability characterization based on K-S statistic (total band):  a = no evidence for variability ($0.05<P_{KS}$); b = possibly variable ($0.005<P_{KS}<0.05$); c = definitely variable ($P_{KS}<0.005$).  No value is reported for sources with fewer than 4 counts or for sources in chip gaps or on field edges.
{\bf Column 16:} Effective exposure time: approximate time the source would have to be observed on-axis (no telescope vignetting) on a nominal region of the detector (no dithering over insensitive regions of the detector) to obtain the reported number of counts. 
{\bf Column 17:} Background-corrected median photon energy (total band).}
\end{deluxetable}

\clearpage

\begin{deluxetable}{rcrrrrrrrrrrrrrccccc}
\centering \rotate \tabletypesize{\tiny} \tablewidth{0pt} 
\tablecolumns{20}

\tablecaption{ Tentative {\em Chandra} Catalog:  Basic Source 
Properties \label{tbl:src_properties_tentative}}

\tablehead{ \multicolumn{2}{c}{Source} &
  &
\multicolumn{4}{c}{Position} &
  &
\multicolumn{5}{c}{Extracted Counts} &
  &
\multicolumn{6}{c}{Characteristics} \\
\cline{1-2} \cline{4-7} \cline{9-13} \cline{15-20}

\colhead{Seq} & \colhead{CXOU J} &
  & 
\colhead{$\alpha_{\rm J2000}$} & \colhead{$\delta_{\rm J2000}$} & 
\colhead{Err} & \colhead{$\theta$} &
  & 
\colhead{Net} & \colhead{$\Delta$Net} & \colhead{Bkgd} & 
\colhead{Net} & \colhead{PSF} &   
  &
\colhead{Signif} & \colhead{$\log P_B$} & \colhead{Anom} & \colhead{Var} &\colhead{EffExp} & \colhead{$E_{median}$}  \\

\colhead{\#} & \colhead{} &
  & 
\colhead{(deg)} & \colhead{(deg)} & \colhead{(\arcsec)} & 
\colhead{(\arcmin)} &
  & 
\colhead{Full} & \colhead{Full} & \colhead{Full} & \colhead{Hard} & 
\colhead{Frac} &
  &
\colhead{} & \colhead{} & \colhead{} & \colhead{} & \colhead{(ks)} & 
\colhead{(keV)}
 \\

\colhead{(1)} & \colhead{(2)} &
  & 
\colhead{(3)} & \colhead{(4)} & \colhead{(5)} & \colhead{(6)} &
  & 
\colhead{(7)} & \colhead{(8)} & \colhead{(9)} & \colhead{(10)} & 
\colhead{(11)} &
  & 
\colhead{(12)} & \colhead{(13)} & \colhead{(14)} & \colhead{(15)} & 
\colhead{(16)} & \colhead{(17)}  }

\startdata
322 & 182024.83$-$161135.3 &  &  275.103482 & -16.193161 &  0.5 &  1.7 &  &     2.9 &   2.5 &   1.1 &     3.3 & 0.90 &  &   0.9 & -1.6 & . & a &   38.2 & 4.7 \\
519 & 182029.48$-$161644.0 &  &  275.122866 & -16.278895 &  0.7 &  6.2 &  &     5.6 &   3.4 &   2.4 &     0.6 & 0.90 &  &   1.4 & -2.5 & . & a &   33.7 & 1.5 \\
\enddata

\tablecomments{Table \ref{tbl:src_properties_tentative} has the same columns as \ref{tbl:src_properties_main} and is published in its entirety in the 
electronic edition of the {\it Astrophysical Journal}.  Interesting sources mentioned in the text are shown here for guidance regarding its form 
and content. 
}

\tablecomments{See notes for Table~\ref{tbl:src_properties_main}}

\end{deluxetable}

\begin{deluxetable}{rcrrrcccrcccccrc}
\centering \rotate \tabletypesize{\scriptsize} \tablewidth{0pt} 
\tablecolumns{16}

\tablecaption{X-ray Spectroscopy for Photometrically Selected 
Sources:  Thermal Plasma Fits \label{tbl:thermal_spectroscopy}}

\tablehead{ \multicolumn{4}{c}{Source\tablenotemark{a}} &  &
\multicolumn{3}{c}{Spectral Fit\tablenotemark{b}} &  &
\multicolumn{5}{c}{X-ray Luminosities\tablenotemark{c}} &  &
\colhead{Notes\tablenotemark{d}} \\ \cline{1-4} \cline{6-8} 
\cline{10-14}

\colhead{Seq} & \colhead{CXOU J} & \colhead{Net} & \colhead{Signif} &  
& \colhead{$\log N_H$} & \colhead{$kT$} & \colhead{$\log EM$} &    &
\colhead{$\log L_s$} & \colhead{$\log L_h$} & \colhead{$\log 
L_{h,c}$} & \colhead{$\log L_t$} & \colhead{$\log L_{t,c}$} &  &
\colhead{}  \\

\colhead{\#} & \colhead{} & \colhead{Counts} & \colhead{} &  &
\colhead{(cm$^{-2}$)} & \colhead{(keV)} & \colhead{(cm$^{-3}$)} &   &
\multicolumn{5}{c}{(ergs s$^{-1}$)} &  &
\colhead{} \\

\colhead{(1)} & \colhead{(2)} & \colhead{(3)} & \colhead{(4)} &  &
\colhead{(5)} & \colhead{(6)} & \colhead{(7)} &  & \colhead{(8)} & 
\colhead{(9)} & \colhead{(10)} &\colhead{(11)} & \colhead{(12)} &  &
\colhead{(13)} }

\startdata
37 & 182009.70$-$161503.7 &     9.7 &   2.1 &  &   23.7  &  1.4  &  55.7  &  &   \nodata & 30.55 & 31.99 & 30.55 &   \nodata &  & \nodata \\
51 & 182012.31$-$160447.7 &    16.4 &   2.9 &  &  {\tiny $-0.3$} 22.3 {\tiny $+0.2$} & {\em  2.0 } & \phantom{{\tiny $+0.2$}} 53.9 {\tiny $+0.2$} &  & 29.54 & 30.27 & 30.41 & 30.35 & 30.85 &  & \nodata \\
123 & 182019.41$-$161329.1 &    30.8 &   4.6 &  &  {\tiny $-0.3$} 22.5 {\tiny $+0.2$} & {\tiny $-1.7$} 3.4 \phantom{{\tiny $-1.7$}} & {\tiny $-0.4$} 54.2 {\tiny $+0.5$} &  & 29.68 & 30.86 & 31.01 & 30.89 & 31.30 &  & \nodata \\
128 & 182019.60$-$161038.8 &     8.6 &   2.1 &  &  {\tiny $-0.5$} 23.2 {\tiny $+0.3$} &  13.6  &  53.8  &  &   \nodata & 30.45 & 30.82 & 30.45 &   \nodata &  & \nodata \\
135 & 182019.88$-$161052.3 &     8.5 &   2.1 &  &   23.8  &  2.0  &  55.3  &  &   \nodata & 30.54 & 31.85 & 30.54 &   \nodata &  & \nodata \\
177 & 182021.21$-$161222.9 &    16.4 &   3.1 &  &  {\tiny $-0.6$} 22.3 {\tiny $+0.4$} & {\tiny $-2.3$} 3.6 \phantom{{\tiny $-2.3$}} & \phantom{{\tiny $+0.7$}} 53.6 {\tiny $+0.7$} &  & 29.40 & 30.32 & 30.42 & 30.37 & 30.71 &  & \nodata \\
182 & 182021.32$-$161140.2 &    11.5 &   2.5 &  &  {\tiny $-1.1$} 21.9 {\tiny $+0.3$} & {\tiny $-0.4$} 0.7 {\tiny $+0.9$} & \phantom{{\tiny $+0.9$}} 53.5 {\tiny $+0.9$} &  & 29.66 & 29.06 & 29.14 & 29.76 & 30.47 &  & \nodata \\
186 & 182021.43$-$160939.1 &    30.3 &   4.6 &  &  {\tiny $-0.6$} 22.1 {\tiny $+0.4$} & {\tiny $-1.7$} 3.0 \phantom{{\tiny $-1.7$}} & {\tiny $-0.4$} 53.8 {\tiny $+0.6$} &  & 29.81 & 30.42 & 30.48 & 30.52 & 30.81 &  & \nodata \\
187 & 182021.43$-$161206.8 &   363.4 &  18.1 &  &  {\tiny $-0.08$} 22.4 {\tiny $+0.07$} & {\tiny $-1.2$} 4.4 {\tiny $+2.6$} & {\tiny $-0.1$} 55.0 {\tiny $+0.1$} &  & 30.56 & 31.76 & 31.89 & 31.79 & 32.14 &  & \nodata \\
230 & 182022.58$-$160251.3 &   365.4 &  18.0 &  &  {\tiny $-0.10$} 22.3 {\tiny $+0.09$} & {\tiny $-1.5$} 4.6 {\tiny $+3.3$} & {\tiny $-0.1$} 55.0 {\tiny $+0.1$} &  & 30.77 & 31.74 & 31.83 & 31.79 & 32.08 &  & \nodata \\
233 & 182022.69$-$160833.9 &    73.1 &   7.6 &  &  {\tiny $-0.4$} 21.7 {\tiny $+0.2$} & {\tiny $-1.5$} 3.5 {\tiny $+10.7$} & {\tiny $-0.2$} 53.9 {\tiny $+0.2$} &  & 30.29 & 30.69 & 30.72 & 30.84 & 31.01 &  & \nodata \\
236 & 182022.74$-$161123.5 &    60.3 &   6.8 &  &  {\tiny $-0.2$} 22.6 {\tiny $+0.2$} & {\tiny $-0.5$} 1.6 {\tiny $+1.8$} & {\tiny $-0.5$} 54.7 {\tiny $+0.4$} &  &   \nodata & 30.86 & 31.12 & 30.90 &   \nodata &  & \nodata \\
239 & 182022.87$-$161148.5 &    19.3 &   3.5 &  &  {\tiny $-0.2$} 23.5 {\tiny $+0.2$} & {\em  2.0 } & {\tiny $-0.3$} 55.2 {\tiny $+0.4$} &  &   \nodata & 30.78 & 31.69 & 30.78 &   \nodata &  & \nodata \\
244 & 182022.93$-$161152.7 &   134.4 &  10.6 &  &  {\tiny $-0.2$} 22.2 {\tiny $+0.1$} & {\tiny $-1.1$} 3.3 {\tiny $+7.5$} & {\tiny $-0.3$} 54.6 {\tiny $+0.2$} &  & 30.42 & 31.24 & 31.33 & 31.30 & 31.63 &  & \nodata \\
246 & 182022.97$-$161131.8 &    12.4 &   2.6 &  &  {\tiny $-0.3$} 23.5 {\tiny $+0.3$} &  2.6  & \phantom{{\tiny $+1.6$}} 54.9 {\tiny $+1.6$} &  &   \nodata & 30.72 & 31.56 & 30.72 &   \nodata &  & \nodata \\
255 & 182023.16$-$161305.7 &    54.4 &   6.4 &  &  {\tiny $-0.3$} 23.5 {\tiny $+0.2$} & {\tiny $-0.7$} 1.6 {\tiny $+8.5$} &  55.8  &  &   \nodata & 31.20 & 32.18 & 31.20 &   \nodata &  & \nodata \\
275 & 182023.80$-$161325.7 &     8.5 &   2.1 &  &  {\tiny $-0.4$} 23.1 {\tiny $+0.3$} &  13.7  &  53.8  &  &   \nodata & 30.44 & 30.79 & 30.44 &   \nodata &  & \nodata \\
281 & 182024.00$-$160818.8 &    29.2 &   4.5 &  &  {\em  20.0 } & {\tiny $-0.3$} 0.6 {\tiny $+0.1$} & {\tiny $-0.2$} 53.3 {\tiny $+0.2$} &  & 30.21 & 28.80 & 28.80 & 30.22 & 30.24 &  & \nodata \\
296 & 182024.39$-$160843.3 &    10.2 &   2.3 &  &  \phantom{{\tiny $+0.4$}} 21.6 {\tiny $+0.4$} & {\em  2.0 } & {\tiny $-0.4$} 53.1 {\tiny $+0.3$} &  & 29.52 & 29.58 & 29.60 & 29.85 & 30.04 &  & \nodata \\
309 & 182024.60$-$161139.2 &    19.0 &   3.4 &  &  {\tiny $-0.3$} 22.8 {\tiny $+0.4$} & {\tiny $-1.1$} 1.7 {\tiny $+3.4$} & \phantom{{\tiny $+1.7$}} 54.5 {\tiny $+1.7$} &  &   \nodata & 30.50 & 30.88 & 30.51 &   \nodata &  & \nodata \\
324 & 182024.87$-$161127.5 &    49.8 &   6.1 &  &  {\tiny $-0.2$} 22.4 {\tiny $+0.2$} & {\tiny $-1.0$} 2.5 {\tiny $+4.0$} & {\tiny $-0.3$} 54.2 {\tiny $+0.4$} &  & 29.88 & 30.74 & 30.87 & 30.80 & 31.24 &  & \nodata \\
329 & 182024.94$-$161131.1 &    15.7 &   3.0 &  &   23.1  & {\em  15.0 } &  54.0  &  &   \nodata & 30.66 & 31.00 & 30.66 &   \nodata &  & \nodata \\
334 & 182025.07$-$161133.9 &    21.8 &   3.7 &  &  {\tiny $-0.3$} 22.4 {\tiny $+0.3$} & {\tiny $-0.8$} 1.7 {\tiny $+2.9$} &  54.1  &  & 29.61 & 30.30 & 30.47 & 30.38 & 30.99 &  & \nodata \\
350 & 182025.50$-$161053.8 &   308.6 &  16.6 &  &  {\tiny $-0.12$} 22.1 {\tiny $+0.10$} & {\tiny $-1.6$} 5.0 {\tiny $+5.6$} & {\tiny $-0.1$} 54.7 {\tiny $+0.1$} &  & 30.72 & 31.55 & 31.62 & 31.61 & 31.85 &  & \nodata \\
370 & 182025.93$-$161114.3 &     8.8 &   2.1 &  &   21.8  &  2.4  &  53.0  &  & 29.28 & 29.60 & 29.64 & 29.77 & 30.03 &  & \nodata \\
374 & 182026.01$-$161022.4 &   102.3 &   9.1 &  &  {\tiny $-0.3$} 22.0 {\tiny $+0.2$} & {\tiny $-1.3$} 3.3 {\tiny $+7.3$} & {\tiny $-0.2$} 54.4 {\tiny $+0.2$} &  & 30.49 & 31.09 & 31.14 & 31.19 & 31.45 &  & \nodata \\
375 & 182026.02$-$161240.0 &     8.4 &   2.0 &  &  {\tiny $-0.6$} 22.6 {\tiny $+0.3$} & {\tiny $-0.7$} 1.3 \phantom{{\tiny $-0.7$}} & \phantom{{\tiny $+1.4$}} 54.1 {\tiny $+1.4$} &  &   \nodata & 29.98 & 30.29 & 30.03 &   \nodata &  & \nodata \\
376 & 182026.04$-$161104.6 &    78.3 &   7.8 &  &  {\tiny $-0.2$} 22.3 {\tiny $+0.3$} & {\tiny $-1.1$} 1.9 {\tiny $+1.6$} & {\tiny $-0.3$} 54.4 {\tiny $+0.7$} &  & 30.20 & 30.81 & 30.93 & 30.91 & 31.38 &  & \nodata \\
377 & 182026.09$-$161039.9 &    13.9 &   2.8 &  &  {\tiny $-0.6$} 22.0 {\tiny $+0.5$} & {\tiny $-1.6$} 2.2 \phantom{{\tiny $-1.6$}} & \phantom{{\tiny $+1.2$}} 53.4 {\tiny $+1.2$} &  & 29.51 & 29.91 & 29.97 & 30.06 & 30.39 &  & \nodata \\
378 & 182026.09$-$160301.3 &    11.5 &   2.3 &  &  \phantom{{\tiny $+1.6$}} 20.4 {\tiny $+1.6$} &  1.4  & {\tiny $-0.3$} 53.1 {\tiny $+0.5$} &  & 29.84 & 29.36 & 29.36 & 29.96 & 29.98 &  & \nodata \\
379 & 182026.11$-$161053.5 &    21.0 &   3.6 &  &  {\tiny $-0.4$} 22.0 {\tiny $+0.3$} & {\tiny $-0.7$} 1.8 {\tiny $+3.1$} & {\tiny $-0.4$} 53.7 {\tiny $+0.4$} &  & 29.73 & 30.07 & 30.14 & 30.23 & 30.62 &  & \nodata \\
396 & 182026.60$-$161055.7 &  2473.5 &  48.7 &  &  {\tiny $-0.04$} 22.1 {\tiny $+0.03$} & {\tiny $-4.1$} 14.5 \phantom{{\tiny $-4.1$}} & {\tiny $-0.02$} 55.6 {\tiny $+0.02$} &  & 31.56 & 32.54 & 32.59 & 32.58 & 32.75 &  & \nodata \\
398 & 182026.62$-$161136.5 &    18.2 &   3.3 &  &  {\tiny $-0.2$} 22.5 {\tiny $+0.2$} & {\tiny $-0.3$} 0.7 {\tiny $+0.5$} & \phantom{{\tiny $+1.0$}} 54.6 {\tiny $+1.0$} &  &   \nodata & 29.99 & 30.30 & 30.17 &   \nodata &  & \nodata \\
433 & 182027.41$-$161331.0 &   263.3 &  15.2 &  &  {\tiny $-0.07$} 22.0 {\tiny $+0.08$} & {\tiny $-0.11$} 0.6 {\tiny $+0.08$} & {\tiny $-0.2$} 55.1 {\tiny $+0.3$} &  & 31.02 & 30.32 & 30.43 & 31.10 & 32.06 &  & \nodata \\
466 & 182028.15$-$161049.3 &   538.1 &  22.2 &  &  {\tiny $-0.10$} 22.3 {\tiny $+0.07$} & {\tiny $-1.4$} 4.8 {\tiny $+5.2$} & {\tiny $-0.1$} 55.1 {\tiny $+0.1$} &  & 30.85 & 31.86 & 31.95 & 31.91 & 32.19 &  & \nodata \\
488 & 182028.65$-$161211.6 &    70.6 &   7.5 &  &  {\tiny $-0.3$} 22.2 {\tiny $+0.2$} & {\tiny $-0.8$} 2.2 {\tiny $+2.7$} & {\tiny $-0.3$} 54.3 {\tiny $+0.3$} &  & 30.15 & 30.78 & 30.89 & 30.87 & 31.30 &  & \nodata \\
511 & 182029.28$-$161041.4 &    53.0 &   6.1 &  &   22.6  & \phantom{{\tiny $+0.5$}} 0.9 {\tiny $+0.5$} & {\tiny $-0.6$} 55.0 \phantom{{\tiny $-0.6$}} &  &   \nodata & 30.55 & 30.89 & 30.65 &   \nodata &  & \nodata \\
512 & 182029.31$-$161047.9 &    77.4 &   7.7 &  &  {\tiny $-0.2$} 22.5 {\tiny $+0.2$} & {\tiny $-0.5$} 1.5 {\tiny $+1.0$} & {\tiny $-0.4$} 54.8 {\tiny $+0.5$} &  &   \nodata & 30.95 & 31.19 & 31.00 &   \nodata &  & \nodata \\
513 & 182029.32$-$161045.2 &    16.8 &   3.0 &  &  {\tiny $-0.2$} 22.5 {\tiny $+0.2$} & {\em  2.0 } & {\tiny $-0.3$} 54.0 {\tiny $+0.2$} &  & 29.43 & 30.34 & 30.52 & 30.39 & 30.96 &  & \nodata \\
515 & 182029.39$-$160943.0 &   155.2 &  11.5 &  &  {\tiny $-0.2$} 22.3 {\tiny $+0.1$} & {\tiny $-0.8$} 2.5 {\tiny $+2.9$} & {\tiny $-0.3$} 54.7 \phantom{{\tiny $-0.3$}} &  & 30.39 & 31.23 & 31.36 & 31.29 & 31.72 &  & \nodata \\
516 & 182029.39$-$161046.5 &    69.5 &   7.2 &  &  {\tiny $-0.2$} 22.5 {\tiny $+0.2$} & {\tiny $-1.5$} 3.0 {\tiny $+7.7$} & {\tiny $-0.3$} 54.5 {\tiny $+0.5$} &  & 29.95 & 31.04 & 31.18 & 31.07 & 31.51 &  & \nodata \\
517 & 182029.39$-$161212.9 &     8.5 &   2.1 &  &  {\tiny $-0.7$} 22.2 {\tiny $+0.3$} &  0.4  &  54.3  &  & 29.48 & 28.94 & 29.14 & 29.59 & 31.13 &  & \nodata \\
518 & 182029.43$-$161050.2 &    47.0 &   5.6 &  &   22.5  &  1.5  & \phantom{{\tiny $+0.1$}} 54.5 {\tiny $+0.1$} &  &   \nodata & 30.66 & 30.90 & 30.72 &   \nodata &  & \nodata \\
536 & 182029.81$-$161045.6 &  1912.2 &  42.7 &  &  {\tiny $-0.03$} 22.4 {\tiny $+0.03$} & {\em  15.0 } & {\tiny $-0.04$} 55.3 {\tiny $+0.04$} &  & 31.63 & 32.37 & 32.48 & 32.44 & 33.16 &  & 2T (0.9 keV) \\
543 & 182029.89$-$161044.5 &  3865.9 &  61.1 &  &  {\tiny $-0.05$} 22.3 {\tiny $+0.04$} & {\tiny $-1.9$} 10.4 {\tiny $+3.1$} & {\tiny $-0.02$} 55.7 {\tiny $+0.02$} &  & 31.85 & 32.78 & 32.85 & 32.83 & 33.28 &  & 2T (0.7 keV)\\
567 & 182030.23$-$161034.9 &    43.6 &   5.5 &  &  {\tiny $-0.1$} 22.3 {\tiny $+0.2$} & {\tiny $-0.3$} 0.6 {\tiny $+0.3$} & {\tiny $-0.4$} 54.8 {\tiny $+1.2$} &  & 30.18 & 30.05 & 30.26 & 30.42 & 31.72 &  & \nodata \\
574 & 182030.44$-$161053.1 &    89.5 &   8.3 &  &   22.4  & \phantom{{\tiny $+0.2$}} 0.6 {\tiny $+0.2$} & {\tiny $-0.4$} 55.3 \phantom{{\tiny $-0.4$}} &  & 30.47 & 30.42 & 30.69 & 30.75 & 32.25 &  & \nodata \\
618 & 182031.35$-$160228.4 &    64.4 &   6.5 &  &   22.5  &  1.3  & {\tiny $-0.59$} 54.9 {\tiny $+0.08$} &  & 30.25 & 30.86 & 31.09 & 30.96 & 31.81 &  & \nodata \\
650 & 182031.89$-$161616.5 &    10.0 &   2.3 &  &   20.6  &  14.9  &  53.0  &  & 29.69 & 30.07 & 30.08 & 30.23 & 30.24 &  & \nodata \\
655 & 182031.97$-$161030.5 &   249.3 &  14.8 &  &  {\tiny $-0.14$} 22.5 {\tiny $+0.09$} & {\tiny $-4.0$} 8.9 \phantom{{\tiny $-4.0$}} & {\tiny $-0.08$} 54.9 {\tiny $+0.13$} &  & 30.46 & 31.83 & 31.95 & 31.85 & 32.13 &  & \nodata \\
706 & 182034.56$-$161523.6 &     9.0 &   2.1 &  &  {\tiny $-0.6$} 23.3 {\tiny $+0.3$} & {\em  15.0 } & {\tiny $-0.6$} 54.0 {\tiny $+0.4$} &  &   \nodata & 30.56 & 31.02 & 30.56 &   \nodata &  & \nodata \\
726 & 182035.63$-$161055.5 &    11.2 &   2.5 &  &  {\tiny $-0.8$} 21.8 {\tiny $+0.3$} & {\em  2.0 } & {\tiny $-0.3$} 53.2 {\tiny $+0.3$} &  & 29.51 & 29.71 & 29.75 & 29.92 & 30.19 &  & \nodata \\
854 & 182053.86$-$160306.5 &   105.5 &   8.9 &  &  \phantom{{\tiny $+0.2$}} 21.0 {\tiny $+0.2$} & \phantom{{\tiny $+0.1$}} 0.6 {\tiny $+0.1$} &  54.0  &  & 30.83 & 29.45 & 29.47 & 30.85 & 30.99 &  & \nodata \\
\enddata

\tablecomments{Table~\ref{tbl:thermal_spectroscopy} is published in 
its entirety in the electronic edition of the {\it Astrophysical 
Journal}.  Interesting sources mentioned in the text are shown here for guidance regarding its form 
and content.}

\tablenotetext{a}{ For convenience, {\bf columns 1--4} reproduce the 
source identification, net counts, and photometric significance data 
from Table~\ref{tbl:src_properties_main}. }

\tablenotetext{b}{ All fits used the ``wabs(apec)'' model in {\it 
XSPEC} and assumed 0.3$Z_{\odot}$ abundances \citep{Imanishi01,Feigelson02}.  {\bf Columns 5 and 6} 
present the best-fit values for the column density and plasma 
temperature parameters. {\bf Column 7} presents the emission measure 
for the model spectrum, assuming a distance of 1.6~kpc. {\it 
Quantities in italics} were frozen in the fit.  Uncertainties 
represent 90\% confidence intervals. More significant digits are used 
for uncertainties $<0.1$ in order to avoid large rounding errors; for 
consistency, the same number of significant digits is used for both 
lower and upper uncertainties. Uncertainties are missing when {\it 
XSPEC} was unable to compute them or when their values were so large 
that the parameter is effectively unconstrained.  
Fits lacking uncertainties, fits with large uncertainties, and fits with frozen parameters should be viewed merely as splines to the data to obtain rough estimates of luminosities; the listed parameter values are unreliable.  
  }

\tablenotetext{c}{ X-ray luminosities are presented in {\bf columns 
8--12}:  s = soft band (0.5--2 keV); h = hard band (2--8 keV); t = 
total band (0.5--8 keV).  Absorption-corrected luminosities are 
subscripted with a $c$; they are omitted when $\log N_H > 22.5$~cm$^{-2}$ since 
the soft band emission is essentially unmeasurable.  }

\tablenotetext{d}{ {\bf 2T} means a two-temperature model was used; 
the second temperature is shown in parentheses.  See spectra in Figure~\ref{OB_spec.fig}. 
 } 

\end{deluxetable}

\clearpage

\begin{deluxetable}{rcrrrcccrccccc}
\centering \rotate 
\tabletypesize{\scriptsize} \tablewidth{0pt}
\tablecolumns{14}

\tablecaption{X-ray Spectroscopy for Photometrically Selected Sources:  Power Law Fits 
\label{tbl:powerlaw_spectroscopy}}

\tablehead{
\multicolumn{4}{c}{Source\tablenotemark{a}} &
  &
\multicolumn{3}{c}{Spectral Fit\tablenotemark{b}} &
  &
\multicolumn{5}{c}{X-ray Fluxes\tablenotemark{c}}  \\ 
\cline{1-4} \cline{6-8} \cline{10-14}

\colhead{Seq} & \colhead{CXOU J} & \colhead{Net} & \colhead{Signif} &
  &
\colhead{$\log N_H$} & \colhead{$\Gamma$} & \colhead{$\log N_{\Gamma}$} &  
  &
\colhead{$\log L_s$} & \colhead{$\log L_h$} & \colhead{$\log L_{h,c}$} & \colhead{$\log L_t$} & \colhead{$\log L_{t,c}$} \\

\colhead{\#} & \colhead{} & \colhead{Counts} & \colhead{} &
  &
\colhead{(cm$^{-2}$)} & \colhead{} & \colhead{} & 
  &
\multicolumn{5}{c}{(photons cm$^{-2}$ s$^{-1}$)} \\

\colhead{(1)} & \colhead{(2)} & \colhead{(3)} & \colhead{(4)} &
  &
\colhead{(5)} & \colhead{(6)} & \colhead{(7)} & 
  &
\colhead{(8)} & \colhead{(9)} &\colhead{(10)} & \colhead{(11)} & \colhead{(12)}
}

\startdata
2 & 181957.67$-$161328.8 &    12.5 &   2.3 &  &   23.1  &  1.1  &  -6.3  &  &    \nodata &  29.82 &  30.10 &  29.82 &    \nodata  \\
5 & 181959.04$-$160350.2 &    55.4 &   5.7 &  &   21.9  &  1.6  &  -5.2  &  &  30.10 &  30.88 &  30.91 &  30.95 &  31.10  \\
13 & 182004.28$-$161128.9 &    19.8 &   3.4 &  &   21.6  & {\tiny $-1.3$} 2.1 \phantom{{\tiny $-1.3$}} &  -5.6  &  &  29.78 &  30.12 &  30.14 &  30.28 &  30.46  \\
40 & 182010.53$-$161401.5 &    24.1 &   3.9 &  &  {\tiny $-1.0$} 21.9 {\tiny $+0.4$} & {\tiny $-1.0$} 1.5 \phantom{{\tiny $-1.0$}} &  -5.6  &  &  29.70 &  30.49 &  30.52 &  30.56 &  30.71  \\
59 & 182013.22$-$160856.5 &    10.3 &   2.3 &  &   22.2  &  0.1  &  -6.4  &  &  28.95 &  30.64 &  30.67 &  30.65 &  30.70  \\
69 & 182015.48$-$160706.0 &    10.1 &   2.3 &  &   22.8  &  5.4  &  -3.6  &  &    \nodata &  29.97 &  30.54 &  30.02 &    \nodata  \\
96 & 182017.83$-$161014.5 &    12.8 &   2.7 &  &  {\tiny $-1.4$} 22.8 {\tiny $+0.5$} &  1.2  &  -5.3  &  &    \nodata &  30.90 &  31.08 &  30.90 &    \nodata  \\
102 & 182018.19$-$160937.0 &    56.5 &   6.6 &  &  {\tiny $-0.3$} 22.5 {\tiny $+0.2$} & {\tiny $-1.0$} 1.5 \phantom{{\tiny $-1.0$}} &  -4.9  &  &    \nodata &  31.07 &  31.18 &  31.08 &    \nodata  \\
112 & 182018.95$-$160736.9 &    17.3 &   3.2 &  &  {\tiny $-0.7$} 22.3 {\tiny $+0.4$} &  1.9  &  -5.4  &  &  29.33 &  30.41 &  30.50 &  30.45 &  30.76  \\
121 & 182019.33$-$161051.9 &     9.5 &   2.2 &  &   23.0  &  3.1  &  -4.4  &  &    \nodata &  30.32 &  30.78 &  30.33 &    \nodata  \\
153 & 182020.49$-$160806.4 &    15.0 &   2.9 &  &  \phantom{{\tiny $+0.5$}} 22.0 {\tiny $+0.5$} &  1.6  &  -5.7  &  &  29.45 &  30.30 &  30.34 &  30.35 &  30.54  \\
175 & 182021.08$-$160738.2 &    20.2 &   3.6 &  &  \phantom{{\tiny $+0.5$}} 22.0 {\tiny $+0.5$} &  1.3  &  -5.8  &  &  29.47 &  30.50 &  30.54 &  30.54 &  30.67  \\
179 & 182021.25$-$160944.3 &     8.4 &   2.0 &  &  \phantom{{\tiny $+0.8$}} 21.5 {\tiny $+0.8$} &  0.6  &  -6.6  &  &  29.11 &  30.18 &  30.19 &  30.21 &  30.24  \\
189 & 182021.59$-$160703.3 &    11.3 &   2.5 &  &   22.5  &  1.6  &  -5.6  &  &    \nodata &  30.30 &  30.43 &  30.32 &    \nodata  \\
196 & 182021.74$-$161317.4 &    10.4 &   2.3 &  &  \phantom{{\tiny $+0.5$}} 22.0 {\tiny $+0.5$} &  1.5  &  -6.0  &  &  29.24 &  30.16 &  30.20 &  30.21 &  30.37  \\
\enddata

\tablecomments{Table~\ref{tbl:powerlaw_spectroscopy} is published in 
its entirety in the electronic edition of the {\it Astrophysical 
Journal}.  A portion is shown here for guidance regarding its form 
and content.}

\tablenotetext{a}{ For convenience, {\bf columns 1--4} reproduce the 
source identification, net counts, and photometric significance data 
from Table~\ref{tbl:src_properties_main}. }

\tablenotetext{b}{ All fits used the ``wabs(powerlaw)'' model in {\it 
XSPEC}.  {\bf Columns 5 and 6} present the best-fit values for the 
column density and power law photon index parameters. {\bf Column 7} 
presents the power law normalization for the model spectrum. {\it 
Quantities in italics} were frozen in the fit.  Uncertainties 
represent 90\% confidence intervals.  Uncertainties are missing when {\it 
XSPEC} was unable to compute them or when their values were so large 
that the parameter is effectively unconstrained.  
Fits lacking uncertainties, fits with large uncertainties, and fits with frozen parameters should be viewed merely as splines to the data to obtain rough estimates of luminosities; the listed parameter values are unreliable.  
  }

\tablenotetext{c}{ X-ray luminosities are presented in {\bf columns 
8--12}:  s = soft band (0.5--2 keV); h = hard band (2--8 keV); t = 
total band (0.5--8 keV).  Absorption-corrected luminosities are 
subscripted with a $c$; they are omitted when $\log N_H > 22.5$~cm$^{-2}$ since 
the soft band emission is essentially unmeasurable.  }

\end{deluxetable}

\clearpage

\begin{deluxetable}{rrrrrrrrrrrrrrrrrr}
\centering \rotate \tabletypesize{\tiny} \tablewidth{0pt} 
\tablecolumns{1}

\tablecaption{Stellar counterparts \label{tbl:counterparts}
}

\tablehead{ \multicolumn{2}{c}{X-ray} & & \multicolumn{6}{c}{Optical/Infrared 
star} &                & \multicolumn{7}{c}{Infrared photometry} \\                                                 
\cline{1-2} \cline{4-9} \cline{11-17}

\colhead{Seq} & \colhead{CXOU} & & \colhead{CEN} & \colhead{B} & 
\colhead{2MASS} & \colhead{SIRIUS} & \colhead{GLIMPSE} & 
\colhead{Notes} & & \colhead{J} & \colhead{H} & \colhead{K} &                                                                                                                                                    
\colhead{[3.6]} & \colhead{[4.5]} & \colhead{[5.8]} & \colhead{[8.0]}  \\

&&&&&&&&&& \colhead{(mag)} & \colhead{(mag)} & \colhead{(mag)} & \colhead{(mag)} & \colhead{(mag)} & \colhead{(mag)} & \colhead{(mag)}  \\

\colhead{(1)} & \colhead{(2)} & & \colhead{(3)} & \colhead{(4)} & 
\colhead{(5)} & \colhead{(6)} & \colhead{(7)} & \colhead{(8)} & & 
\colhead{(9)} & \colhead{(10)} & \colhead{(11)} & \colhead{(12)} & 
\colhead{(13)} & \colhead{(14)} & \colhead{(15)}    }

\startdata
37 & 182009.70$-$161503.7 &&      \nodata &\nodata&                \nodata  &             \nodata &              \nodata & ---                   &&   \nodata &   \nodata &   \nodata &\nodata &\nodata &\nodata &\nodata \\ 
51 & 182012.31$-$160447.7 &&      \nodata &\nodata&  18201227$-$1604466 0bc &               25643 &  G015.1109$-$00.5795 & ---\tablenotemark{051}&&     15.89 &     12.27 &     10.35 &   8.70 &   8.39 &   8.31 &   8.36 \\ 
97 & 182017.84$-$161117.0 &&      \nodata &\nodata&                \nodata  &               13045 &              \nodata & ---\tablenotemark{097}&&   \nodata &    17.81: &    16.47: &\nodata &\nodata &\nodata &\nodata \\ 
123 & 182019.41$-$161329.1 &&      \nodata &\nodata&                \nodata  &               15128 &              \nodata & ---\tablenotemark{123}&&     12.75 &      9.68 &      7.45 &\nodata &\nodata &\nodata &\nodata \\ 
128 & 182019.60$-$161038.8 &&      \nodata &   353 &  18201960$-$1610382 000 &               12415 &              \nodata & ---\tablenotemark{128}&&     16.57 &     14.92 &     11.92 &\nodata &\nodata &\nodata &\nodata \\ 
135 & 182019.88$-$161052.3 &&      \nodata &\nodata&                \nodata  &             \nodata &              \nodata & ---                   &&   \nodata &   \nodata &   \nodata &\nodata &\nodata &\nodata &\nodata \\ 
163 & 182020.73$-$160717.9 &&      \nodata &\nodata&  18202075$-$1607176 000 &               28479 &              \nodata & ---                   &&     17.81 &     13.53 &     11.30 &\nodata &\nodata &\nodata &\nodata \\ 
177 & 182021.21$-$161222.9 &&      \nodata &\nodata&  18202121$-$1612231 000 &               12302 &              \nodata & ---\tablenotemark{177}&&     16.88 &     14.62 &     13.42 &\nodata &\nodata &\nodata &\nodata \\ 
182 & 182021.32$-$161140.2 &&      17 (B3) &   336 &  18202133$-$1611403 000 &               12345 &              \nodata & ---                   &&     12.96 &     12.51 &     12.33 &\nodata &\nodata &\nodata &\nodata \\ 
186 & 182021.43$-$160939.1 &&      35 (B3) &   333 &  18202142$-$1609392 00c &               14179 &  G015.0565$-$00.6501 & ---                   &&     11.79 &     10.90 &     10.42 &   9.77 &   9.79 &\nodata &\nodata \\ 
187 & 182021.43$-$161206.8 &&      \nodata &   335 &  18202143$-$1612068 000 &               13003 &  G015.0203$-$00.6694 & ---                   &&     14.14 &     12.17 &     11.27 &  10.24 &  10.16 &\nodata &\nodata \\ 
230 & 182022.58$-$160251.3 &&      \nodata &\nodata&  18202258$-$1602510 000 &   \tablenotemark{a} &  G015.1587$-$00.6007 & ---                   &&     15.57 &     13.72 &   \nodata &  11.39 &  10.94 &  10.29 &\nodata \\ 
233 & 182022.69$-$160833.9 && 16 (H:O9-B2) &   311 &  18202270$-$1608342 000 &               28651 &              \nodata & ---                   &&      9.99 &      9.44 &      9.10 &\nodata &\nodata &\nodata &\nodata \\ 
236 & 182022.74$-$161123.5 &&      \nodata &   310 &  18202274$-$1611234 00c &               11686 &  G015.0335$-$00.6683 & ---\tablenotemark{236}&&     14.29 &     12.73 &     11.97 &  10.56 &  10.42 &\nodata &\nodata \\ 
239 & 182022.87$-$161148.5 &&      \nodata &\nodata&                \nodata  &             \nodata &              \nodata & ---\tablenotemark{239}&&   \nodata &   \nodata &   \nodata &\nodata &\nodata &\nodata &\nodata \\ 
244 & 182022.93$-$161152.7 &&      \nodata &\nodata&  18202294$-$1611528 000 &               12326 &              \nodata & ---\tablenotemark{244}&&     14.11 &     12.95 &     12.34 &\nodata &\nodata &\nodata &\nodata \\ 
246 & 182022.97$-$161131.8 &&      \nodata &\nodata&                \nodata  &             \nodata &              \nodata & ---\tablenotemark{246}&&   \nodata &   \nodata &   \nodata &\nodata &\nodata &\nodata &\nodata \\ 
255 & 182023.16$-$161305.7 &&      \nodata &\nodata&                \nodata  &             \nodata &              \nodata & ---\tablenotemark{255}&&   \nodata &   \nodata &   \nodata &\nodata &\nodata &\nodata &\nodata \\ 
275 & 182023.80$-$161325.7 &&      \nodata &\nodata&                \nodata  &               12260 &              \nodata & ---\tablenotemark{275}&&     15.99 &     15.40 &     15.15 &\nodata &\nodata &\nodata &\nodata \\ 
281 & 182024.00$-$160818.8 &&       7 (F8) &   293 &  18202401$-$1608187 000 &               27470 &              \nodata & ---\tablenotemark{281}&&     10.00 &      9.70 &      9.58 &\nodata &\nodata &\nodata &\nodata \\ 
296 & 182024.39$-$160843.3 &&  31 (H:O9.5) &   289 &  18202439$-$1608434 000 &               27457 &              \nodata & ---                   &&     10.56 &      9.84 &      9.40 &\nodata &\nodata &\nodata &\nodata \\ 
309 & 182024.60$-$161139.2 &&      \nodata &   284 &  18202460$-$1611394 000 &               14322 &              \nodata & ---\tablenotemark{309}&&     13.14 &     10.62 &      9.34 &\nodata &\nodata &\nodata &\nodata \\ 
322 & 182024.83$-$161135.3 &&      \nodata &\nodata&                \nodata  &               14776 &              \nodata & ---\tablenotemark{322}&&   \nodata &   \nodata &   \nodata &\nodata &\nodata &\nodata &\nodata \\ 
324 & 182024.87$-$161127.5 &&      \nodata &   279 &  18202488$-$1611276 000 &               14553 &              \nodata & ---\tablenotemark{324}&&     13.02 &     11.85 &     10.75 &\nodata &\nodata &\nodata &\nodata \\ 
329 & 182024.94$-$161131.1 &&      \nodata &\nodata&                \nodata  &               14878 &              \nodata & ---\tablenotemark{329}&&    15.88: &    14.46: &    12.46: &\nodata &\nodata &\nodata &\nodata \\ 
334 & 182025.07$-$161133.9 &&      \nodata &   273 &  18202508$-$1611339 000 &               14856 &              \nodata & ---\tablenotemark{334}&&     13.40 &     11.58 &     10.29 &\nodata &\nodata &\nodata &\nodata \\ 
350 & 182025.50$-$161053.8 &&      83 (B3) &   267 &  18202550$-$1610537 000 &               12337 &              \nodata & ---                   &&     13.45 &     12.08 &     11.28 &\nodata &\nodata &\nodata &\nodata \\ 
370 & 182025.93$-$161114.3 &&      \nodata &\nodata&  18202589$-$1611142 cpc &               14914 &              \nodata & ---                   &&    14.64: &     14.12 &     13.41 &\nodata &\nodata &\nodata &\nodata \\ 
371 & 182025.93$-$160949.4 &&      \nodata &\nodata&                \nodata  &               13929 &              \nodata & ---                   &&     15.57 &     14.06 &     13.44 &\nodata &\nodata &\nodata &\nodata \\ 
372 & 182025.95$-$160938.2 &&      \nodata &\nodata&                \nodata  &               13173 &  G015.0653$-$00.6659 & ---                   &&     16.33 &     14.49 &     13.63 &  11.80 &  11.30 &\nodata &\nodata \\ 
373 & 182026.00$-$161253.0 &&      \nodata &\nodata&                \nodata  &             \nodata &              \nodata & ---                   &&   \nodata &   \nodata &   \nodata &\nodata &\nodata &\nodata &\nodata \\ 
374 & 182026.01$-$161022.4 &&      \nodata &   252 &  18202602$-$1610225 c00 &               11830 &              \nodata & ---                   &&     14.14 &     12.66 &     11.97 &\nodata &\nodata &\nodata &\nodata \\ 
375 & 182026.02$-$161240.0 &&      \nodata &   256 &  18202603$-$1612398 000 &               11404 &              \nodata & ---                   &&     14.92 &     12.87 &     11.66 &\nodata &\nodata &\nodata &\nodata \\ 
376 & 182026.04$-$161104.6 &&      26 (B3) &   253 &  18202603$-$1611045 000 &               13500 &  G015.0443$-$00.6776 & ---                   &&     11.53 &     10.87 &     10.53 &   9.85 &   9.90 &\nodata &\nodata \\ 
377 & 182026.09$-$161039.9 &&      \nodata &\nodata&  18202609$-$1610404 00c &               13925 &              \nodata & ---                   &&     15.00 &     13.65 &     13.05 &\nodata &\nodata &\nodata &\nodata \\ 
378 & 182026.09$-$160301.3 &&      \nodata &\nodata&  18202614$-$1603007 00c &   \tablenotemark{a} &  G015.1630$-$00.6145 & ---                   &&     13.92 &     13.58 &     13.63 &  13.49 &  13.45 &\nodata &\nodata \\ 
379 & 182026.11$-$161053.5 &&      \nodata &\nodata&                \nodata  &               14870 &              \nodata & ---                   &&     14.66 &     13.53 &     12.80 &\nodata &\nodata &\nodata &\nodata \\ 
396 & 182026.60$-$161055.7 &&      \nodata &\nodata&                \nodata  &               14346 &              \nodata & ---                   &&     15.02 &     13.79 &     13.16 &\nodata &\nodata &\nodata &\nodata \\ 
398 & 182026.62$-$161136.5 &&      \nodata &   246 &  18202661$-$1611369 c00 &             \nodata &              \nodata & ---\tablenotemark{398}&&   \nodata &     12.01 &     10.56 &\nodata &\nodata &\nodata &\nodata \\ 
399 & 182026.62$-$160822.9 &&      74 (B3) &   245 &  18202662$-$1608229 c00 &               27036 &              \nodata & ---                   &&     13.37 &     12.34 &     11.77 &\nodata &\nodata &\nodata &\nodata \\ 
433 & 182027.41$-$161331.0 &&      (H: O6) &     0 &  18202742$-$1613309 000 &               14703 &              \nodata & -s-\tablenotemark{433}&&      8.30 &      7.89 &     9.18: &\nodata &\nodata &\nodata &\nodata \\ 
466 & 182028.15$-$161049.3 &&      85 (B2) &   223 &  18202815$-$1610493 000 &                6104 &              \nodata & ---\tablenotemark{466}&&     12.67 &     11.33 &     10.62 &\nodata &\nodata &\nodata &\nodata \\ 
488 & 182028.65$-$161211.6 &&      \nodata &   215 &  18202866$-$1612115 000 &                5988 &              \nodata & ---\tablenotemark{488}&&     11.35 &     10.55 &     10.08 &\nodata &\nodata &\nodata &\nodata \\ 
495 & 182028.86$-$161042.0 &&      84 (B2) &   207 &  18202887$-$1610421 0cc &                6109 &              \nodata & ---\tablenotemark{495}&&     12.69 &     11.58 &     11.11 &\nodata &\nodata &\nodata &\nodata \\ 
510 & 182029.21$-$161425.6 &&      \nodata &\nodata&  18202924$-$1614254 000 &                1967 &              \nodata & ---                   &&     14.74 &     14.21 &     13.88 &\nodata &\nodata &\nodata &\nodata \\ 
511 & 182029.28$-$161041.4 &&      \nodata &   202 &  18202930$-$1610411 ccc &                3559 &              \nodata & ---                   &&     14.46 &     12.86 &     12.08 &\nodata &\nodata &\nodata &\nodata \\ 
512 & 182029.31$-$161047.9 &&      \nodata &\nodata&                \nodata  &                6070 &              \nodata & ---                   &&     14.77 &     12.93 &     12.11 &\nodata &\nodata &\nodata &\nodata \\ 
513 & 182029.32$-$161045.2 &&      \nodata &\nodata&                \nodata  &                6071 &              \nodata & ---                   &&   \nodata &     14.29 &     13.15 &\nodata &\nodata &\nodata &\nodata \\ 
514 & 182029.36$-$160919.2 &&      \nodata &\nodata&  18202937$-$1609192 0cc &               21537 &              \nodata & ---                   &&   \nodata &     14.99 &     14.27 &\nodata &\nodata &\nodata &\nodata \\ 
515 & 182029.39$-$160943.0 &&      98 (B4) &   200 &                \nodata  &               13962 &  G015.0707$-$00.6786 & ---                   &&     13.43 &     12.20 &     11.63 &  10.97 &  10.89 &\nodata &\nodata \\ 
516 & 182029.39$-$161046.5 &&      \nodata &\nodata&                \nodata  &                9296 &              \nodata & ---                   &&    14.79: &     13.26 &     12.35 &\nodata &\nodata &\nodata &\nodata \\ 
517 & 182029.39$-$161212.9 &&      \nodata &   201 &  18202940$-$1612127 000 &                6619 &              \nodata & ---                   &&     13.04 &     12.10 &     11.72 &\nodata &\nodata &\nodata &\nodata \\ 
518 & 182029.43$-$161050.2 &&      \nodata &\nodata&                \nodata  &                6069 &              \nodata & ---                   &&     13.58 &     12.26 &     11.61 &\nodata &\nodata &\nodata &\nodata \\ 
519 & 182029.48$-$161644.0 &&      \nodata &\nodata&                \nodata  &             \nodata &              \nodata & ---                   &&   \nodata &   \nodata &   \nodata &\nodata &\nodata &\nodata &\nodata \\ 
536 & 182029.81$-$161045.6 &&    1 (O4+O4) &   189 &                \nodata  &             \nodata &              \nodata & ---\tablenotemark{536}&&   \nodata &   \nodata &   \nodata &\nodata &\nodata &\nodata &\nodata \\ 
543 & 182029.89$-$161044.5 &&    1 (O4+O4) &   189 &  18202986$-$1610449 00d &             \nodata &              \nodata & ---\tablenotemark{536}&&      7.24 &      6.29 &      5.75 &\nodata &\nodata &\nodata &\nodata \\ 
567 & 182030.23$-$161034.9 && 61 (H:O9-B2) &   181 &  18203022$-$1610349 000 &                7833 &              \nodata & ---                   &&     10.72 &      9.70 &      9.23 &\nodata &\nodata &\nodata &\nodata \\ 
574 & 182030.44$-$161053.1 && 37 (H:O3-O6) &   174 &  18203044$-$1610530 000 &               10398 &              \nodata & -s-\tablenotemark{574}&&     10.11 &      8.56 &      7.88 &\nodata &\nodata &\nodata &\nodata \\ 
600 & 182030.95$-$161039.4 &&      \nodata &   163 &  18203095$-$1610393 c00 &                6139 &              \nodata & ---\tablenotemark{600}&&     12.86 &     10.99 &      9.76 &\nodata &\nodata &\nodata &\nodata \\ 
618 & 182031.35$-$160228.4 &&      \nodata &\nodata&                \nodata  &   \tablenotemark{a} &              \nodata & ---\tablenotemark{618}&&   \nodata &   \nodata &   \nodata &\nodata &\nodata &\nodata &\nodata \\ 
647 & 182031.84$-$161138.1 &&      28 (B0) &   150 &  18203184$-$1611383 000 &                8849 &  G015.0470$-$00.7023 & ---                   &&     12.24 &     11.71 &     11.40 &  10.40 &   9.94 &   9.36 &  7.68 \\ 
650 & 182031.89$-$161616.5 &&      \nodata &\nodata&  18203183$-$1616170 ss0 &                6300 &  G014.9787$-$00.7389 & ---\tablenotemark{650}&&     12.61 &     10.95 &      9.57 &   7.75 &\nodata &   6.84 &  6.38 \\ 
655 & 182031.97$-$161030.5 &&      \nodata &   148 &  18203198$-$1610305 c00 &                6784 &  G015.0639$-$00.6940 & ---                   &&     15.63 &     13.58 &     12.63 &  11.55 &  11.61 &\nodata &\nodata \\ 
660 & 182032.06$-$160200.0 &&      \nodata &\nodata&                \nodata  &   \tablenotemark{a} &              \nodata & ---\tablenotemark{660}&&   \nodata &   \nodata &   \nodata &\nodata &\nodata &\nodata &\nodata \\ 
665 & 182032.34$-$160154.1 &&      \nodata &\nodata&                \nodata  &   \tablenotemark{a} &              \nodata & ---\tablenotemark{665}&&   \nodata &   \nodata &   \nodata &\nodata &\nodata &\nodata &\nodata \\ 
706 & 182034.56$-$161523.6 &&      \nodata &\nodata&                \nodata  &             \nodata &              \nodata & ---                   &&   \nodata &   \nodata &   \nodata &\nodata &\nodata &\nodata &\nodata \\ 
726 & 182035.63$-$161055.5 &&      45 (B1) &    93 &  18203561$-$1610555 ccc &                4442 &  G015.0646$-$00.7101 & ---                   &&     11.32 &     10.97 &     10.81 &  10.43 &  10.45 &\nodata &\nodata \\ 
854 & 182053.86$-$160306.5 &&      \nodata &\nodata&  18205385$-$1603063 000 &               20977 &  G015.2139$-$00.7131 & ---\tablenotemark{854}&&      8.69 &      8.40 &      8.30 &   8.15 &   8.25 &   8.20 &   8.07 \\ 
\enddata

\tablecomments{Table~\ref{tbl:counterparts} is published in its 
entirety in the electronic edition of the {\it Astrophysical 
Journal}.  Interesting sources mentioned in the text are shown here for guidance regarding its form 
and content.}

\tablecomments{{\bf Columns 1--2:} Source identification from Table~\ref{tbl:src_properties_main}.
{\bf Column 3:} Source number from \citet{Chini80} and spectral type.  Spectral types preceded by ``H:'' come from \citet{Hanson97}; others come from SIMBAD, then \citet{Chini80}.  
{\bf Column 4:} Source number from \citet{Hanson97} derived from \citet{Bumgardner92}.  A reference frame offset of 1.9\arcsec\ between ACIS and \citet{Hanson97} positions was removed before matching the catalogs.
{\bf Column 5:} 2MASS designation.  Three characters following the designation report the 2MASS Contamination and Confusion Flag defined in the Explanatory Supplement to the 2MASS All Sky Data Release at http://www.ipac.caltech.edu/2mass/releases/allsky/doc/sec2\_2a.html .
{\bf Column 6:} SIRIUS catalog running sequence number from \citet{Jiang02}.  Sources outside the SIRIUS field of view are indicated by table note  \tablenotemark{a}.
{\bf Column 7:} GLIMPSE (a {\em Spitzer} legacy science program) designation.   Sources are taken from the Highly Reliable Catalog, Version 1 at http://www.astro.wisc.edu/sirtf/glimpsedata.html .
{\bf Column 8:} Match ambiguity flags and notes on matches to other catalogs.  An ``s'' in the first/second/third position indicates there are ``secondary'' (i.e. multiple) matches with the 2MASS/SIRIUS/GLIMPSE catalog respectively. 
{\bf Columns 9--11:} JHK photometry. Values are generally taken from SIRIUS when a match is identified, or from 2MASS otherwise.
However the bright sources \#366, 701, 720, and 854 suffer saturation in the SIRIUS photometry so the 2MASS values are reported instead.  
Magnitudes have been corrected to the CIT photometry system (Elias et al. 1982).  
Annotation with {\bf :} indicates reported photometry errors exceed 0.1 mag.  
2MASS values are annotated by \boldmath ${>}$ \unboldmath when the 2MASS Photometric Quality Flag indicates a flux upper limit (flag=U), and are omitted when the flag indicates poor photometry (flag=E). 
{\bf Columns 12--15:} Far infrared photometry taken from GLIMPSE.  Median photometric errors for the reported counterparts are 0.13, 0.12, 0.16, 0.16 mag in the four bands.  Maximum errors are 0.34, 0.32, 0.34, 0.35 mag. 
}

\end{deluxetable}


\subsubsection{Sources Without Stellar Counterparts \label{sec:isolated}}

IR associations could not be found for 115 of the 886 X-ray sources.
Their spatial locations are plotted as 
small diamonds in Figure~\ref{fig:K_17x17}{\it a}. 

Visual examination of the SIRIUS and 2MASS images suggests that at 
least 14 of these probably have NIR associations that are not 
cataloged, either because the counterpart is only marginally resolved 
from a brighter star or because it lies near the limit of the NIR 
survey.  Visual examination of the GLIMPSE image shows many 
point-like objects that are not listed in the Highly Reliable Catalog; thus some ACIS 
sources unidentified in Table~\ref{tbl:counterparts} may have 
$3.6-8$ $\mu$m counterparts.  

Extragalactic sources, mostly AGN at moderate 
redshifts, dominate X-ray source counts at high Galactic latitudes 
and some should be detected in Galactic plane fields despite the 
heavy obscuration.  \citet{Getman06a} and \citet{Wang06} have 
quantitatively calculated the expected background contamination in 
ACIS observations of the Cepheus~B and NGC~6357 star formation 
regions.  In both cases, they predict that $20 \pm 10$ sources should 
appear in a $\sim$40 ks ACIS exposure with dense Galactic cloud 
material, and only a few of those should be detectable in the IR.  
Some of these extragalactic sources are probably among the 
sources with spectra best fit by power law rather than thermal models 
(Table~\ref{tbl:powerlaw_spectroscopy}); as noted above, twelve of those sources lack IR counterparts.  The spatial 
distribution of unassociated sources in 
Figure~\ref{fig:K_17x17}{\it a} is consistent with a random sprinkling 
of $\sim$20 extragalactic sources.  

The remaining $\sim$80 ACIS sources without counterparts are likely 
to be new members of the M17 cloud complex.  
The high median energy of this unassociated population ($E_{median}=3.0$~keV) indicates that these sources are deeply embedded ($A_V \ga 15$~mag).
Some are probably low mass 
stars with strong magnetic activity; several dozen such X-ray 
discovered stars were found in the cloud behind the Orion Nebula in 
the COUP study \citep{Getman05a}.  This interpretation is supported by 
their spatial distribution: Figure~\ref{fig:K_17x17}{\it a} shows that they 
are concentrated in the IR-bright North Bar and South Bar where IR 
surveys are most confused by nebular emission.  
Some of these may be very young with the local absorption in an envelope or disk, as suggested by Getman et al.\ for the COUP sources.

\subsubsection{NIR Properties of ACIS Sources \label{sec:NIRprop}}

\begin{figure}
\includegraphics[angle=0.,height=6.5in]{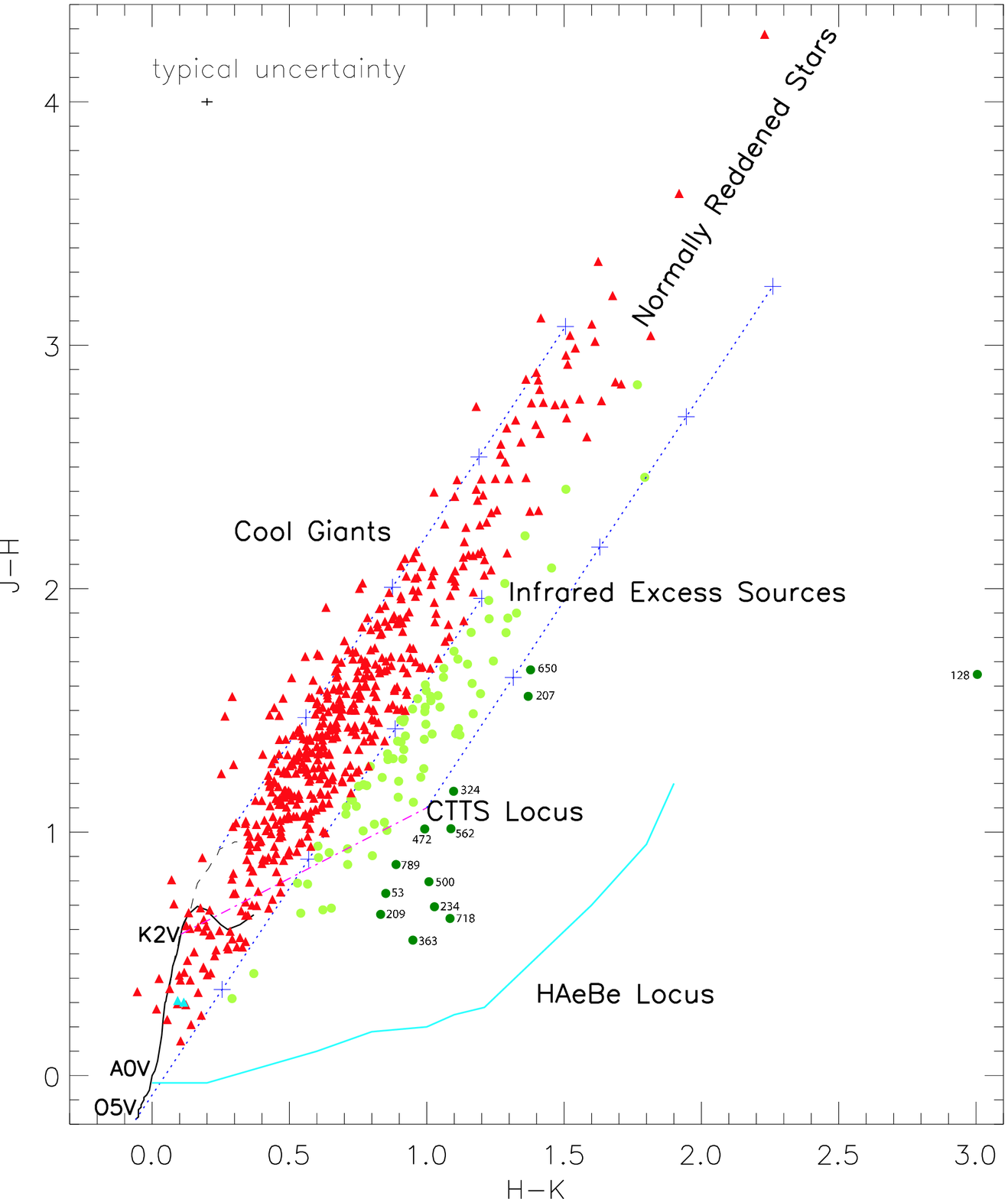} 
\caption{NIR color-color diagram. 
The (light and dark) green circles and red triangles represent sources with and without $K$-band
excess, respectively.  The dark green circles represent Class I objects, labeled by their ACIS sequence numbers.
The two blue triangles are foreground stars (\#281, \#854, \S \ref{sec:new_OB}).  
The black solid and gray long-dash lines denote the loci of main sequence stars and giants, respectively, from \citet{Bessell88}. 
The purple dash dotted line is the locus for classical T Tauri stars (CTTS; Class II pre-main sequence stars) from \citet{Meyer97}.
The cyan solid line is the locus for HAeBe stars from \citet{Lada92}. 
Three blue dashed lines (marked every 5 mag) are drawn parallel to a reddening vector with length $A_V=20$~mag, originating from the O5 dwarf, the M4 giant, and the end of the pre-main sequence locus.
\label{fig:ccd}}
\end{figure}

Following \citet[][Figure 2]{Hanson97}, Figure~\ref{fig:ccd} shows the NIR $J-H$ vs.\ $H-K$ color-color diagram for 609 out of 886 {\em Chandra} stars with high-quality (errors $<0.1$ mag) $JHK$ photometry listed in Table~\ref{tbl:counterparts}.  This plot is valuable for estimating the disk/envelope evolutionary class \citep[I, II, or III, from][]{Lada87} of protostars and young stellar objects.  
Most {\em Chandra} sources occupy the color space defined by the two leftmost dashed lines (rightward of the locus of main sequence stars, between these two reddening lines and including the extension of those lines to the upper right); this region of color space is associated with young stars lacking inner disks, referred to as Class III objects, which are reddened by interstellar extinction.  

The $\sim$100 stars to the right of this reddened band are NIR excess sources; most occupy the color space between the middle and right-most dashed lines (including sources below the classical T Tauri star locus that would fall to the left of the rightmost reddening line if it was extended downward); this space is associated with pre-main sequence stars with circumstellar accretion disks, referred to as Class II objects.
A handful of stars occupy the color space (rightward of the right-most dashed line and its downward extension) associated with protostars, referred to as Class I objects.  The Herbig AeBe locus shows where these intermediate-mass pre-main sequence stars lie in color space; some IR excess sources could be HAeBe stars.  The outlier \#128 is probably an interesting Class I protostar; it is a faint but very hard X-ray source.

Above the leftmost reddening line is the region of color space occupied by background giant stars \citep{Hanson97}.  Simulations of X-ray detected background giants from the Cepheus~B study \citep{Getman06a} showed that typically only $\sim 3$ such objects should be detected in our M17 {\em Chandra} observation; the fact that more than 30 points occupy this region of color space in Figure~\ref{fig:ccd} shows that such color-color diagrams are not definitive for classifying sources.  NIR photometry can be contaminated by diffuse emission or crowding, causing pre-main sequence members of M17 to scatter into this region of color space.  

As these classes are based solely on the NIR ($JHK$) spectral energy distributions (SEDs) of the sources, they reflect the evolutionary stage of inner disks.
Our analysis is insensitive to the evolutionary status of centrally-cleared, outer disks.  Thus our Class III status based on $JHK$ colors for the known O9--B3 stars \#186, 233, 296, and 488 is not in conflict with the Class I status given to these sources by \citet{Nielbock01}, whose classification is based on IR SEDs that extend out to 20~$\mu$m, where outer disks can be detected.

\begin{figure}
\includegraphics[angle=0.,height=6.5in]{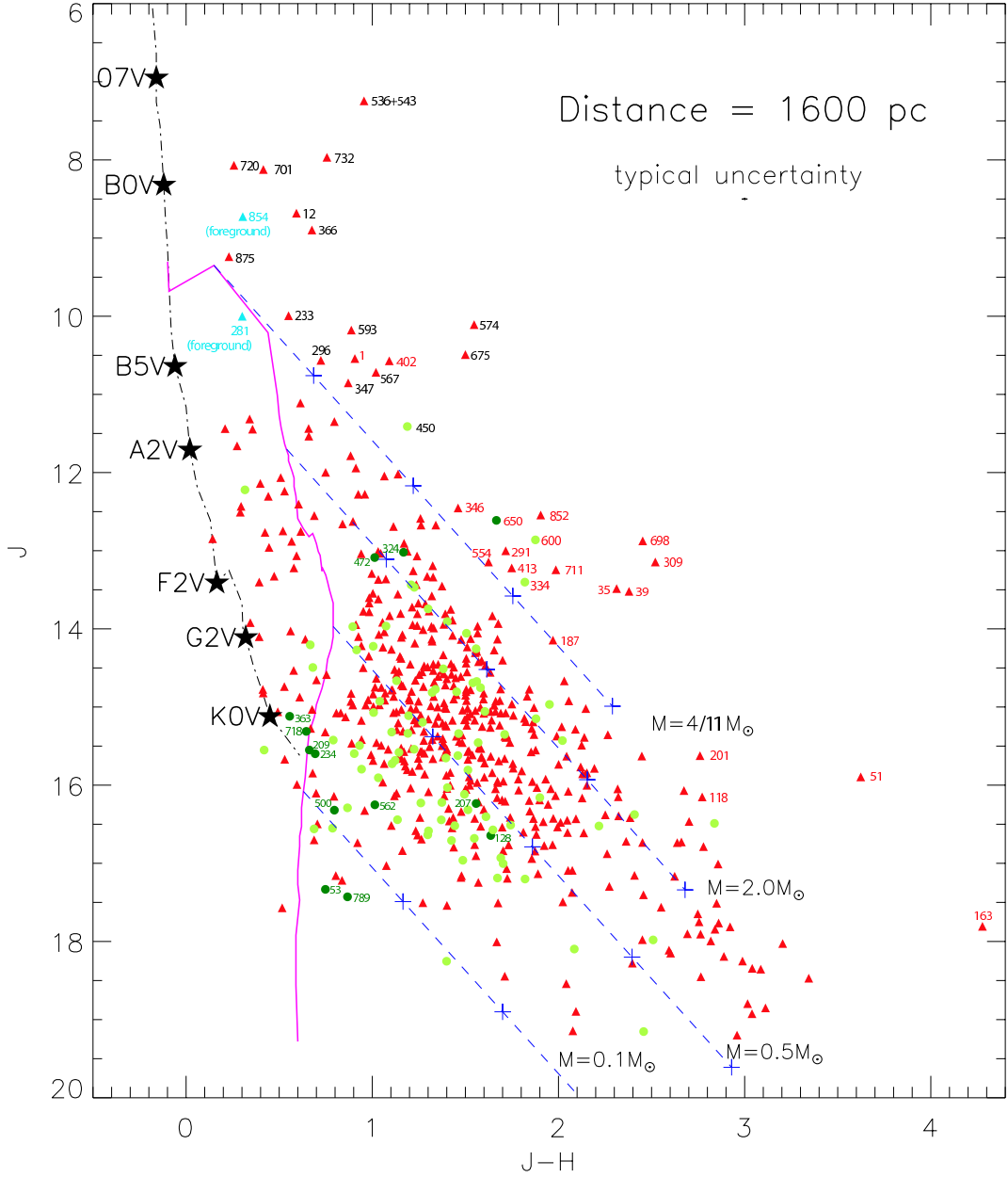} 
\caption{NIR $J$ vs.\ $J-H$ color-magnitude diagram. 
The  (light and dark) green circles and red triangles represent sources with and without $K$-band excess, respectively.  
The dark green circles represent Class I objects, labeled by their ACIS sequence numbers.
The two blue triangles are foreground stars (\S \ref{sec:new_OB}).  
The purple solid line is the 1 Myr isochrone from \citet{Siess00} and \citet{Baraffe98}. 
The gray dash dotted line marks 
the location of Zero Age Main Sequence (ZAMS) stars; various spectral types are marked by black stars. 
The blue dashed lines represent the standard reddening vectors for various stellar masses, marked for every $A_V=5$ mag by the blue crosses. 
Sequence numbers for ACIS sources down to 4\Msol are shown in black for cataloged OB stars (from Table~\ref{OB.tab}) and in red otherwise.
The O star \#433 is among those omitted due to poor photometry.
Sources \#51 and \#163 are discussed in \S \ref{sec:new_OB}.
Typical uncertainties are shown graphically; the small median $J$-band uncertainty (0.02~mag) may be difficult to see.
\label{fig:cmd}}
\end{figure}

Figure~\ref{fig:cmd} shows the NIR $J$ vs.\ $J-H$ color-magnitude 
diagram for the same stars shown in Figure~\ref{fig:ccd}.  This plot is 
valuable for estimating individual masses and reddening.  Assuming that
they lie on the 1~Myr isochrone, the majority of ACIS stars appear to 
be G and F stars ($0.5 \la M \la 2$\Msol) reddened by $3 \la A_V 
\la 15$~mag.  Sources to the left of the 1~Myr isochrone may be older than the NGC~6618 population (\S \ref{sec:contamination}) or may have contaminated photometry.

The drop off of stars towards higher masses is due to the stellar Initial Mass Function (IMF).
The peak of the mass distribution in the figure is reasonably consistent with the mass completeness limit of 1.7\Msol estimated in \S \ref{sec:XLF} for the lightly obscured population.
Since the stellar IMF rises steeply towards a peak around $M \sim$~0.3-0.5\Msol, most of our X-ray sources are associated with stars having masses below our completeness limit.
The minimum absorption of $A_V \sim 3$~mag is consistent with the absorption reported 
in the M17 cluster \citep{Chini80}.  Approximately 40 stars are highly  
obscured with $A_V \geq 20$~mag; other X-ray discovered members without NIR associations undoubtedly are even more obscured. 

Table~\ref{tbl:highmass} lists 138 X-ray stars (from the sample of 609 high-quality counterparts) with dereddened 
mass estimates $M \ge 2.0$\Msol, ordered by mass.  
Appended to the table are five sources whose photometry has large errors but nevertheless strongly suggests that the stars are massive.
Some massive stars cataloged in the literature are included in this table; they are marked with table notes.
Some cataloged massive stars (from Tables~\ref{Nielbock_HM.tab} and \ref{OB.tab}) detected by {\em Chandra} are missing:
ACIS \#239, \#322, \#536, and \#543 have no photometry available; ACIS \#128, \#182, \#399, \#647, and \#726 have dereddened mass estimates less than $2.0$\Msol.
We adopt the reddening law of \citet{Rieke85}\footnote{
The abnormal reddening law of $R_V=4.8$ derived for M17 cloud material by \citet{Chini98} should have little or no impact on the NIR extinction \citep{Cardelli89}; thus we use the standard reddening law from \citet{Rieke85}.
}. 
Note that inference of stellar masses and visual extinction from
dereddening involves an ambiguity for stars within the mass range 2-4\Msol
(Figure~\ref{fig:cmd}) because the reddening vector intercepts the 1 Myr isochrone twice.
Table~\ref{tbl:highmass} reports the lower of the two mass estimates.  
As a result there is an apparent gap between 4-10\Msol in the reported mass distribution, and some reported masses in the range of 2-4\Msol may be approximately one half their true values. 
The table also reports a NIR ($JHK$) SED classification for each object (Class I, II, or III described above) obtained from the color-color diagram. 
Of course, these class estimates refer only to hot inner disks and are subject to photometric uncertainties, so misclassifications are present.  These approximate masses and classifications are used in \S \ref{sec:spatial}, \S \ref{sec:new_OB}, and \S \ref{sec:disks}.


\begin{deluxetable}{rcrrr}
\centering 

\tabletypesize{\tiny} \tablewidth{0pt} \tablecolumns{5} 

\tablecaption{Photometrically Selected Intermediate- and High-Mass Stars 
\label{tbl:highmass}} 

\tablehead{ \colhead{ACIS \#} & \colhead{CXOU } & \colhead{$A_V$} & 
\colhead{Mass} & \colhead{Class} \\

&& \colhead{(mag)} & \colhead{($M_{\odot}$)} & \colhead{(est.)}\\

\colhead{(1)} & \colhead{(2)} & \colhead{(3)} & \colhead{(4)} & \colhead{(5)}  
}
\startdata
         732\tnm{a} & 182035.87$-$161542.5 &  8.2 & 34.8 & III  \\
         574\tnm{a} & 182030.44$-$161052.8 & 15.6 & 34.2 & III  \\
          51        & 182012.31$-$160447.4 & 35.0 & 31.0 & III  \\
         698        & 182034.40$-$160938.8 & 24.1 & 30.4 & III  \\
         309\tnm{b} & 182024.60$-$161139.0 & 24.7 & 29.4 & III  \\
         675\tnm{a} & 182033.06$-$161121.3 & 15.2 & 29.0 & III  \\
         163        & 182020.74$-$160717.7 & 41.1 & 29.0 & III  \\
          12\tnm{a} & 182003.04$-$160206.5 &  6.7 & 23.0 & III  \\
          39        & 182010.50$-$160528.6 & 23.4 & 21.8 & III  \\
          35        & 182009.19$-$161430.9 & 22.7 & 20.5 & III  \\
         701\tnm{a} & 182034.49$-$161011.6 &  4.2 & 20.5 & III  \\
         852        & 182053.43$-$161210.1 & 18.9 & 19.5 & III  \\
         366\tnm{a} & 182025.86$-$160832.1 &  6.4 & 18.8 & III  \\
         720\tnm{a} & 182035.39$-$161048.2 &  6.4 & 18.6 & III  \\
         402        & 182026.66$-$160708.6 & 11.3 & 18.3 & III  \\
         593\tnm{a} & 182030.84$-$161007.2 &  9.4 & 17.3 & III  \\
         600\tnm{c} & 182030.95$-$161039.1 & 18.6 & 16.8 &  II  \\
         711        & 182034.85$-$160626.2 & 19.6 & 16.2 & III  \\
         567\tnm{a} & 182030.23$-$161034.7 & 10.6 & 16.1 & III  \\
           1        & 181957.56$-$161204.3 &  9.5 & 15.3 & III  \\
         650        & 182031.89$-$161616.2 & 16.6 & 14.9 &   I  \\
         450\tnm{a} & 182027.80$-$161101.4 & 12.2 & 14.6 &  II  \\
         201        & 182021.82$-$161123.4 & 26.9 & 14.2 & III  \\
         291        & 182024.33$-$161352.8 & 17.1 & 13.6 & III  \\
         347\tnm{a} & 182025.44$-$161115.5 &  9.2 & 13.2 & III  \\
         233\tnm{a} & 182022.70$-$160833.6 &  6.2 & 13.2 & III  \\
         413        & 182026.96$-$160347.3 & 17.4 & 12.9 & III  \\
         346        & 182025.42$-$161118.6 & 14.7 & 12.9 & III  \\
         334        & 182025.08$-$161133.7 & 18.0 & 12.9 &  II  \\
         875\tnm{a} & 182101.07$-$160546.7 &  3.2 & 12.8 & III  \\
         296\tnm{a} & 182024.39$-$160843.0 &  7.8 & 12.7 & III  \\
         118        & 182019.25$-$161326.4 & 26.9 & 11.8 & III  \\
         554        & 182030.03$-$161034.5 & 16.2 & 11.7 & III  \\
         187        & 182021.44$-$161206.6 & 19.4 & 11.3 & III  \\
         757        & 182037.62$-$160332.3 & 21.7 &  3.7 & III  \\
          27        & 182008.25$-$160552.5 &  7.4 &  3.7 & III  \\
         691        & 182034.02$-$160800.6 & 23.5 &  3.7 &  II  \\
         817        & 182043.77$-$161330.1 & 18.7 &  3.4 & III  \\
         813        & 182043.42$-$161245.9 &  5.7 &  3.4 & III  \\
         488\tnm{a} & 182028.66$-$161211.3 &  3.2 &  3.4 & III  \\
         466\tnm{a} & 182028.15$-$161049.1 &  8.4 &  3.4 & III  \\
         608\tnm{a} & 182031.12$-$160929.6 &  7.5 &  3.3 & III  \\
         582        & 182030.64$-$161028.3 & 16.8 &  3.3 & III  \\
         560        & 182030.15$-$161055.6 & 15.7 &  3.3 & III  \\
         507\tnm{a} & 182029.17$-$160942.8 &  7.5 &  3.3 & III  \\
         438        & 182027.62$-$161043.0 & 20.9 &  3.3 & III  \\
         307        & 182024.56$-$161127.5 & 14.7 &  3.3 & III  \\
         588        & 182030.73$-$161101.4 & 22.3 &  3.2 & III  \\
         456        & 182027.86$-$160955.3 &  1.3 &  3.2 & III  \\
         375        & 182026.02$-$161239.8 & 14.8 &  3.2 & III  \\
         288        & 182024.23$-$161109.3 & 11.2 &  3.2 & III  \\
         229        & 182022.58$-$161024.5 & 10.9 &  3.2 & III  \\
         186\tnm{a} & 182021.43$-$160938.8 &  3.9 &  3.2 & III  \\
          83        & 182016.84$-$160729.0 & 21.6 &  3.2 & III  \\
         506        & 182029.14$-$161054.0 &  4.1 &  3.1 & III  \\
         206        & 182021.92$-$161412.3 & 25.5 &  3.1 & III  \\
         763        & 182038.02$-$160310.5 & 26.7 &  3.0 & III  \\
         687        & 182033.76$-$161304.6 &  7.8 &  3.0 & III  \\
         678        & 182033.21$-$161058.1 & 11.0 &  3.0 & III  \\
         677        & 182033.17$-$160746.3 & 15.4 &  3.0 & III  \\
         526        & 182029.61$-$160603.5 & 20.3 &  3.0 & III  \\
         508        & 182029.20$-$161110.7 &  1.6 &  3.0 & III  \\
         495\tnm{a} & 182028.87$-$161041.7 &  5.8 &  3.0 & III  \\
         343        & 182025.32$-$160939.4 &  4.4 &  3.0 & III  \\
         649        & 182031.86$-$161047.1 & 20.0 &  2.9 & III  \\
         638\tnm{a} & 182031.70$-$160945.6 &  7.0 &  2.9 & III  \\
         376\tnm{a} & 182026.04$-$161104.3 &  1.5 &  2.9 & III  \\
         350\tnm{a} & 182025.50$-$161053.6 &  8.2 &  2.9 & III  \\
         312\tnm{a} & 182024.62$-$161108.4 &  6.3 &  2.9 & III  \\
         253        & 182023.13$-$161131.8 & 14.9 &  2.9 & III  \\
         682        & 182033.43$-$161042.5 & 13.6 &  2.9 &  II  \\
         783        & 182039.71$-$161312.5 &  7.8 &  2.8 & III  \\
         595        & 182030.89$-$160904.9 & 10.0 &  2.8 & III  \\
         505        & 182029.12$-$161247.2 & 15.2 &  2.8 & III  \\
         496\tnm{a} & 182028.87$-$161109.8 &  6.5 &  2.8 & III  \\
         423        & 182027.20$-$160420.6 & 10.8 &  2.8 & III  \\
         408        & 182026.82$-$160749.3 &  4.0 &  2.8 & III  \\
         645        & 182031.82$-$161123.6 & 11.2 &  2.7 & III  \\
         564        & 182030.19$-$161213.4 & 17.0 &  2.7 & III  \\
         512        & 182029.32$-$161047.7 & 12.6 &  2.7 & III  \\
         406\tnm{a} & 182026.79$-$161057.8 &  7.4 &  2.7 & III  \\
         273        & 182023.76$-$161101.2 & 10.0 &  2.7 & III  \\
         105        & 182018.44$-$161418.7 & 14.2 &  2.7 & III  \\
         324        & 182024.87$-$161127.2 &  6.2 &  2.7 &   I  \\
         771        & 182038.82$-$160604.6 & 21.9 &  2.6 & III  \\
         727        & 182035.67$-$160242.5 &  8.6 &  2.6 & III  \\
         607        & 182031.09$-$161125.9 & 17.0 &  2.6 & III  \\
         549        & 182029.95$-$160953.5 &  6.7 &  2.6 & III  \\
         489        & 182028.67$-$160926.0 &  2.3 &  2.6 & III  \\
         179        & 182021.25$-$160944.1 & 15.8 &  2.6 & III  \\
         150        & 182020.32$-$160237.1 & 12.4 &  2.6 & III  \\
          57        & 182012.96$-$161308.1 & 17.8 &  2.6 &  II  \\
         800        & 182041.79$-$160241.7 &  9.5 &  2.5 & III  \\
         518        & 182029.44$-$161049.9 &  7.5 &  2.5 & III  \\
         305        & 182024.54$-$160737.4 & 12.7 &  2.5 & III  \\
         710        & 182034.83$-$160750.9 &  9.3 &  2.5 &  II  \\
         592        & 182030.80$-$161041.8 & 14.1 &  2.5 &  II  \\
         758        & 182037.78$-$160427.1 & 22.5 &  2.4 & III  \\
         568        & 182030.28$-$161131.2 &  8.6 &  2.4 & III  \\
         538        & 182029.82$-$160957.0 &  9.7 &  2.4 & III  \\
         401        & 182026.65$-$160310.3 & 17.5 &  2.4 & III  \\
          30        & 182008.44$-$161409.4 &  8.2 &  2.4 & III  \\
          16        & 182004.72$-$160937.1 & 13.8 &  2.4 & III  \\
         285        & 182024.10$-$160839.9 &  9.7 &  2.4 &  II  \\
         628        & 182031.52$-$160910.1 &  9.6 &  2.3 & III  \\
         515        & 182029.39$-$160942.7 &  6.7 &  2.3 & III  \\
         368        & 182025.87$-$160922.9 & 15.2 &  2.3 & III  \\
         236        & 182022.75$-$161123.2 &  9.7 &  2.3 & III  \\
         137        & 182019.95$-$160805.0 & 14.9 &  2.3 & III  \\
          21        & 182006.33$-$160453.5 & 15.1 &  2.3 & III  \\
         537        & 182029.82$-$161057.2 & 12.7 &  2.3 &  II  \\
         335        & 182025.10$-$161129.4 &  8.2 &  2.3 &  II  \\
         668        & 182032.48$-$161047.0 & 21.8 &  2.2 & III  \\
         655        & 182031.98$-$161030.2 & 14.1 &  2.2 & III  \\
         555        & 182030.04$-$161233.9 & 14.9 &  2.2 & III  \\
         511        & 182029.29$-$161041.2 & 10.1 &  2.2 & III  \\
         437        & 182027.59$-$161133.3 & 10.3 &  2.2 & III  \\
         435        & 182027.48$-$161013.2 &  3.4 &  2.2 & III  \\
         374        & 182026.01$-$161022.1 &  8.9 &  2.2 & III  \\
         310        & 182024.60$-$161046.1 &  9.1 &  2.2 & III  \\
         264        & 182023.44$-$160953.1 &  9.9 &  2.2 & III  \\
         113        & 182019.06$-$160629.3 & 19.6 &  2.2 & III  \\
          94        & 182017.82$-$160453.1 & 23.9 &  2.2 & III  \\
         680        & 182033.39$-$160918.6 &  6.3 &  2.2 &  II  \\
         654        & 182031.97$-$160924.9 &  6.5 &  2.2 &  II  \\
         717        & 182035.29$-$161118.4 &  7.9 &  2.1 & III  \\
         694        & 182034.08$-$161044.3 &  4.6 &  2.1 & III  \\
         681        & 182033.41$-$161115.6 &  9.6 &  2.1 & III  \\
         534        & 182029.76$-$160936.3 &  7.0 &  2.1 & III  \\
         434        & 182027.47$-$161358.4 &  4.7 &  2.1 & III  \\
         200        & 182021.79$-$161041.8 & 16.7 &  2.1 & III  \\
         159        & 182020.55$-$160738.0 & 12.4 &  2.1 & III  \\
         142        & 182020.03$-$161330.8 & 17.9 &  2.1 & III  \\
         804        & 182042.04$-$161107.2 & 15.0 &  2.0 & III  \\
         262        & 182023.34$-$160512.8 & 23.3 &  2.0 & III  \\
         225        & 182022.49$-$160951.5 & 10.7 &  2.0 & III  \\
         192        & 182021.62$-$160924.5 &  8.3 &  2.0 & III  \\
         359        & 182025.73$-$161045.8 &  7.0 &  2.0 &  II  \\  
  7\tnm{d}          & 182000.65$-$161112.0 &\nodata&\nodata&\nodata \\ 
  9\tnm{e}          & 182001.73$-$160529.0 &\nodata&\nodata&\nodata \\ 
398\tnm{bf}         & 182026.62$-$161136.5 &\nodata& 4:    &\nodata \\ 
433\tnm{af}         & 182027.41$-$161331.0 &\nodata& $>$4\phantom{:}&\nodata \\ 
539\tnm{f}          & 182029.84$-$161041.3 &\nodata& 4:    &\nodata \\ 
\enddata

\tablecomments{{\bf Columns 1--2:} Source identification from Table~\ref{tbl:src_properties_main}.\\
{\bf Columns 3--4:} Visual absorption and mass, derived from dereddened location in 
color-magnitude diagram along the standard interstellar reddening vector.\\
{\bf Column 5:} Suggested disk/envelope evolutionary class (see text) based on source's position in NIR color-color diagram (Figure~\ref{fig:ccd}). 
ACIS \#281 and \#854 are likely foreground stars (\S \ref{sec:new_OB}) and are omitted.
ACIS \#123 and \#543 have inaccurate photometry and are omitted.
}

\tablenotetext{a}{Historically recognized OB stars from Table~\ref{OB.tab}.}

\tablenotetext{b}{High-mass Class I source from Table~\ref{Nielbock_HM.tab}.}

\tablenotetext{c}{ACIS \#600 is B~163 in \citet{Hanson97} and is listed in their Table~6 as a candidate young stellar object.}

\tablenotetext{d}{$J$-band photometry is not available, however source is bright in $K$-band (10.9 mag) and is likely massive.}
\tablenotetext{e}{$J$-band photometry is not available, however source is bright in $K$-band (9.9 mag) and is likely massive.}

\tablenotetext{f}{Photometry has large errors.}

\end{deluxetable}


\subsection{X-ray sources not associated with M17 \label{sec:contamination}}

Approximately 20 extragalactic sources are expected to appear in our ACIS catalog (\S \ref{sec:isolated}).
An additional $\sim$20 foreground stars are expected to appear in the catalog, based on extrapolation of foreground contamination simulations presented by \citet{Getman06a} and \citet{Wang06}.  This is roughly consistent with the population of sources observed between the 1My isochrone and the zero-age main sequence (ZAMS) track in the color-magnitude diagram discussed in \S \ref{sec:NIRprop} (Figure~\ref{fig:cmd}).
Additionally, some of those sources show large proper motions in the UCAC2 catalog \citep{Zacharias04} and/or are faint, soft, minimally-absorbed X-ray sources.
Thus, in total $\sim$5\% of our catalog of 886 ACIS sources are expected to be unrelated to the M17 star forming region.  
We postpone a detailed membership study until we have a longer ACIS observation of M17, where the X-ray spectra will be better characterized and more useful for identifying the unrelated populations.

\section{Global Properties of the Stellar Populations \label{sec:stellarpop}}

The COUP study of the Orion Nebula 
Cluster, whose membership was reliably known in advance from many 
optical and NIR studies, shows a strong statistical association 
between X-ray luminosity and stellar mass in lower mass ($M \leq 
3$\Msol) pre-main sequence stars \citep{Preibisch05a}. This 
relationship is confirmed in other young stellar clusters such as Cep 
OB3b and Cep~B \citep{Getman06a} and NGC~2264 \citep{Flaccomio06, 
Rebull06}. The astrophysical origin of the relation is uncertain: it 
may arise from internal dynamo processes \citep{Preibisch05a} or 
constraints on the external magnetic coronae \citep{Jardine06}.

However, the empirical $L_x - M$ correlation means that X-ray catalogs 
should give a surprisingly complete census of lightly obscured cloud 
members down to a mass limit dependent on the $L_x$ limit of the observation \citep{Feigelson05b,Getman06a}.  
Our census of the lightly obscured M17 population is complete down to a hard-band X-ray luminosity of
$\log L_{h,c} \sim 30.5$~ergs~s$^{-1}$ (\S \ref{sec:XLF}).
Assuming that the M17 stars follow the same empirical $L_{h,c} - M$ 
relation as the Orion population \citep{Preibisch05a,Getman06a}, that completeness limit roughly corresponds to a mass of 1.7\Msol.
This is achieved with $<$5\% contamination from non-cluster members (\S \ref{sec:contamination}).  

A second critical point is that X-ray emission arises primarily from 
magnetic flaring and thus X-ray selected samples are relatively unbiased 
with respect to the presence or absence of circumstellar disks 
\citep{Feigelson06}. Some systematic effects are present--- 
 Class II stars (systems with accretion disks) are systematically $\sim$2 times fainter 
than Class III stars (systems with weak or absent disks) in the same mass stratum, and a 
soft X-ray component from accretion may be occasionally seen---but these 
are relatively small \citep{Preibisch05a}. 
Thus the X-ray sample complements IR 
population studies that often rely on the presence of warm dusty disks, 
although X-ray studies are less complete for the lowest mass objects. 
From the $JHK$ photometric study of the SIRIUS field (somewhat smaller 
than the ACIS field), \citet{Jiang02} identified 454 Class I 
protostars, 2798 Class II/III pre-main sequence stars, and 281 OB 
candidates.  The Class I objects are concentrated along the North Bar and 
South Bar in the \hii region.  The Class II/III stars are distributed 
widely across the field with a deficit in the M17-SW region 
attributable to heavy absorption.

\subsection{Spatial Distribution of X-ray Stars in M17
\label{sec:spatial}}

\begin{figure}
 \centering
\plotone{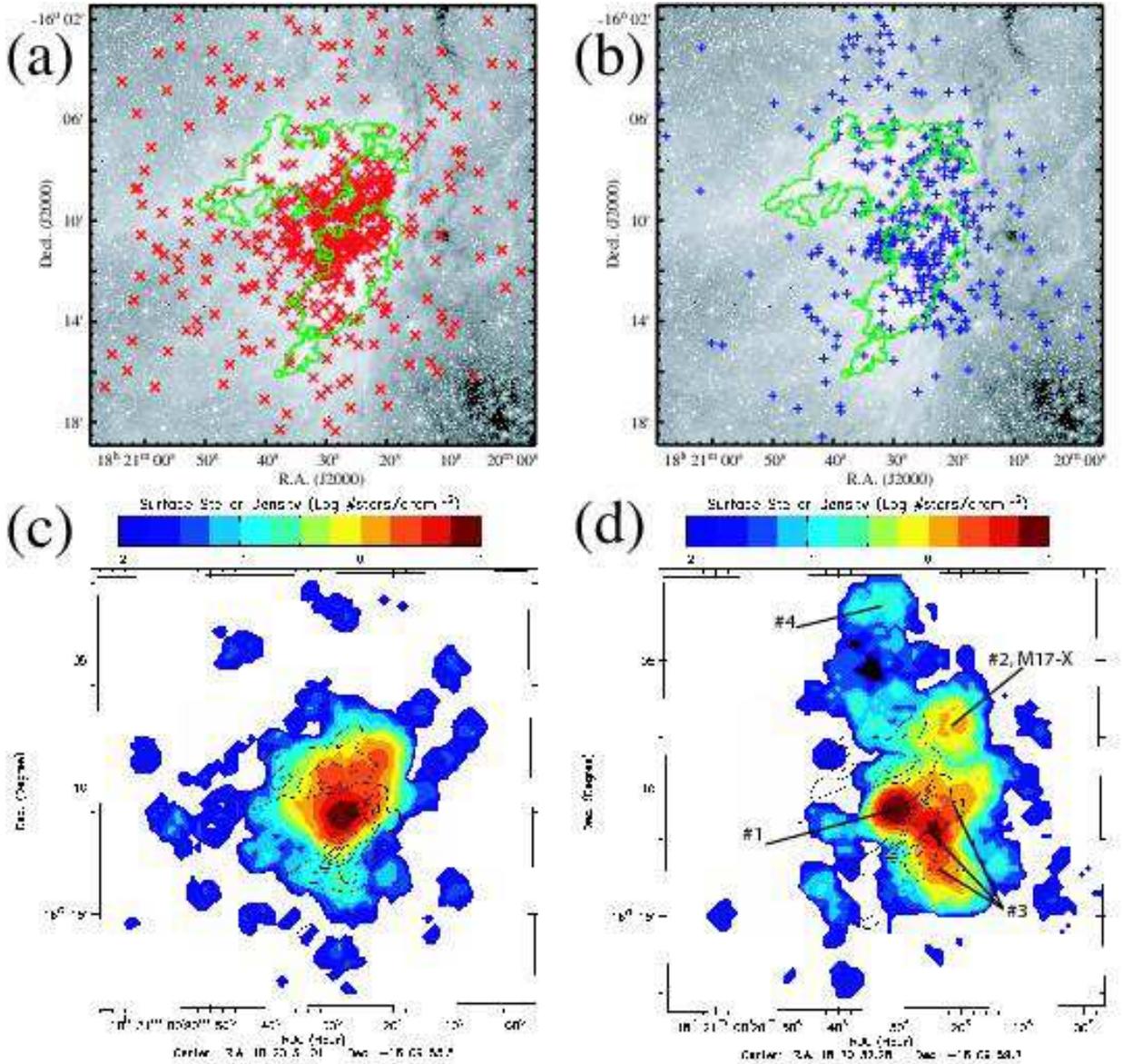} 

\caption{Spatial distribution of ACIS X-ray sources.  The left-hand 
panels show the lightly obscured population ($E_{median} < 2.5$~keV or $A_V 
\leq 10$~mag), and the right-hand panels show the heavily obscured 
population ($E_{median} > 2.5$~keV or $A_V \geq 10$~mag).  
The top panels show the individual star positions superposed on the {\em Spitzer} GLIMPSE 3.6 $\mu$m grayscale image, with green contours outlining the North Bar and South Bar in the \hii region.
The bottom panels show a map of the stellar surface density after smoothing with a 
30\arcsec\/ radius kernel; contours from the {\em MSX} 8 $\mu$m map are overlaid to show the North Bar and South Bar. 
Four structures discussed in \S \ref{sec:spatial} are labeled in 
panel (d); structure \#2 is the newly identified M17-X embedded 
cluster.  
\label{fig:stellar_density}}
\end{figure}

Figure~\ref{fig:stellar_density} shows the distribution of X-ray 
stars in the ACIS field, both as individual stars and smoothed with a 
box kernel of width 30\arcsec. The low 
obscuration and high obscuration sources are discriminated by their spectra as those with 
median energy below and above 2.5~keV respectively.  This criterion corresponds to an 
absorption of $\log N_H \sim 22.3$~cm$^{-2}$ or $A_V \sim 10$~mag 
\citep{Feigelson05b}.  


The lightly obscured population shows a simple, centrally 
concentrated structure centered at $(\alpha,\delta) = (18^h20^m28^s, 
-16^\circ10\arcmin50\arcsec)$. This lies about 30\arcsec\/ (0.2 pc 
projected) west of the O4+O4 binary CEN~1 (ACIS \#536 and 543; the field center is shown in \S \ref{sec:OB}).  The 
B2 star CEN~85 (ACIS \#466) coincides with the center of the cluster 
distribution.  The inner $2$\arcmin\/ (0.9 pc) radius appears 
spherical, but the distribution becomes elongated towards the 
northwest around 2\arcmin-4\arcmin\/ from the center.  
The central concentration of the 
cluster (surface stellar density $> 1$ per arcmin$^2$) traced by the ACIS sources is $\sim$8$\arcmin \times 6\arcmin$\/ 
($\sim$4 $\times 3$~pc).  

Several features are seen in the stellar surface density map for the 
heavily obscured population (Figure~\ref{fig:stellar_density}{\it b,d}). 
\begin{enumerate}

\item The highest concentration of stars ($\sim$40) coincides with the main 
centrally concentrated cluster seen in 
Figure~\ref{fig:stellar_density}{\it a,c}.  This is undoubtedly the same 
physical structure exhibiting a wide range of obscuration.  

\item 
A secondary stellar concentration ($\sim$14 stars) is located northwest of the cluster
center at \\($18^h20^m22^s,-16^\circ07\arcmin30\arcsec$) with 
3\arcmin\/ (1.4 pc) diameter.  This may be part of the northwest elongation 
seen in the lightly obscured population, but there is a distinct dip 
in stellar density between it and the central cluster.  This cannot 
be attributed to absorption: there is a band of CO-emitting molecular 
material west and north of the secondary stellar concentration, but none is 
seen in the dip region, as shown in Figure~\ref{fig:K_17x17}{\it b}.
We label this secondary stellar group 
M17-X in Figure~\ref{fig:stellar_density}{\it d}, where ``X'' indicates 
X-ray discovered, as it was not previously noted in IR studies. 
As this concentration is based on only 14 stars, its reliability is uncertain.

\item High-density enhancements ($\sim$40 stars) in the X-ray star density extend
from the main cluster 3\arcmin\/ towards the west and 4\arcmin\/ to 
the southwest.  The densest portion of this enhancement coincides with 
M17-UC1 (\S \ref{sec:emb_UC1}) and it ends around the 
KWO (\S \ref{sec:emb_KW}). This arcuate  
distribution runs along the eastern edge of the M17-SW molecular 
core, following a very similar distribution of Class I protostellar 
candidates found by \citet{Jiang02}.  This is an independent 
detection of this elongated stellar structure; only a handful of 
the ACIS stars are SIRIUS Class I candidates.  The ACIS enhancement 
thus gives a clear view of the spatial distribution of the triggered 
stellar population along the shock front between the M17 \hii region 
and the M17-SW molecular core. 

\item A stellar concentration (10--15 stars) is present at the northern edge of the 
ACIS field, corresponding to the embedded M17-North group of 
protostars (\S \ref{sec:emb_M17N}). 

\end{enumerate}

\begin{figure}
\plotone{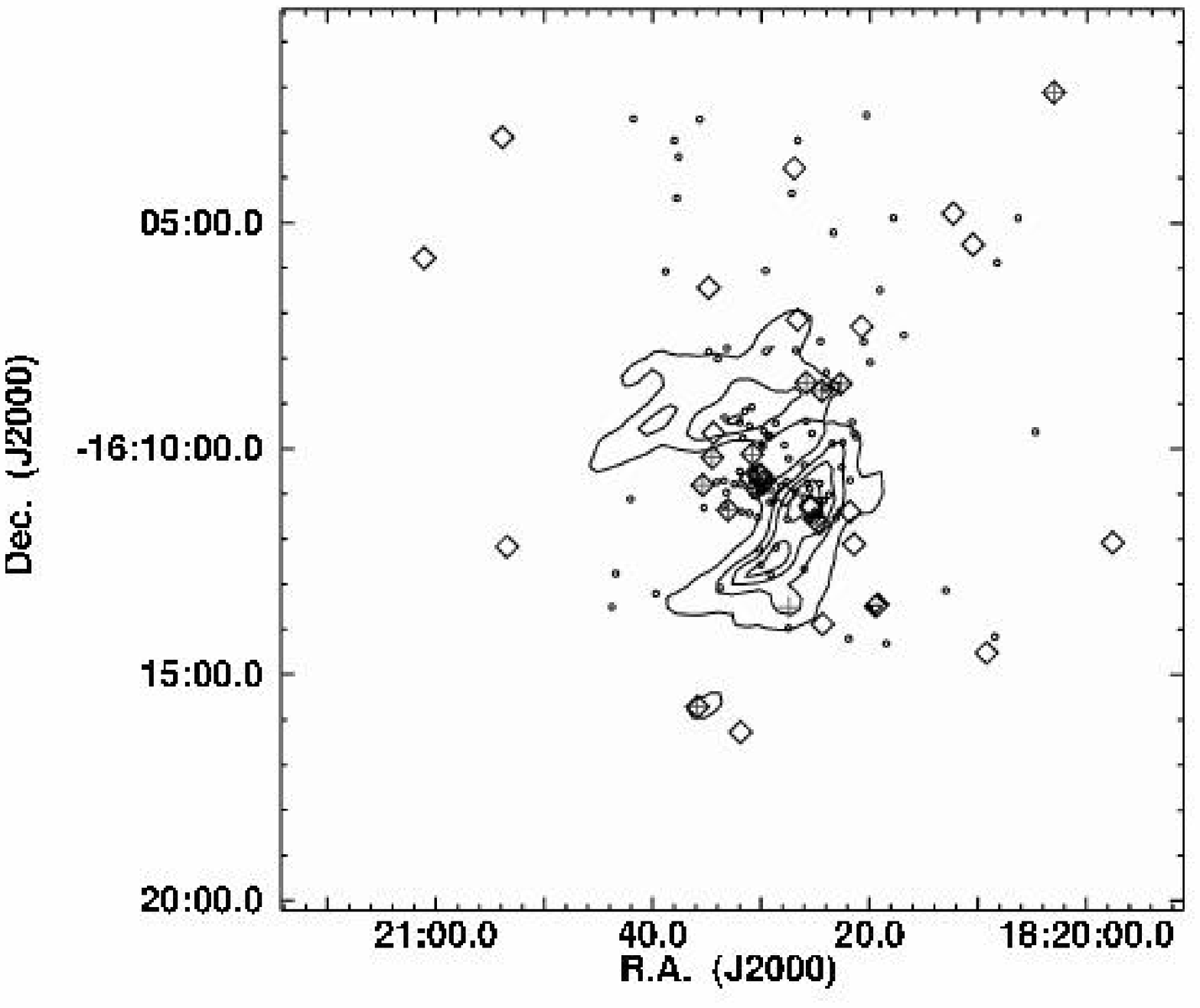} 

\caption{Spatial distribution of X-ray selected intermediate-mass stars 
($2 \lesssim M \lesssim 8$\Msol, small circles)  and high-mass stars 
($M \ga 8$\Msol, diamonds). The {\em MSX} 8$\mu m$ contours are 
overlaid.  
The pluses mark previously known O stars (\S \ref{sec:OB} and Table~\ref{OB.tab});
the known O star not marked as high-mass (missing its diamond) was detected (ACIS \#433 $\Leftrightarrow$ OI~345) but no mass estimate is reported in Table~\ref{tbl:highmass} due to large photometry errors.
\label{fig:highmass_spatial}}
\end{figure}

The spatial distributions of the intermediate- and high-mass stars detected by {\em Chandra} are shown in Figure~\ref{fig:highmass_spatial} based on mass estimates described in \S \ref{sec:NIRprop} and Table~\ref{tbl:highmass}.
As expected, the highest concentration of massive stars is located 
around the known bright O stars in NGC~6618. Other 
concentrations of interest include: M17-SW centered on 
($18^h20^m24^s,-16^\circ11\arcmin 30\arcsec$) around UC1 (\S \ref{sec:emb_UC1}), 
and a tight grouping of three stars around the KWO (\S \ref{sec:emb_KW}).  


It is also important to note that several dozen intermediate- and 
high-mass stars are distributed widely throughout the ACIS field.  
These mostly are lightly obscured and several were identified as 
possible OB stars from the early $UBV$ photometric study of 
\citet{Ogura76}.  Most of these stars do not appear to be extremely 
young.  Among the intermediate-mass stars, $\sim$100 have NIR colors 
consistent with reddened Class III objects while about a dozen have $K$-band 
excesses characteristic of Herbig Ae/Be stars (Figure \ref{fig:ccd}).

\subsection{Quantifying the Stellar Population with the XLF
\label{sec:XLF}}

Establishing and understanding the Initial Mass Function (IMF) that
arises from star formation processes are major goals underlying
studies of young stellar clusters \citep{Corbelli05}. This effort is
subject to a variety of difficulties. One of the major problems is
the huge level of background and foreground contamination seen in
optical and NIR observations of most clusters. Statistical correction for contamination
using observations of nearby control fields or using Galactic stellar
models can give uncertain results. \citet{Jiang02} used Galactic
models of foreground and (extinguished) background stellar populations
to correct the observed $K$-band luminosity function (KLF) in regions of the
SIRIUS field.  The results were satisfactory in the limited mass
range $0.5 \lesssim M \lesssim 2$\Msol, but uncertainties in contamination
affected the IMF measurement at higher masses.
For example, Jiang et al.\ found that
it was not possible to derive a corrected KLF for region
2C (Figure~\ref{fig:K_17x17}) where the estimated NIR contamination was higher than the
number of detected stars.

X-ray surveys, on the other hand, suffer relatively little
contamination (\S \ref{sec:contamination}), and the statistical
link between X-ray luminosities and masses (\S
\ref{sec:stellarpop}) permits an association between the X-ray
Luminosity Function (XLF) and the IMF.  This idea was recently
proposed \citep{Feigelson05a} and applied to {\em Chandra}
populations in the OMC-1 \citep{Grosso05}, Cep B/Cep OB3b
\citep{Getman06a} and NGC~6357 \citep{Wang06} regions. Through
their independent analyses, Getman et al.\ further confirm the
tight connection between the XLF and the IMF.

As the total band luminosity systematically underestimates the
true emission for obscured objects due to the absence of soft band
photons, we restrict our analysis to the X-ray luminosities in the
hard (2--8)~keV energy band: the observed luminosity $L_h$ and
that corrected for absorption $L_{h,c}$. These quantities
are available for $\sim$600 of the brighter sources in M17 (\S
\ref{sec:fluxes}).

\begin{figure}[t]
\centering
\includegraphics[angle=0.,width=6.5in]{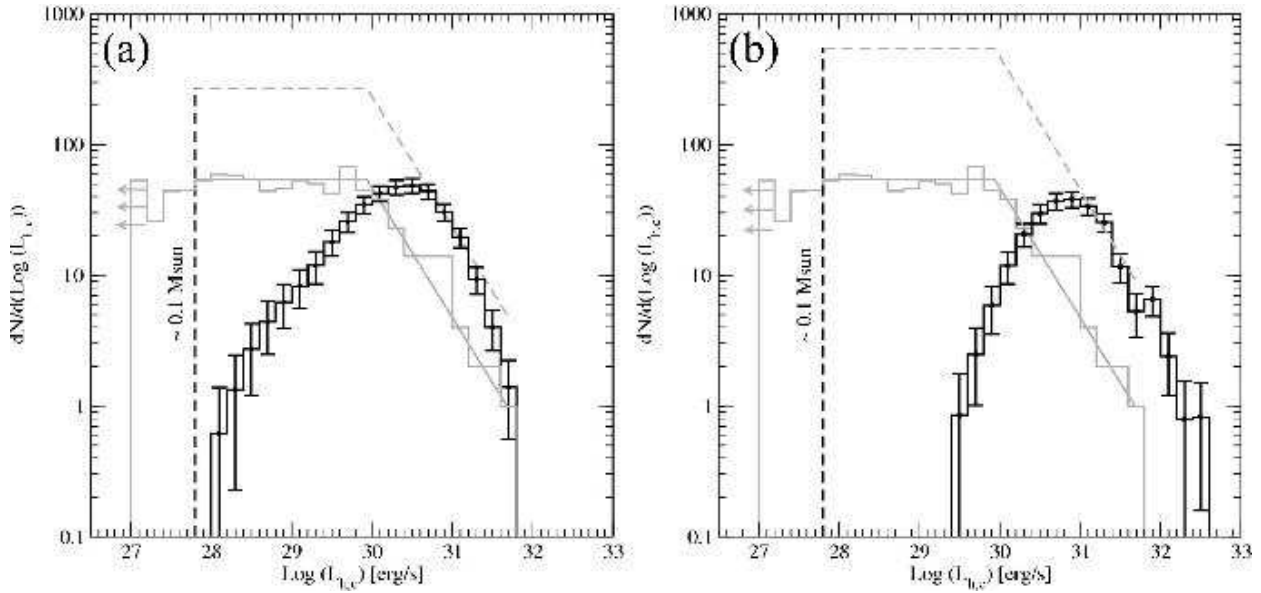} 

\caption{Hard band X-ray luminosity functions (corrected for
absorption) of the lightly obscured (black histogram with error
bars, left panel) and heavily obscured (right panel) samples of X-ray sources in M17. A calibration
XLF from the Orion Nebula Cluster is shown as the gray histogram; 
luminosities below $10^{27}$ ergs/s occupy the left-most bin.
Regression lines to the ONC XLF (gray, solid lines) were scaled
upward to match the M17 XLFs (gray, dashed lines). The ONC sample
is $100\%$ complete down to $\sim 0.1$\Msol (black dashed
line). \label{fig:xlf}}
\end{figure}

Following the approach taken in \citet{Getman06a} and in \S
\ref{sec:spatial}, we divide the 886 M17 ACIS X-ray sources into
514 lightly obscured and 372 heavily obscured samples using a
threshold of median energy $E_{median} = 2.5$~keV, corresponding
to $A_V \sim 10$~mag. Figure~\ref{fig:stellar_density} shows the
spatial distributions of these two samples and
Figure~\ref{fig:xlf} shows the XLF comparison analysis for
these samples using the absorption corrected X-ray luminosity
$L_{h,c}$. 
For brevity, the similar analysis using $L_h$ is not described in detail.
The $\sim$5\% contamination due to unrelated Galactic or extragalactic sources (\S \ref{sec:contamination}) is ignored in the following analysis.


The lightly obscured Orion Nebula Cluster (ONC) has both the XLF and IMF
accurately measured \citep{Feigelson05b,Getman06a}. 
We assume here, as in \citet{Getman06a} and \citet{Wang06}, that the XLFs of young stellar clusters have the same powerlaw slope at high luminosity and thus the ONC can reasonably serve as a calibrator for more distant clusters.  
XLFs should be relatively insensitive to age effects; \citet{PreibischFeigelson05} found only a slight ($\sim$0.3 in log $L_t$) decline in luminosity between 1 and 10 Myr old ONC populations.
Specifically, the ONC comparison sample consists
of 839 COUP cloud members, excluding OB stars and sources with
absorbing column density $\log N_H > 22.0$~cm$^{-2}$. The ONC sample
is nearly complete down to 0.1\Msol and about 50\% complete within
the low-mass stellar-sub-stellar range of 0.03-0.1\Msol
\citep{Preibisch05b,Getman06a}.

Luminosities are available for 70\% (361 out of 514) and 63\% (234
out of 372) of the lightly and heavily obscured M17 samples,
respectively. Uncertainties, $\Delta \log
L_{h,c}$, are estimated for each luminosity based on the number of
detected counts, using the simulated flux vs.\ flux
uncertainty correlation from Figure 12 of \citet{Getman06a}. 

Examination of Figure~\ref{fig:xlf} shows that the ONC and M17
XLFs are nearly parallel at the high luminosity end, and the M17 source
counts turn over due to completeness limits around $\log L_{h,c}
\sim 30.5$~ergs~s$^{-1}$ ($M \sim 1.7$\Msol) for the lightly
obscured sample and $\log L_{h,c} \sim$30.9~ergs~s$^{-1}$ ($M \sim
2.6$\Msol) for the heavily obscured sample. We thus scale
the ONC XLF regression line vertically upward to estimate the
total population of M17 in the ACIS field.  Due to systematic
uncertainties, the XLF-$L_h$ and XLF-$L_{h,c}$ analyses provide
slightly different estimates for the total population of the lightly
obscured sample in M17 down to 0.1\Msol, so we report a range for this quantity of 
$2500-3500$ stars. About 60\% of this population lies in the concentrated
NGC~6618 cluster. Similar uncertainty for the heavily obscured population gives an
estimated range of $5000-7000$ obscured objects. Thus the total
estimated population of young objects in M17 down to 0.1\Msol
within the $17\arcmin \times 17\arcmin$ ACIS-I field
is 7500--10500 stars.

\section{X-rays from embedded populations \label{sec:embedded}}

\subsection{The M17-UC1 region \label{sec:emb_UC1}}

Radio and IR studies have shown that star formation around the M17 
\hii region is most active in the M17-SW cloud.  The best-studied 
subregion is around the massive binary IRS~5N and 5S with separation 
5\arcsec\/ \citep[e.g.][]{Felli84, Johnson98, Chini00, Nielbock01}.  
IRS~5N has bolometric luminosity $L_{bol} \sim 5000$~L$_\odot$ 
peaking in the MIR and ionizes the rapidly evolving 
ultracompact \hii region M17-UC1 $\Leftrightarrow$ G15.04-0.68.  IRS~5S with $L_{bol} 
\sim 2000$~L$_\odot$ is radio-quiet and less absorbed.  They appear 
to be massive Class I protostars, probably early-B stars. The 
interstellar matter around these central stars is complex with dense 
dusty cloud clumps and an arc-shaped ionization front.  

\begin{figure}
\centering 
\plottwo{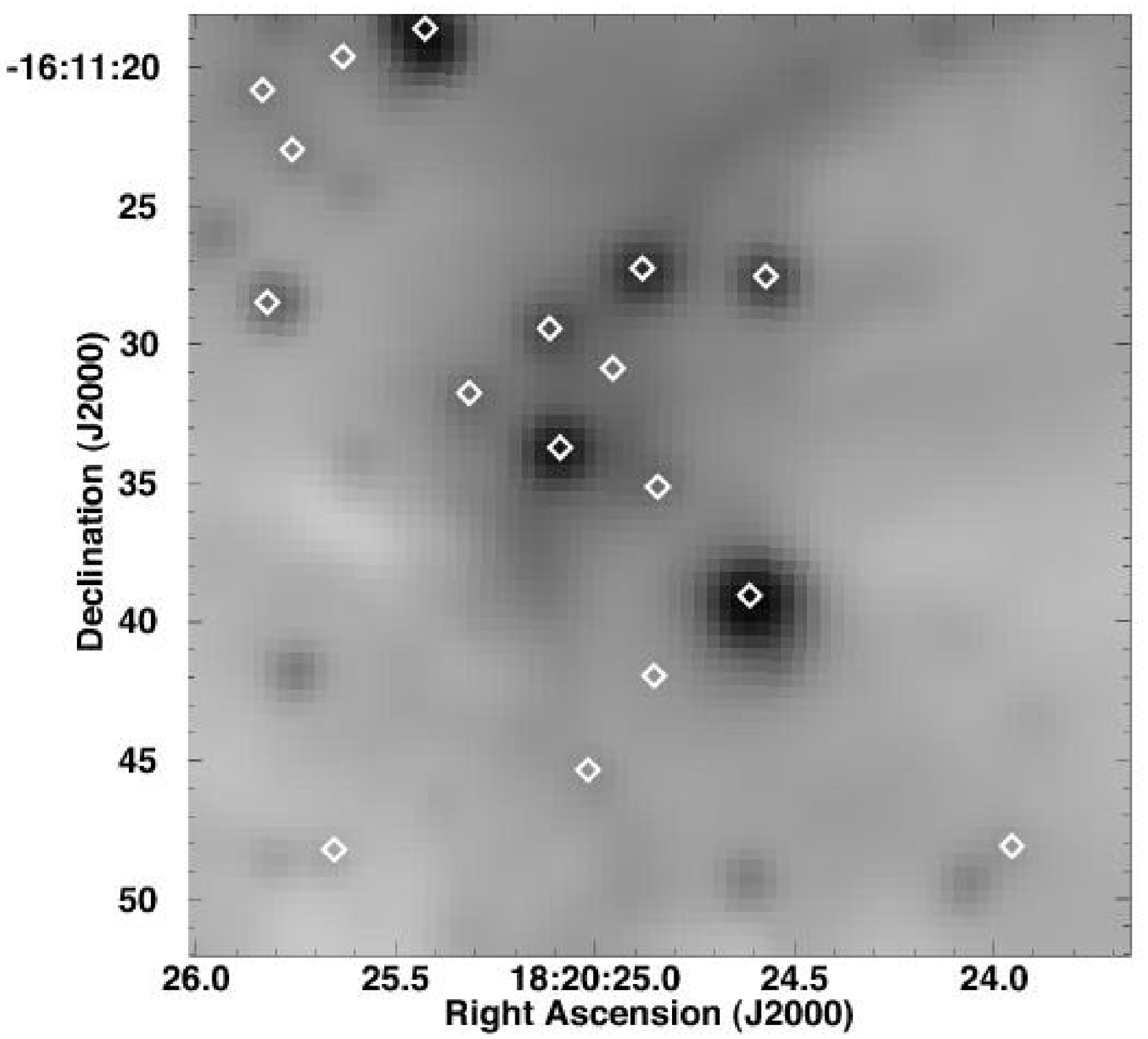}{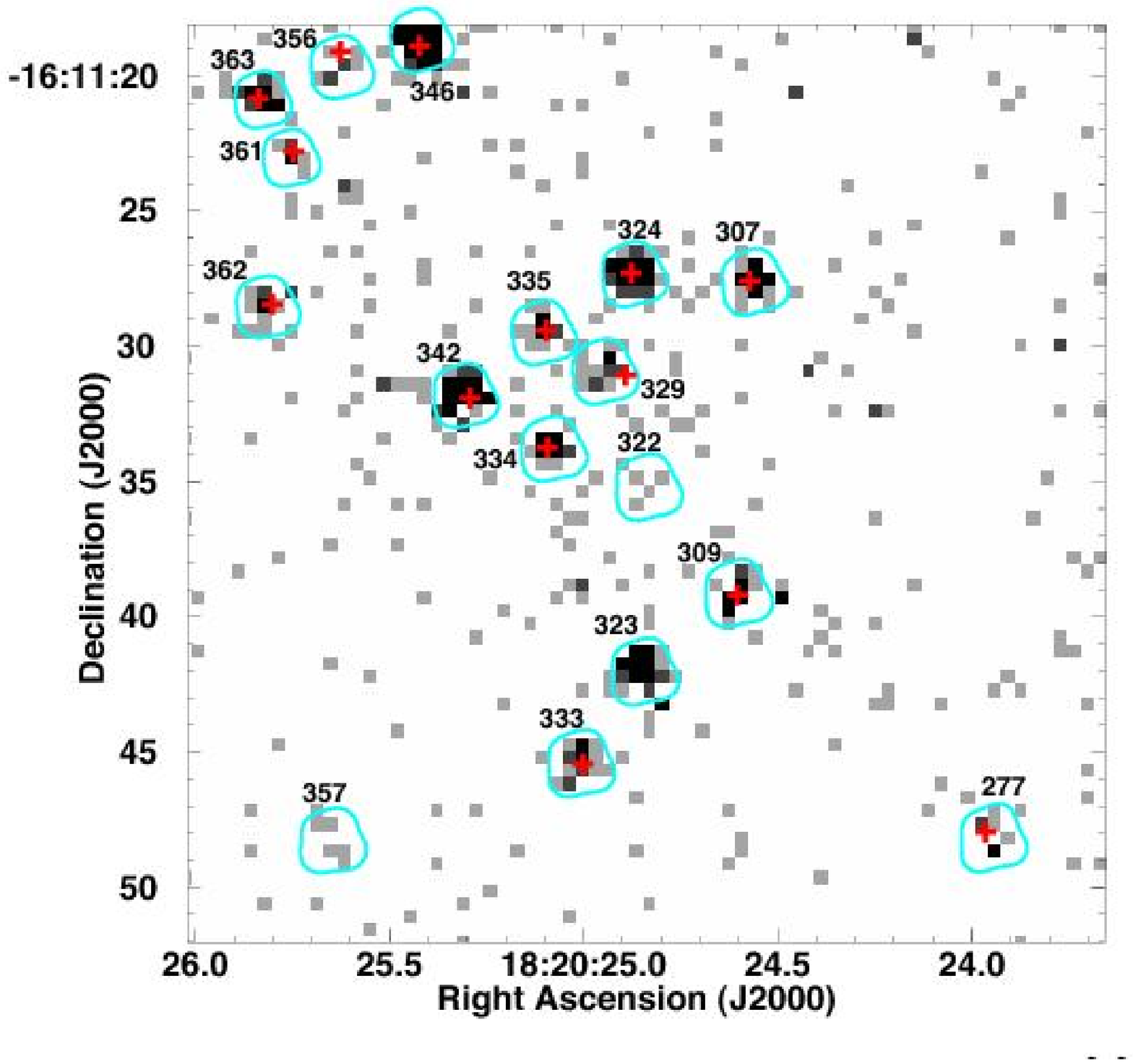} 
\plottwo{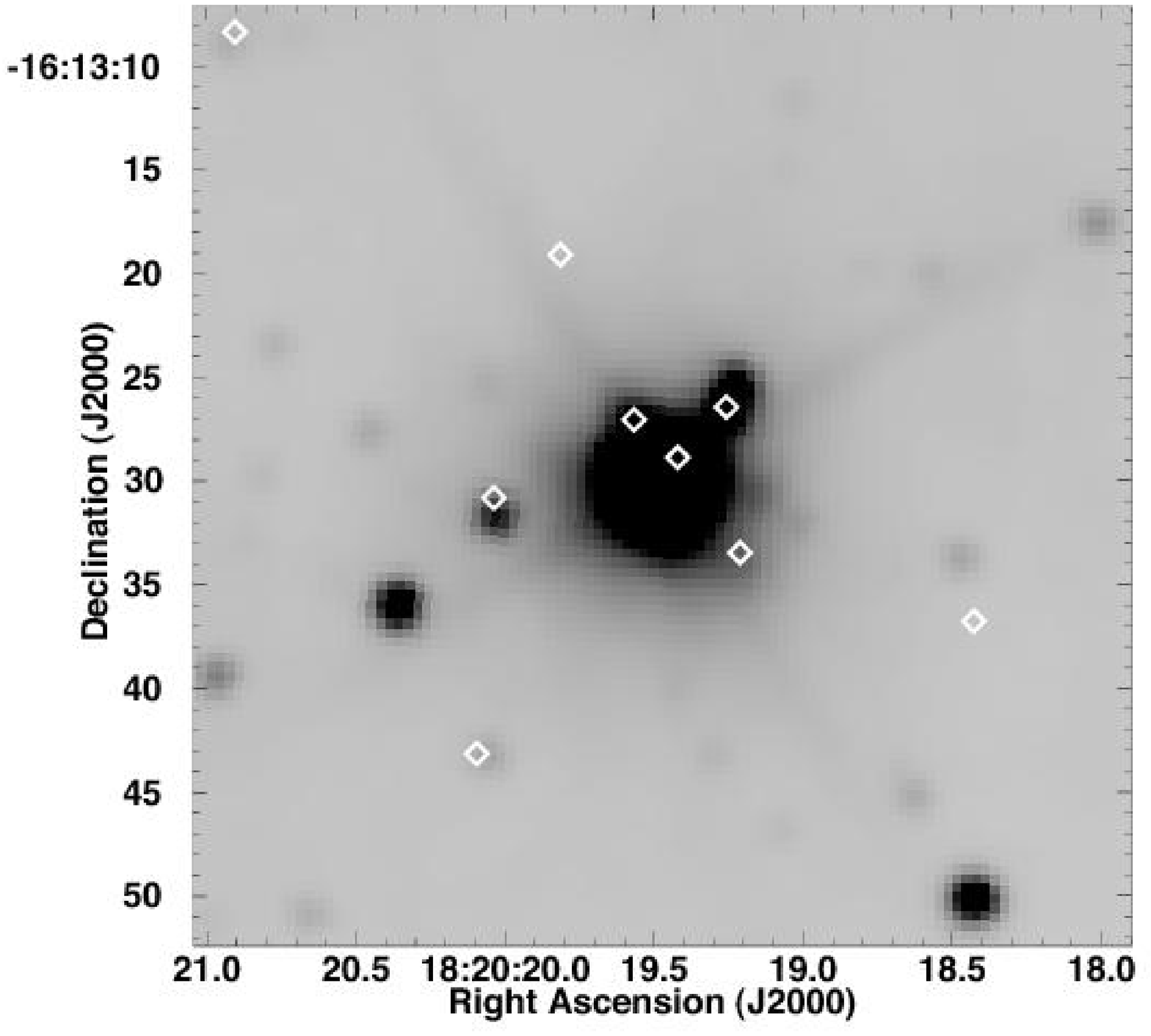}{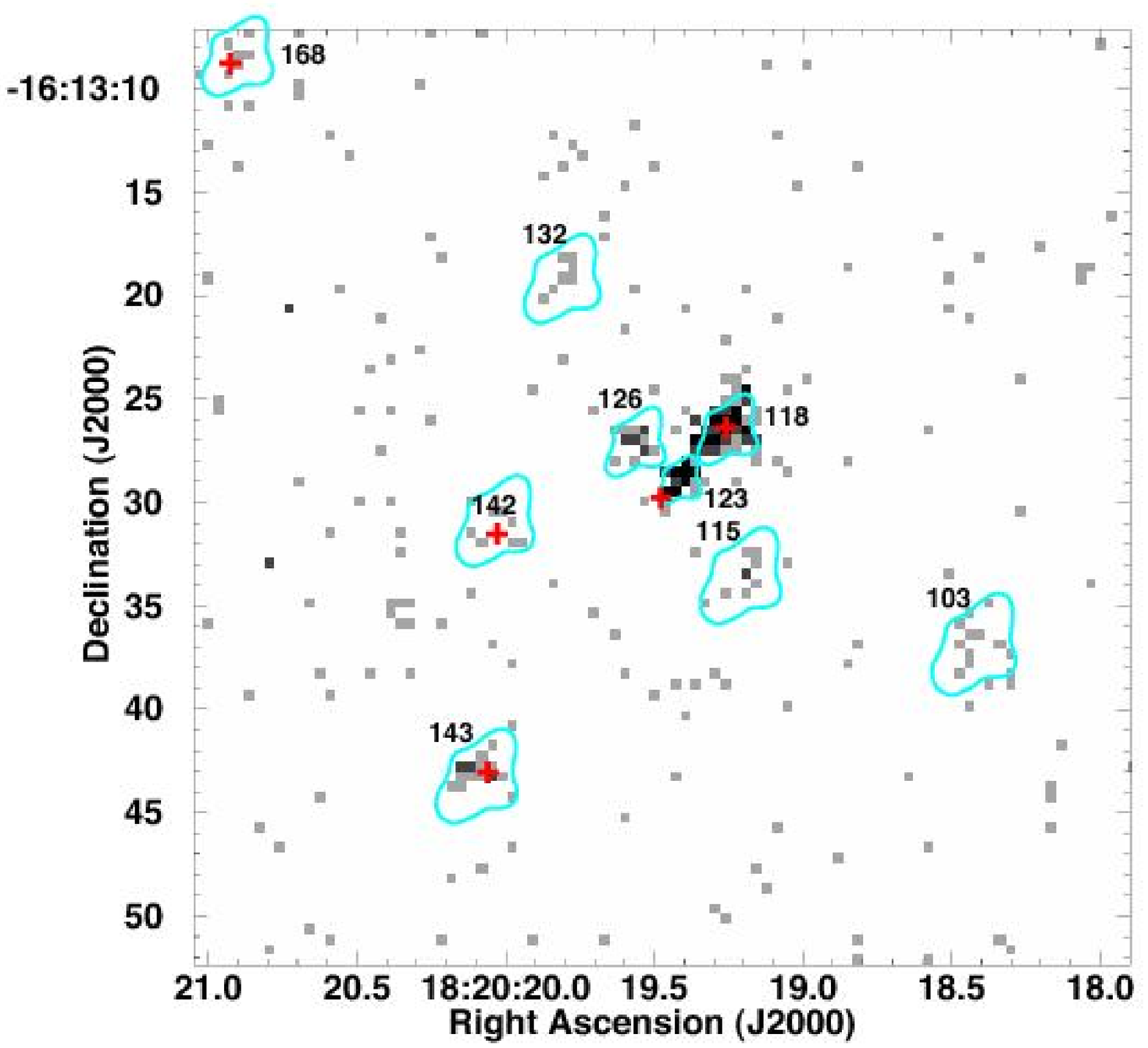} 

\caption{M17-UC1 (upper panels, \S \ref{sec:emb_UC1}) and the Kleinmann-Wright Object (lower panels, \S \ref{sec:emb_KW}) seen in the SIRIUS $K$-band (left) and {\em Chandra} X-ray (right) images.  
The X-ray gray scales span 1 to 21 counts (upper right) and 1 to 9 counts (lower right) in $0.5\arcsec \times 0.5\arcsec$\/ pixels.  
ACIS sources are marked by diamonds (left) and sequence numbers (right); source extraction regions matching the {\em Chandra} Point Spread Function are shown in blue. 
SIRIUS counterparts (+) range from 9 to 14 mag (upper left) and from 7.5 to 15.5 mag (lower left) in K.
The radio arc of ionized gas associated with M17-UC1 can been faintly seen in the $K$-band image (upper left).  
ACIS \#309 matches the brightest IR component of IRS~5S.
\label{fig:UC1_K-X} } 
\end{figure}

The top panels of Figure~\ref{fig:UC1_K-X} show the deep $K$ and {\em Chandra} images of 
IRS~5 and its vicinity.  
ACIS \#322 did not arise from the source finding process but was manually added to the catalog at the known position of the MIR source IRS~5N $\Leftrightarrow$ UC~1.
Although its formal detection significance 
(column 13 of Table~\ref{tbl:src_properties_tentative}) is low (0.9), confidence that X-ray emission is seen 
at that position is bolstered by the fact that the 5 nearby X-ray 
photons are harder ($E_{median}=4.7$~keV) than the local background ($E_{median}=3.5$~keV). 
IRS~5S (\#309) is 
definitely present with 19 photons.  Its X-ray spectrum indicates 
absorption of $\log N_H \sim 22.8$~cm$^{-2}$ and intrinsic 
(absorption-corrected) hard band luminosity $\log L_{h,c} \sim 
30.9$~ergs~s$^{-1}$.  This source was recently resolved into 3 components using a high-resolution $K$-band image \citep{Chini05}; the X-ray source is spatially associated with the brightest $K$-band component.


A cluster of $\sim$20 X-ray stars is present within 20\arcsec\/ 
(0.15~pc) radius of the IRS~5 binary, most with NIR counterparts.   
These sources are heavily absorbed and have intrinsic 
 X-ray luminosities near the top of the XLF. 
Source \#324 is associated with the NIR star B~279 ($K = 
10.75$~mag) which is likely responsible for powerful methanol and OH masers 
\citep{Norris87}.  It emits $\log L_{h,c} \sim 30.9$~ergs~s$^{-1}$ 
with absorption $\log N_H \sim 22.4$~cm$^{-2}$.  Source \#334 is 
associated with the IR star B~273 with $K=10.3$~mag and 
strong CO bandhead absorption but no emission lines \citep{Chini05}.  
It emits $\log L_{h,c} \sim 30.5$~ergs~s$^{-1}$ with $\log N_H \sim 
22.4$~cm$^{-2}$.  Chini et al.\ interpret this source as an FU Orionis type object, a massive young star with heavy accretion.

This concentration of $\sim$20 X-ray luminous stars around IRS~5 has 
about 1/5 the number of similarly luminous stars seen in the lightly 
obscured NGC~6618 cluster (\S \ref{sec:XLF}), implying that the total population of the 
embedded cluster is $\sim$300 stars.  Source \#329, one of the 
weaker X-ray sources in the region, showed flaring 
behavior and is probably one of these lower mass stars.  However, 
this inference is quite uncertain, and inconsistent with the 
relatively small population found in the Becklin-Neugebauer region 
behind the Orion Nebula which is similar to the M17-SW IRS~5/UC1 
complex.  In Orion, the deep COUP image shows only a sparse cluster 
of lower mass stars accompanying the IR-luminous group of proto-B 
stars \citep{Grosso05}.  We look forward to more directly measuring 
the population around IRS~5 in an upcoming deeper {\em Chandra} observation.

A group of dense molecular cores  
\citep[Peak 3 in the survey of][]{Vallee96} and masers are concentrated about 30\arcsec\/ west 
of M17-UC1. \citet{Johnson98} argue that this region is the one most 
recently triggered by the shock from the main M17 \hii region.   
Two ACIS sources (\#239 and \#244) are positionally coincident with the luminous Class I IR source 
Anon~1 (\S \ref{sec:protostars}), reported by \citet{Nielbock01} to have an extremely steep spectral index.
The properties of source \#239 suggest it is the more likely true association; it is unusually X-ray luminous, constant, and heavily absorbed ($\log L_{h,c} \sim 31.7$~ergs~s$^{-1}$ and $\log N_H \sim 23.5$~cm$^{-2}$), and has no identified NIR or MIR counterpart.   
The observed absorption is roughly half of the estimated total absorption through Peak 3 
\citep{Hobson93}, consistent with the star lying at the center of the core.  
The second ACIS source, \#244, shows flare-like variability (Figure~\ref{fig:lightcurve}) and more typical properties ($\log L_{h,c} \sim 31.1$~ergs~s$^{-1}$, $\log N_H \sim 22.2$~cm$^{-2}$, $K \sim 12$~mag counterpart), suggestive of a young star lying in front of the star forming cores though still inside the M17 giant molecular cloud.

\subsection{The Kleinman-Wright Object and its cluster \label{sec:emb_KW}}

The lower panels of Figure~\ref{fig:UC1_K-X} show the deep $K$ and {\em Chandra} images 
around the IR-luminous star known as the Kleinmann-Wright 
Object (KWO) or M17-IRS1 \citep{Kleinmann73b}.  IR study reveals a 
photosphere with $T_{eff} = 25,000$ K and $L_{bol} = 5 \times 
10^3$~L$_\odot$ with strong variable hydrogen emission lines 
\citep{Chini04a}.  It is modeled as a Herbig Be star of spectral type B0V lying inside a 
dusty envelope with $M \sim 10$\Msol, $r \sim 0.1$~pc, and $N_H \sim 2 \times 10^{22}$~cm$^{-2}$.  It may be the most heavily 
absorbed and youngest Herbig Be star known, with a Class I 
protostellar IR spectral energy distribution.  Chini et al.\ resolve a nearby bright NIR source (KWO component 2) lying 1.2\arcsec\/ northwest of the Herbig Be star (KWO component 1).  Chini et al.\ 
further find a concentrated cluster of about 150 heavily reddened 
$JHK$ stars within 0.7 pc (70\arcsec) of the KWO.

ACIS source \#123, with 31 counts, lies 1.2\arcsec\/ northwest of SIRIUS source 15128; SIRIUS does not resolve the two components described by \citet{Chini04a}, although the SIRIUS J-band image appears elongated to the northwest, in the direction of the ACIS source.  Comparing the SIRIUS and ACIS positions to components 1 and 2 of the KWO, 
based on positions obtained by Chini et al.\ in the $i$-band with 
the ESO NTT telescope, it appears that the SIRIUS source closely matches KWO component 1 (the Herbig Be star, which is brighter in the NIR), while ACIS \#123 closely matches KWO component 2.  The Chini et al.\ positions are systematically offset from the ACIS position and the SIRIUS position (after registration to ACIS/2MASS) by $\sim 1\arcsec$ to the southwest. 
The KWO Herbig Be star is not detected in this {\em Chandra} observation.
\citet{Chini04a} report that component 2 ($\Leftrightarrow$ ACIS \#123) is probably a member of the KWO cluster.  
It exhibits $\log L_{h,c} \sim 31.0$~ergs~s$^{-1}$ and $\log 
N_H \sim 22.5$~cm$^{-2}$ equivalent to $A_V \sim 20$~mag.
The situation resembles that of the Becklin-Neugebauer Object in the OMC-1 where a previously unknown lower-mass companion greatly outshines the young B star in X-rays \citep{Grosso05}.

A dozen other ACIS sources are present within 30\arcsec\/ of the KWO, a 
higher density than generally seen in the M17-SW cloud though less 
rich than seen around IRS~5/UC~1 (\S \ref{sec:emb_UC1}).  Most of these 
sources have $11<K<15$~mag counterparts consistent with 
intermediate-mass stars in the embedded cluster.  As with UC1, an upcoming 
deeper {\em Chandra} observation should detect more members of the cluster around 
the KWO and perhaps the Herbig Be star itself.

\subsection{The M17-North cloud \label{sec:emb_M17N}}

The M17-North cloud cores are dusty molecular concentrations 
extending over 2\arcmin\/ at a location around 10\arcmin\/ north  of 
the bright M17 \hii region \citep{Wilson03}.  Star formation here is 
probably occurring without triggering by the \hii region.  It has five 
embedded IR sources; radio and ionized line emission suggest 
that IRS 3 is a late-O star behind the cloud core \citep{Klein99}. 
IRS 1 and 3 are off the ACIS field; IRS 2 is at the field edge; IRS 4 and 5 are not detected in the {\em Chandra} image.  However, the ACIS 
image (Figure~\ref{fig:ACIS_17x17}) and the smoothed map of 
obscured ACIS sources (Figure~\ref{fig:stellar_density}, right panel) 
show a density enhancement of a dozen absorbed sources.  For example, 
source \#618 lying 10\arcsec\/ from M17-North~IRS~4 has $\log L_h \sim 
30.8$~ergs~s$^{-1}$ and $\log N_H \sim 22.5$~cm$^{-2}$ without any 
known counterpart.  Two sources (\#660 and 665) lie 6\arcsec\/ from 
each other, again without $K$-band counterparts.  

Assuming a standard XLF (\S \ref{sec:XLF}, Figure~\ref{fig:xlf}), the 
presence of this group of X-ray luminous stars implies that dozens or 
hundreds of lower mass stars are present.  As with the IRS~5/UC1 and 
KWO clusters, the population of the clusters cannot be 
reliably discerned from the present shallow X-ray exposure.  However, 
we note that the (probably untriggered) M17-North group appears to 
be spread over a larger region ($\sim$2\arcmin or 1 pc) than 
the (probably triggered) clusters in the M17-SW cloud.

\subsection{Protostar candidates \label{sec:protostars}}

{\em Chandra} studies of nearby low-mass star forming regions (such 
as the Ophiuchus, Taurus, and Serpens clouds) have established that 
Class I protostars are X-ray emitters with typical luminosities of 
$\sim$$10^{30}$~ergs~s$^{-1}$, flaring light curves, and hard spectra 
\citep[e.g.][]{Imanishi01, Getman05a, Preibisch04}.  The case for 
X-ray emission from Class 0 sources is problematic 
\citep{Tsujimoto05,Getman06b}.  Systematic studies of X-ray protostars in 
MSFRs, however, have not yet emerged.

From the $JHK$ study of \citet{Jiang02}, 157 candidate Class 
I protostars were reliably 
identified from strong $K$-band photometric excesses.  
About half lie in the North Bar and South Bar in the \hii 
region, while the other half are widely distributed across their 
field.  Others can be identified by MIR excesses in the GLIMPSE 
survey.

Table~\ref{protostars.tab} lists the 64 {\em Chandra} sources associated 
with protostellar candidates satisfying one or more of the following 
criteria: 
\begin{enumerate}

\item Strong $K$-band excess in the $(J-H)$ vs.\ $(H-K)$ diagram
beyond the locus of reddened Class II T Tauri stars (Figure~\ref{fig:ccd}).
\item Excess in the shortest wavelength band of the {\em Spitzer} IRAC
detector.  Observations of Taurus-Auriga young stars with reliable 
classifications show that the color criterion \mbox{$K-[3.6]>1.5$~mag} isolates 
Class 0-I protostellar systems from Class II-III T Tauri systems 
\citep{Hartmann05}.
\item Excess in the second IRAC band with \mbox{$[3.6]-[4.5]>0.7$~mag} also
separates protostars from T Tauri stars \citep{Hartmann05}.

\end{enumerate}
Uncertainties on the $K-[3.6]$ and $[3.6]-[4.5]$ colors were calculated, and only sources exceeding the stated color thresholds by more than 1 standard deviation are reported.
X-ray luminosities in the hard band are reproduced from 
Table~\ref{tbl:thermal_spectroscopy} when available; for the fainter 
sources, we approximate $\log L_h \sim 29$~ergs~s$^{-1}$, based on the ratio between counts and $L_h$ for heavily absorbed brighter stars in Table~\ref{tbl:thermal_spectroscopy}.
The luminosities range up to $10^{31.0}$~ergs~s$^{-1}$, similar to the distribution of luminosities of 
Class I sources in the $\rho$ Ophiuchi cloud \citep{Imanishi01}. 

Only 13 out of 157 Class I candidates identified by $K$-band excess (criterion 1 above) are 
detected. The paucity of extreme $K$-band excess sources can also be 
seen in the NIR color-color diagram discussed in \S \ref{sec:NIRprop} (Figure~\ref{fig:ccd}). 
The median K magnitude ($\sim$13.5) for the 13 X-ray detected sources is 1 magnitude brighter than
the median for the 157 Class I NIR candidates, suggesting that, at the sensitivity limits of our
observation, only the higher-mass end of the protostar XLF is detected.
An upcoming longer {\em Chandra} observation should detect lower-mass Class I candidates.

An additional 51 Class I candidates in Table~\ref{protostars.tab} identified by colors from longer wavebands (criteria 2 and 3 above), more sensitive to circumstellar disks, are detected.
Many of these are highly obscured; half have $E_{median} > 3$~keV (corresponding to $A_V \ga 15$~mag).
These 51 sources (and perhaps some of the $\sim$80 X-ray sources without counterparts described in \S \ref{sec:isolated}) represent a significant population of Class I candidates that were not previously available from NIR studies.


\begin{deluxetable}{rcccccccc}
\tablecaption{Photometrically selected X-ray emitting candidate
protostars 
\label{protostars.tab}} 

\tablecolumns{9}
\tabletypesize{\small}

\tablehead{ \colhead{Seq} & \colhead{CXOU} & \colhead{K} & 
\colhead{$\log L_h$} & \colhead{$\log N_H$} & \colhead{$E_{median}$} &
\multicolumn{3}{c}{Photometric Criteria} \\ \cline{7-9}

&& \colhead{(mag)} & \colhead{(erg s$^{-1}$)} & \colhead{(cm$^{-2}$)} & \colhead{(keV)} & \colhead{JH,HK} & 
\colhead{K-[3.6]} & \colhead{[3.6]-[4.5]} \\

\colhead{(1)} & \colhead{(2)} & \colhead{(3)} & \colhead{(4)} & 
\colhead{(5)} & \colhead{(6)} & \colhead{(7)} & \colhead{(8)} & 
\colhead{(9)}
}

\startdata
   7\tnm{a} & 182000.65$-$161112.0  & 10.88   & 30.40       & 22.0    & 2.0 & \nodata & $\surd$   & \nodata \\
  21\tnm{a} & 182006.33$-$160453.7  & 12.52   & 29:\phn\phn & \nodata & 3.3 & \nodata & $\surd$   & \nodata \\
  26        & 182008.21$-$161040.9  & 13.72   & 30.33       & 22.9    & 2.8 & \nodata & $\surd$   & \nodata \\
  30\tnm{a} & 182008.44$-$161409.6  & 11.62   & 30.71       & 21.8    & 1.9 & \nodata & $\surd$   & \nodata \\
  47        & 182011.77$-$160512.2  & 14.35   & 29:\phn\phn & \nodata & 2.3 & \nodata & $\surd$   & \nodata \\
  51\tnm{a} & 182012.31$-$160447.7  & 10.35   & 30.27       & 22.3    & 2.3 & \nodata & $\surd$   & \nodata \\
  53        & 182012.44$-$160610.0  & 15.73   & 29:\phn\phn & \nodata & 1.1 & $\surd$ & \nodata   & \nodata \\
  57\tnm{a} & 182012.96$-$161308.4  & 12.46   & 29:\phn\phn & \nodata & 2.0 & \nodata & $\surd$   & \nodata \\
  58        & 182013.09$-$161238.1  &\nodata  & 29:\phn\phn & \nodata & 3.4 & \nodata & \nodata   & $\surd$ \\
  60        & 182013.48$-$160912.2  & 14.45   & 29:\phn\phn & \nodata & 1.8 & \nodata & $\surd$   & \nodata \\
  74        & 182015.80$-$161033.2  & 14.99   & 29:\phn\phn & \nodata & 2.3 & \nodata & $\surd$   & \nodata \\
  81        & 182016.54$-$161003.0  & 13.83   & 30.99       & 23.1    & 4.1 & \nodata & $\surd$   & \nodata \\
  94\tnm{a} & 182017.82$-$160453.4  & 13.67   & 31.11       & 22.3    & 3.3 & \nodata & $\surd$   & \nodata \\
  95        & 182017.82$-$161401.2  & 13.75   & 30.79       & 22.8    & 3.7 & \nodata & $\surd$   & \nodata \\
 105\tnm{a} & 182018.44$-$161419.0  & 12.22   & 30.52       & 22.6    & 2.0 & \nodata & $\surd$   & \nodata \\
 106        & 182018.51$-$161421.4  & 12.76   & 30.38       & 22.2    & 2.6 & \nodata & $\surd$   & \nodata \\
 108        & 182018.65$-$160607.8  & 14.07   & 29:\phn\phn & \nodata & 4.1 & \nodata & $\surd$   & \nodata \\
 109        & 182018.83$-$161424.8  &\nodata  & 30.45       & 22.8    & 3.6 & \nodata & \nodata   & $\surd$   \\
 113\tnm{a} & 182019.06$-$160629.6  & 12.96   & 30.84       & 22.9    & 3.1 & \nodata & $\surd$   & \nodata \\
 117        & 182019.24$-$160800.3  & 13.01   & 30.44       & 22.5    & 2.7 & \nodata & $\surd$   & $\surd$ \\
 128        & 182019.60$-$161038.8  & 11.92   & 30.45       & 23.2    & 4.8 & $\surd$ & \nodata   & \nodata \\
 137\tnm{a} & 182019.95$-$160805.2  & 12.44   & 29:\phn\phn & \nodata & 3.1 & \nodata & $\surd$   & \nodata \\
 156        & 182020.53$-$160456.3  & 14.15   & 30.84       & 22.6    & 3.1 & \nodata & $\surd$   & \nodata \\
 179\tnm{a} & 182021.25$-$160944.3  & 12.45   & 29:\phn\phn & \nodata & 2.5 & \nodata & $\surd$   & \nodata \\
 200\tnm{a} & 182021.79$-$161042.0  & 12.81   & 30.72       & 22.6    & 2.6 & \nodata & $\surd$   & \nodata \\
 201\tnm{a} & 182021.82$-$161123.6  & 11.36   & 29:\phn\phn & \nodata & 4.4 & \nodata & $\surd$   & \nodata \\
 207        & 182021.97$-$161100.6  & 13.31   & 29:\phn\phn & \nodata & 3.1 & $\surd$ & \nodata   & \nodata \\
 209        & 182022.01$-$161024.8  & 14.06   & 29.99       & 22.3    & 2.0 & $\surd$ & \nodata   & \nodata \\
 221        & 182022.40$-$161403.9  & 13.29   & 29:\phn\phn & \nodata & 3.2 & \nodata & $\surd$   & \nodata \\
 234        & 182022.70$-$160951.5  & 13.88   & 30.36       & 22.6    & 2.7 & $\surd$ & \nodata   & \nodata \\
 235        & 182022.70$-$161613.8  & 14.90   & 29:\phn\phn & \nodata & 3.5 & \nodata & $\surd$   & \nodata \\
 304        & 182024.51$-$160644.3  & 15.06   & 30.56       & 22.7    & 2.7 & \nodata & $\surd$   & \nodata \\
 305\tnm{a} & 182024.54$-$160737.6  & 12.00   & 30.18       & 22.2    & 3.7 & \nodata & $\surd$   & \nodata \\
 324\tnm{ab}& 182024.87$-$161127.5  & 10.75   & 30.74       & 22.4    & 2.7 & $\surd$ & \nodata   & \nodata \\
 328        & 182024.93$-$160459.5  & 14.38   & 29:\phn\phn & \nodata & 3.2 & \nodata & $\surd$   & \nodata \\
 363\tnm{b} & 182025.82$-$161121.0  & 13.61   & 30.32       & 22.2    & 2.1 & $\surd$ & \nodata   & \nodata \\
 372        & 182025.95$-$160938.2  & 13.63   & 29:\phn\phn & \nodata & 2.5 & \nodata & $\surd$   & \nodata \\
 380        & 182026.12$-$160455.9  & 14.24   & 31.07       & 22.8    & 3.6 & \nodata & $\surd$   & \nodata \\
 438\tnm{a} & 182027.62$-$161043.3  & 12.25   & 29:\phn\phn & \nodata & 3.8 & \nodata & $\surd$   & \nodata \\
 472        & 182028.28$-$161130.5  & 11.08   & 30.52       & 22.4    & 2.6 & $\surd$ & \nodata   & \nodata \\
 498        & 182028.95$-$160957.7  & 13.26   & 30.43       & 22.4    & 2.2 & \nodata & $\surd$   & \nodata \\
 500        & 182029.01$-$160831.5  & 14.52   & 29:\phn\phn & \nodata & 2.9 & $\surd$ & \nodata   & \nodata \\
 535        & 182029.80$-$161125.4  & 12.46   & 29.72       & 22.2    & 1.8 & \nodata & $\surd$   & \nodata \\
 562        & 182030.18$-$161216.5  & 14.15   & 29.86       & 22.4    & 2.3 & $\surd$ & \nodata   & \nodata \\
 595\tnm{a} & 182030.88$-$160905.1  & 11.47   & 30.55       & 21.9    & 2.1 & \nodata & $\surd$   & \nodata \\
 596        & 182030.90$-$160954.7  & 12.69   & 29:\phn\phn & \nodata & 2.3 & \nodata & $\surd$   & \nodata \\
 607\tnm{a} & 182031.09$-$161126.1  & 12.42   & 30.43       & 22.0    & 2.6 & \nodata & $\surd$   & \nodata \\
 610        & 182031.23$-$160218.2  & 12.71   & 30.83       & 22.7    & 3.1 & \nodata & $\surd$   & \nodata \\
 637        & 182031.70$-$160612.1  & 14.64   & 29:\phn\phn & \nodata & 4.3 & \nodata & $\surd$   & \nodata \\
 643        & 182031.79$-$160349.3  & 14.26   & 29:\phn\phn & \nodata & 2.9 & \nodata & $\surd$   & \nodata \\
 650\tnm{a} & 182031.89$-$161616.5  &  9.57   & 30.07       & 20.6    & 1.7 & $\surd$ & $\surd$   & \nodata \\
 652        & 182031.94$-$160239.7  & 13.13   & 31.30       & 22.8    & 3.8 & \nodata & $\surd$   & \nodata \\
 654\tnm{a} & 182031.97$-$160925.1  & 11.26   & 30.50       & 21.9    & 2.6 & \nodata & $\surd$   & \nodata \\
 668\tnm{a} & 182032.47$-$161047.2  & 13.54   & 29:\phn\phn & \nodata & 4.4 & \nodata & $\surd$   & \nodata \\
 677\tnm{a} & 182033.17$-$160746.6  & 12.10   & 29:\phn\phn & \nodata & 2.7 & \nodata & $\surd$   & \nodata \\
 679        & 182033.32$-$161050.9  & 13.63   & 29:\phn\phn & \nodata & 3.1 & \nodata & $\surd$   & \nodata \\
 682\tnm{a} & 182033.42$-$161042.7  & 11.79   & 29:\phn\phn & \nodata & 3.4 & \nodata & $\surd$   & \nodata \\
 709        & 182034.79$-$161053.8  & 11.66   & 30.47       & 21.8    & 1.9 & \nodata & $\surd$   & \nodata \\
 718        & 182035.34$-$161325.4  & 13.58   & 29:\phn\phn & \nodata & 2.3 & $\surd$ & \nodata   & \nodata \\
 757\tnm{a} & 182037.62$-$160332.6  & 12.00   & 30.95       & 22.3    & 2.7 & \nodata & $\surd$   & $\surd$ \\
 762        & 182037.99$-$160925.8  & 12.78   & 30.62       & 21.8    & 2.0 & \nodata & $\surd$   & \nodata \\
 789        & 182040.93$-$161111.7  & 15.67   & 29:\phn\phn & \nodata & 1.3 & $\surd$ & \nodata   & \nodata \\
 782        & 182039.62$-$161146.5  & 12.41   & 30.70       & 22.8    & 3.1 & \nodata & $\surd$   & \nodata \\
 784        & 182039.83$-$160500.1  & 15.41   & 30.46       & 22.8    & 3.3 & \nodata & $\surd$   & \nodata \\
\enddata

\tablecomments{{\bf Columns 1--2:} Source identification from Table~\ref{tbl:src_properties_main}.
{\bf Column 3:} $K$-band magnitudes from \citet{Jiang02}.
{\bf Columns 4--5:} X-ray spectroscopic parameters from Table~\ref{tbl:thermal_spectroscopy}, when available.  For faint sources, a value of $\log L_h \sim 29$~ergs~s$^{-1}$ is assumed (see text).
{\bf Column 6:} Median X-ray energy from Table~\ref{tbl:src_properties_main}.
{\bf Columns 7--9}: Selection criteria from \S \ref{sec:protostars}.}

\tablenotetext{a}{Source is also classified as an intermediate- or high-mass star (Table~\ref{tbl:highmass}).}

\tablenotetext{b}{These protostars lie in the M17-SW~IRS~5 cluster 
around the ultracompact \hii region UC1.}  

\end{deluxetable}


The two sources from Table~\ref{tbl:highmass} identified as Class I (\#324 and \#650) are two of the $K$-band excess sources in Table~\ref{protostars.tab}.  An additional 21 sources in Table~\ref{protostars.tab} also appear in Table~\ref{tbl:highmass}.  Assuming that the protostar criteria defined above supersede the estimated disk/envelope evolutionary classes in Table~\ref{tbl:highmass}, these 23 sources (marked by note {\it a} in Table~\ref{protostars.tab}) define a new sample of X-ray selected intermediate- and high-mass protostars; none are known OB stars.  These stars especially warrant further study.
 
\citet{Nielbock01} report an independent survey for high-mass Class 
I stars in M17 based on ground-based 10~$\mu$m and 20~$\mu$m 
imaging.  Of their 22 luminous IR sources, we detect the 9 listed in Table~\ref{Nielbock_HM.tab} (nominally including the tentative {\em Chandra} detection of IRS 5N).  The X-ray non-detections are not systematically fainter in the IR or more obscured.
Four of the sources in Table~\ref{Nielbock_HM.tab} are cataloged OB stars (ACIS \#186, 233, 296, and 488), but the remaining five stars (ACIS \#128, 239, 309, 322, and 398) should be added to the 23 described above as X-ray detected intermediate- and high-mass protostar candidates. 
Only the first of those, \#128, also appears in our list of candidate protostars (Table~\ref{protostars.tab}).  However it is not identified (in Table~\ref{tbl:highmass}) as massive based on its NIR colors; it is a distinct outlier in the $JHK$ color-color diagram (Figure~\ref{fig:ccd}) and is a hard X-ray source.
Source \#239, has no NIR counterpart; \#309 and \#398 appear in Table~\ref{tbl:highmass} but not in Table~\ref{protostars.tab}.
Source \#322($\Leftrightarrow$ IRS~5S; \S \ref{sec:emb_UC1}), has a SIRIUS counterpart but unreliable $JHK$ photometry.

All the sources in Table~\ref{Nielbock_HM.tab} have X-ray luminosities around $10^{30}$~ergs~s$^{-1}$.  This is 
typical of late-O and early-B stars without strong radiatively 
accelerated winds  \citep{Stelzer05}.  In some cases, the X-ray 
emission probably arises from unresolved low mass companions.  None 
of these 9 stars are powerful wind sources like the central O4+O4 binary (\S \ref{sec:Kleinmann}).
A useful comparison can be made to the cluster 
of proto-B stars in the Becklin-Neugebauer region of the Orion 
Molecular Cloud 1 (OMC-1).  In OMC-1, one star (Source n) has $\log 
L_h \ga 30$~ergs~s$^{-1}$ and three others are detected around $\log 
L_h \sim 29$~ergs~s$^{-1}$ \citep{Grosso05}.  In M17, there are a 
dozen stars similar to Source n.   These stars are 
not confined to the well-studied M17-UC1 and KWO clusters; 
indeed, several appear on the western side of the image away from the 
M17 \hii region.
This supports the claim that star formation in the M17 cloud is widely distributed, as indicated by the widespread  distribution of $JHK$ protostars seen by \citet{Jiang02}.

Finally, three interesting high-mass sources merit comment.
Source \#488 ($\Leftrightarrow$ IRS~15) has recently been described by \citet{Chini06} as a newly-formed 26\Msol star (spectral type B0.5V) that has stopped accreting but is still surrounded by a huge remnant disk.  Its X-ray spectral properties are similar to other early-B stars in M17.
No X-ray source is found associated with 
M17-SO1 (M17 silhouette object 1), an intermediate- or high-mass 
star with a dusty envelope and bipolar reflection nebula 
\citep{Chini04b, Sako05}.  CEN~92, a reddened, massive binary system with 
$L_{IR} > 10^5$~L$_\odot$ and a circumstellar IR-bright shell 
\citep{Chini05}, is also undetected in X-rays.  Chini et al.\ note that CEN~92 and IRS~5N (ionizing M17-UC1) might be similar objects; both are emission-line stars and IRS~5N might be binary.  


\begin{deluxetable}{rclccrccc}
\tablecaption{X-ray emitting high-mass Class I sources 
from \citet{Nielbock01} 
\label{Nielbock_HM.tab}} 

\tablecolumns{9}

\tablehead{\colhead{Seq} & \colhead{CXOU} & \colhead{Name} & 
\colhead{K} & \colhead{$\alpha_{K,N}$} & \colhead{$L_{IR}$} & \colhead{SpTy} &  
\colhead{$\log L_h$} &                          
\colhead{$\log N_H$} \\

&&& \colhead{(mag)} && \colhead{(L$_\odot$)} && \colhead{(erg s$^{-1}$)} & 
\colhead{(cm$^{-2}$)} \\

\colhead{(1)} & \colhead{(2)} & \colhead{(3)} & \colhead{(4)} & 
\colhead{(5)} & \colhead{(6)} & \colhead{(7)}& \colhead{(8)}& \colhead{(9)} 
}

\startdata 
128 & 182019.60$-$161038.8 & B 353        & 11.92   & 2.1 &   75 &\nodata& 30.45       & 23.2    \\
186 & 182021.43$-$160939.1 & CEN 35       & 10.42   & 0.8 &$<$140& B3    & 30.42       & 22.1    \\
233 & 182022.69$-$160833.9 & CEN 16\tnm{a}&  9.10   & 1.3 &$<$275& B0:   & 30.69       & 21.7 \\
239 & 182022.87$-$161148.5 & Anon 1       &\nodata  & 3.3 &  590 &\nodata& 30.78       & 23.5    \\
296 & 182024.39$-$160843.3 & CEN 31\tnm{a}&  9.40   & 2.3 &  910 & O9.5  & 29.58       & 21.6 \\
309 & 182024.60$-$161139.2 & IRS 5S       &  9.34   & 2.5 & 1900 &\nodata& 30.50       & 22.8 \\
322 & 182024.83$-$161135.3 & IRS 5N\tnm{b}&\nodata  & 4.2 & 4775 &\nodata& 29:\phn\phn & \nodata \\
398 & 182026.62$-$161136.5 & IRS 10       & 10.56   & 1.5 &   55 &\nodata& 29.99       & 22.5    \\
488 & 182028.65$-$161211.6 & IRS 15\tnm{a}& 10.08   & 3.0 & 1405 & B0.5  & 30.78       & 22.2 \\
\enddata

\tablecomments{{\bf Columns 1--2:} Source identification from Table~\ref{tbl:src_properties_main}.
{\bf Column 3:} Counterpart identification from \citet{Nielbock01}.
{\bf Column 4:} $K$-band magnitudes from \citet{Jiang02}.
{\bf Column 5:} Spectral index from \citet{Nielbock01}.
{\bf Column 6:} Infrared luminosity from \citet{Nielbock01}.
{\bf Column 7:} Spectral type.
{\bf Columns 8--9:} X-ray spectroscopic parameters from Table~\ref{tbl:thermal_spectroscopy}.}

\tablenotetext{a}{These stars are surrounded by resolved 
infrared-bright dusty disks \citep{Chini05,Chini06}.}

\tablenotetext{b}{Source \#322 is tentative (\S \ref{sec:emb_UC1}) and its luminosity is assumed.}
\end{deluxetable}


\subsection{Distributed star formation across the molecular cloud
\label{sec:emb_distributed}}

A number of heavily obscured X-ray stars lie in other dense molecular 
regions of M17-SW.  \citet{Vallee96} identify six peaks in dust and 
molecular emission maps as dense cores with sizes $10-30$\arcsec\/ 
($0.1-0.3$~pc), masses $70-470$\Msol, densities $1-8 \times 
10^5$~cm$^{-3}$ and absorbing columns of $1.5-3 \times 10^{23}$~cm$^{-2}$. 
ACIS sources \#236 in Peak 1, \#97 in Peak 2, \#177 in Peak 4, and \#275 in 
Peak 5 all have intrinsic X-ray luminosities around $\log L_{h,c} 
\sim 30.5-31$~ergs~s$^{-1}$ and absorptions around $\log N_H \sim 
22.3-22.6$~cm$^{-2}$.  The luminous source \#239 $\Leftrightarrow$ Anon 1 in Peak 3 
was discussed earlier in \S \ref{sec:emb_UC1} and 
Table~\ref{Nielbock_HM.tab}.  None of these molecular gas 
concentrations appears to have produced rich stellar clusters, but 
the high X-ray luminosities of the detected stars implies that more 
sources with lower luminosities are present.

We can examine the X-ray source list for heavily obscured stars that 
have not been studied at other wavelengths.  Due to the short 
exposure, only the stars with $\log L_h \ga 30.5$~ergs~s$^{-1}$ are 
sufficiently bright to permit quantitative estimation of high levels 
of obscuration. Six ACIS sources have spectral fits with $\log N_H \geq 
23.3$~cm$^{-2}$ equivalent to $A_V \ga 100$~mag:  \#37, 135, 239, 
246, 255, and 706.   Two of these (\#239 and 246) reside in the M17-UC1 
cluster while the others are not in known molecular cores.  

Altogether, we support the evidence of \citet{Jiang02} and others that star 
formation is widely distributed across the cloud in addition to 
concentrations around M17-UC1, the KWO, and M17-North.  This 
implies either that star formation triggered by the passage of the 
ionization front of the large M17 \hii region does not always produce 
concentrated clusters and/or that much of the star formation in the 
cloud was not triggered.

\section{X-rays from massive stars} \label{sec:OB}

Individual O stars have long been known to emit soft X-rays at 
levels $L_x \propto 10^{-7}$~L$_{bol} \sim 10^{31-33}$~ergs~s$^{-1}$ \citep{Chlebowski89,Berghoefer97,Sana06}, 
arising from a myriad of weak shocks within their radiatively accelerated 
winds \citep{Lucy80, Pallavicini81, Owocki99}.  While this soft ($kT \sim 0.5$~keV), 
slowly varying X-ray component dominated early studies of O stars, 
recent studies reveal discrepancies with X-ray line widths, inferred 
densities, and temperatures \citep{Waldron01}.  In particular, a 
moderately hard component ($kT \sim 2-3$~keV) may be present which is sometimes rotationally 
modulated  and/or shows rapid variability \citep{Feigelson02, 
Schulz03, Stelzer05, Schulz06}.  While not fully understood, this component has been 
attributed to a large-scale magnetically confined (or at least channeled) wind shock that 
diverts radial outflow into an equatorial disk, causing an X-ray emitting shock \citep{Babel97, 
Gagne05}. A luminous hard X-ray component, sometimes reaching $L_h 
\sim 10^{33}$~ergs~s$^{-1}$ and $kT > 10$~keV, can be produced by wind-wind collisions 
in close massive binaries, particularly if one component is a 
Wolf-Rayet star \citep[e.g.][]{Skinner02,Pollock05,Skinner06}, or possibly by a non-thermal inverse Compton process \citep{Chen91}.
An emerging class of late-O/early-B emission line stars known as ``$\gamma$~Cas analogs'' \citep{Smith06,Rakowski06} show very hard X-ray spectra ($kT > 8$~keV) and variable X-ray lightcurves, perhaps due to some combination of winds, magnetic fields, and disk processes \citep{Smith04}.
Wind-generated X-rays become insignificant in stars cooler than early-B types, but these systems are sometimes detected in X-rays due to lower mass pre-main sequence companions \citep[e.g.][]{Stelzer05,Sana06} or perhaps due to their own intrinsic emission \citep[e.g.][]{Stelzer06}.

M17 provides an excellent laboratory for study of these OB 
wind-generated X-rays due to the large population of coeval and 
codistant massive stars in a single field.  The main limitation is that we 
cannot uniformly measure their soft X-ray components due to heavy 
absorption towards some systems.  

Table~\ref{OB.tab} gives the X-ray properties of all published O and early B 
stars in the region \citep[from TFM03, with IRS~15 added from][]{Chini06}
\footnote{
The two O stars listed in TFM03 as {\em Chandra} non-detections (CEN~102 and CEN~34) were determined by \citet{Hanson97} to be background red giants based on their NIR brightness and the presence of CO bandhead absorption in their IR spectra.  The B stars CEN~95 and CEN~33, also not detected in X-rays, were similarly identified by Hanson et al.\ to be background sources.  These four sources are omitted from Table~\ref{OB.tab}.
}.  
The stars are listed in 
order of decreasing mass; see the table notes for descriptions of the 
tabulated quantities. 
While all of the O stars in Table~\ref{OB.tab} are detected with {\em Chandra}, only 19 of the 34 B0--B3 stars are detected.  

\begin{figure}
\plotone{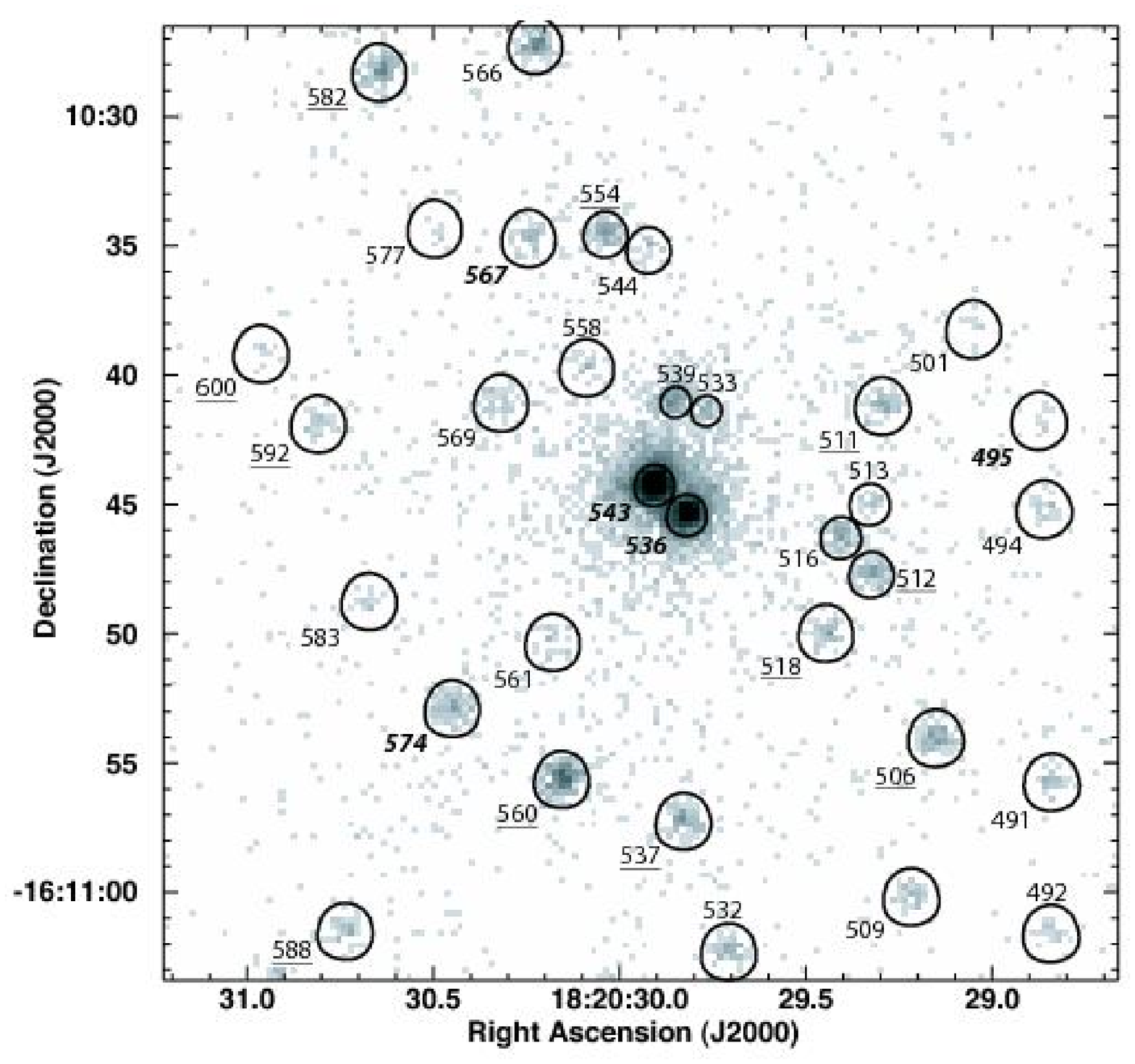} 

\caption{ACIS image of the central region of the NGC~6618 cluster with 
source sequence numbers and extraction regions indicated.  This image is shown at 
high-resolution with $0.25\arcsec \times 0.25\arcsec$\/ pixels. 
Sources \#536 and \#543 (bold-italic) are the two O4 components of Kleinmann's Anonymous Star (\S \ref{sec:Kleinmann}).
Sources \#495, 567, and 574 (bold-italic) are also known OB stars (Table~\ref{OB.tab}).
Sources with underlined labels are listed in Table~\ref{tbl:highmass} as candidate new intermediate- or high-mass stars.
\label{central_img.fig}} 
\end{figure}

\begin{figure}
\centering 
\plotone{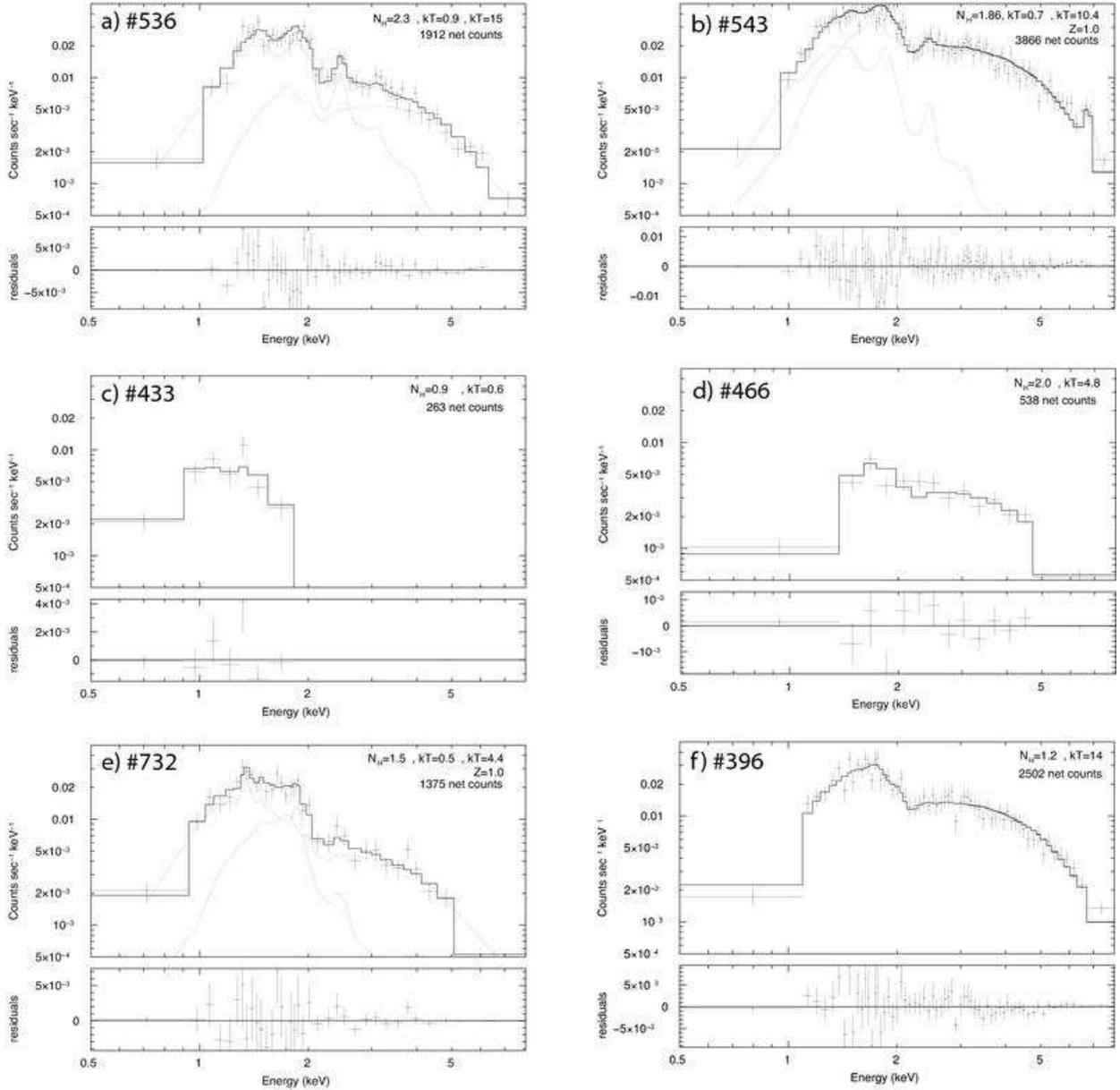} 

\caption{Thermal plasma models for the X-ray spectra of selected OB stars in M17.  The absorbing column density $N_H$ is shown in units of $10^{22}$~cm$^{-2}$; plasma temperature is shown in units of keV. Elemental abundances were $Z = 0.3 Z_\odot$ unless otherwise specified.  Panels a, b, and e are two-temperature models; model components are shown separately as dotted curves.
{\bf (a,b)} Spectral models for the binary components of Kleinmann's 
Anonymous Star (\#536 and \#543, \S \ref{sec:Kleinmann}).  In these 
two-temperature models, both O4 stars show a typical 
soft plasma; both also show a much harder plasma component most likely caused by colliding winds 
from an unresolved binary component \citep[e.g.][]{Pollock05}. 
{\bf (c)} A soft spectrum from the O6 star OI~345 (\#433). 
{\bf (d)} A B2 star (\#466) with a hard spectrum; this star lies at the center of NGC~6618 (\S \ref{sec:spatial}).  
{\bf (e)} A hard spectrum from the O8 star OI~352 (\#732). 
{\bf (f)} A hard spectrum from the unidentified luminous source \#396 (\S \ref{sec:new_OB}).
\label{OB_spec.fig}} 
\end{figure}

Figure~\ref{central_img.fig} shows a close-up view of the cluster 
center where many of the OB stars are concentrated.  
Figure~\ref{OB_spec.fig} shows the X-ray spectra of the components of the O4+O4 binary 
and four other OB stars, illustrating the diversity of plasma 
properties. 
Note that all but one of these spectra show very hard thermal plasma components with $kT > 4$~keV.  As mentioned above, such hard X-ray emission is most commonly seen in colliding wind binaries; a famous example is $\eta$ Carinae \citep{Corcoran04}.


\begin{deluxetable}{lcrrcccccrcccc} 

\tablecolumns{13}                                                        
\tablecaption{X-ray properties of cataloged OB stars in NGC 6618 
\label{OB.tab}
} 
\tabletypesize{\tiny} 
\tablehead{ 
\multicolumn{5}{c}{Optical Properties} && \multicolumn{7}{c}{X-ray Properties} \\
\cline{1-5} \cline{7-13} \colhead{Name} & \colhead{SpTy}  & 
\colhead{SIRIUS} &  \colhead{K}  & \colhead{$\log \tilde{L_{bol}}$} 
&&  \colhead{Seq} & \colhead{$\Delta \phi$} & \colhead{NetCts} & 
\colhead{$\log N_H$} & \colhead{kT} & \colhead{$\log L_h$} & \colhead{$\log L_{t,c}$} \\

&&& \colhead{(mag)} & \colhead{(L$_\odot$)} && \colhead{\#} & \colhead{($\arcsec$)} && 
\colhead{(cm$^{-2}$)} & \colhead{(keV)} & \colhead{(erg s$^{-1}$)} & 
\colhead{(erg s$^{-1}$)} \\

\colhead{(1)} & \colhead{(2)} & \colhead{(3)} & \colhead{(4)} & \colhead{(5)} && 
\colhead{(6)} & \colhead{(7)} & \colhead{(8)} & 
\colhead{(9)} &\colhead{(10)} & \colhead{(11)} & \colhead{(12)} 
}                                                          
\startdata                                                                                       
CEN  43 & O3-O4  &                  10396 &  8.18 &  5.8  && 675    & 0.2   &   82  &  22.1 &   2.0  &  31.02 & 31.53 \\ 
CEN  1a & O4     &\nodata\tablenotemark{a}&  6.59:&  5.7  && 536    &\nodata& 1912  &  22.4 &0.9+15  &  32.37 & 33.16 \\ 
CEN  1b & O4     &\nodata\tablenotemark{a}&  6.59:&  5.7  && 543    &\nodata& 3866  &  22.3 &0.7+10.4&  32.78 & 33.28 \\ 
CEN   2 & O5     &                  10399 &  7.49 &  5.5  && 701    & 0.1   &  250  &  21.2 &   1.6  &  31.14 & 31.69 \\ 
CEN  37 & O3-O6  &                  10398 &  7.88 &  5.5  && 574    & 0.2   &   90  &  22.4 &   0.6  &  30.42 & 32.25 \\ 
OI  345 & O6     &                  14703 &  9.18 &  5.3  && 433    & 0.0   &  263  &  22.0 &   0.6  &  30.32 & 32.06 \\ 
CEN  18 & O7-O8  &                  28446 &  7.84 &  5.1  && 366    & 0.2   &  128  &  21.8 &   4.4  &  31.02 & 31.31 \\ 
CEN  25 & O7-O8  &                   7877 &  8.94 &  5.1  && 593    & 0.1   &   16  &  21.9 &   0.7  &  29.53 & 31.00 \\ 
OI 352  & O8     &\nodata\tablenotemark{b}&  6.82 &  5.0  && 732    &\nodata& 1375  &  22.2 &0.5+4.4 &  32.00 & 32.95 \\ 
CEN  16 & O9-B2  &                  28651 &  9.10 &  4.8  && 233    & 0.4   &   73  &  21.7 &   3.5  &  30.69 & 31.01 \\ 
CEN  61 & O9-B2  &                   7833 &  9.23 &  4.8  && 567    & 0.1   &   44  &  22.3 &   0.6  &  30.05 & 31.72 \\ 
CEN   3 & O9     &                  10401 &  7.59 &  4.8  && 720    & 0.2   &  108  &  21.6 &   0.6  &  29.72 & 31.34 \\ 
OI 174  &O9      &\nodata\tablenotemark{c}&  7.80 &  4.8  &&  12    & 0.5   &  121  &  22.1 &   0.5  &  30.10 & 32.04 \\ 
CEN  31 & O9.5   &                  27457 &  9.40 &  4.7  && 296    & 0.1   &   10  &  22.0 &   2:~  &  28.81 & 30.31 \\ 
CEN  92 & B0     &                  14714 &  9.10 &  4.5  &&\nodata &\nodata&\nodata&\nodata& \nodata&$<$29.0~&$<$29.5\\ 
CEN  28 & B0     &                   8849 & 11.40 &  4.5  && 647    & 0.1   &    4  &  21:~ &   2:~  &  28.5: & 28.8: \\ 
OI  582 & B0     &\nodata\tablenotemark{d}&\nodata&  4.5  && 875    &\nodata&   19  &  20:~ &   1.7  &  29.69 & 30.20 \\ 
IRS 15  & B0.5   &                   5988 & 10.08 &  4.3  && 488    & 0.0   &   71  &  22.2 &   2.2  &  30.78 & 31.30 \\ 
CEN  57 & B1     &                  14554 &  9.58 &  4.0  && 347    & 0.2   &  117  &  22.1 &   2.5  &  31.00 & 31.44 \\ 
CEN 101 & B1     &                   9090 & 11.39 &  4.0  && 496    & 0.0   &  111  &  22.0 &   1.8  &  30.76 & 31.32 \\ 
CEN  97 & B1     &                  13963 & 10.78 &  4.0  && 507    & 0.0   &  134  &  22.2 &   2.9  &  31.11 & 31.52 \\ 
CEN 100 & B1     &                  17147 & 10.85 &  4.0  && 608    & 0.0   &  223  &  22.1 &   3.9  &  31.41 & 31.75 \\ 
CEN  45 & B1     &                   4442 & 10.81 &  4.0  && 726    & 0.2   &   11  &  21.8 &   2:~  &  29.71 & 30.19 \\ 
CEN  51 & Early B&                  14187 &  9.79 &  3.5  &&\nodata &\nodata&\nodata&\nodata& \nodata&$<$29.0~&$<$29.5\\ 
CEN  24 & B2     &                  15126 &  8.42 &  3.5  &&\nodata &\nodata&\nodata&\nodata& \nodata&$<$29.0~&$<$29.5\\ 
CEN  49 & B2     &                  14649 &  9.72 &  3.5  &&\nodata &\nodata&\nodata&\nodata& \nodata&$<$29.0~&$<$29.5\\ 
CEN  48 & B2     &                  13124 & 10.98 &  3.5  &&\nodata &\nodata&\nodata&\nodata& \nodata&$<$29.0~&$<$29.5\\ 
CEN  90 & B2     &                  14345 & 11.51 &  3.5  && 406    & 0.1   &79\tnm{e}& 22.2&   1.8  &  30.79 & 31.39 \\ 
CEN  85 & B2     &                   6104 & 10.62 &  3.5  && 466    & 0.0   &  538  & 22.3  &   4.8  &  31.86 & 32.19 \\ 
CEN  84 & B2     &                   6109 & 11.11 &  3.5  && 495    & 0.2   &    8  & 21:~  &   2:~  &  28.8: & 29.1: \\ 
CEN  96 & B2     &                  14492 & 10.81 &  3.5  &&\nodata &\nodata&\nodata&\nodata& \nodata&$<$29.0~&$<$29.5\\ 
CEN  89 & B2     &                   7390 & 11.33 &  3.5  &&\nodata &\nodata&\nodata&\nodata& \nodata&$<$29.0~&$<$29.5\\ 
CEN  99 & B2     &                   9160 & 11.58 &  3.5  && 638    & 0.3   &   88  & 21.8  &   2.2  &  30.67 & 31.12 \\ 
CEN  17 & B3     &                  12345 & 12.33 &  3.2  && 182    & 0.2   &12\tnm{e}& 21.9&  0.7   &  29.06 & 30.47 \\ 
CEN  93 & B3     &                  14349 &  9.59 &  3.2  &&\nodata &\nodata&\nodata&\nodata& \nodata&$<$29.0~&$<$29.5\\ 
CEN  35 & B3     &                  14179 & 10.42 &  3.2  && 186    & 0.1   &   30  & 22.1  &   3.0  &  30.42 & 30.81 \\ 
CEN  75 & B3     &                  23809 & 11.99 &  3.2  &&\nodata &\nodata&\nodata&\nodata& \nodata&$<$29.0~&$<$29.5\\ 
CEN  94 & B3     &                  12342 & 11.47 &  3.2  &&\nodata &\nodata&\nodata&\nodata& \nodata&$<$29.0~&$<$29.5\\ 
CEN  91 & B3     &                  14141 & 11.25 &  3.2  && 312    & 0.1   &   86  & 22.1  &   3.7  &   30.94& 31.29 \\ 
CEN  83 & B3     &                  12337 & 11.28 &  3.2  && 350    & 0.0   &309\tnm{f}&22.1&   5.0  &  31.55 & 31.85 \\ 
CEN  14 & B3     &                  14893 & 12.16 &  3.2  &&\nodata &\nodata&\nodata&\nodata& \nodata&$<$29.0~&$<$29.5\\ 
CEN  26 & B3     &                  13500 & 10.53 &  3.2  && 376    & 0.1   &   78  & 22.3  &   1.9  &  30.81 & 31.38 \\ 
CEN  74 & B3     &                  27036 & 11.77 &  3.2  && 399    & 0.2   &    6  & 21.:  &   2:~  &  28.7: & 29.0: \\ 
CEN  46 & B3     &                   7804 &  9.47 &  3.2  && 450    & 0.8   &    6  & 21.:  &   2:~  &  28.7: & 29.0: \\ 
CEN  47 & B3     &                   9089 & 10.80 &  3.2  &&\nodata &\nodata&\nodata&\nodata& \nodata&$<$29.0~&$<$29.5\\ 
CEN  65 & B3     &                   3592 & 11.50 &  3.2  &&\nodata &\nodata&\nodata&\nodata& \nodata&$<$29.0~&$<$29.5\\ 
CEN  27 & B3     &                   7284 &  9.90 &  3.2  &&\nodata &\nodata&\nodata&\nodata& \nodata&$<$29.0~&$<$29.5\\ 
CEN  44 & B3     &                   1183 & 13.86 &  3.2  &&\nodata &\nodata&\nodata&\nodata& \nodata&$<$29.0~&$<$29.5\\ 

\enddata

\tablecomments{ {\bf Column 1:}  This list is obtained from Appendix A of 
TFM03 (with IRS~15 added) which gives optical cross-identifications,           
positions and spectral types. The stars are listed first in order of 
decreasing mass, and then by right ascension.  CEN~=~\citet{Chini80},  
OI~=~\citet{Ogura76}.
The CEN~1 O4+O4 binary 
(\S\ref{sec:Kleinmann}) is also known as Kleinmann's Anonymous Star 
\citep{Kleinmann73a} and CEN~3 is BD~-16$^\circ$4818.  
The B3 stars CEN~50, CEN~58, CEN~81, and CEN~82 are omitted because accurate positions are not available.
The stars CEN~33~$\Leftrightarrow$~B~324, CEN~34~$\Leftrightarrow$~B~358, CEN~95~$\Leftrightarrow$~B~266, and CEN~102~$\Leftrightarrow$~B~305 are omitted because they are likely to be background red giants \citep{Hanson97}; none of these sources was detected in X-rays. \\
{\bf Column 2:}  Spectral types are from \citet{Hanson97}, \citet{Chini80}, and SIMBAD. \\
{\bf Columns 3--4:}  Source numbers and $K$-band magnitudes 
are from \citet{Jiang02}.\\
{\bf Column 5:}  Bolometric luminosities are estimated from calibrations of $L_{bol}$ with spectral type, 
\citet{Martins05} for O3--O9.5 stars and \citet{deJager87} for B stars.  
No use is made of available photometry. \\                                       
{\bf Column 6:} {\em Chandra} source number, from 
Table~\ref{tbl:src_properties_main}.\\
{\bf Column 7:}  Offset (in arcseconds) between the {\em 
Chandra} and $SIRIUS$ sources. Both fields are aligned to the 2MASS 
astrometric frame.  \\
{\bf Columns 8--12:}  X-ray properties from 
Table~\ref{tbl:thermal_spectroscopy}: extracted counts after                        
background subtraction; column density and plasma energy from fits to 
the ACIS spectra (: in $\log N_H$ values are approximated from median 
energy \citep{Feigelson05b}; : in $kT$ are assumed); observed hard band luminosity ($2-8$ 
keV); inferred total band luminosity corrected for absorption ($0.5-8$~keV).
Upper limits to luminosities are matched to the fainter sources
in Table~\ref{tbl:thermal_spectroscopy}.
}

\tablenotetext{a}{We adopt the nomenclature CEN~1a and CEN~1b for the O4+O4 binary components of Kleinmann's Anonymous Star (\S \ref{sec:Kleinmann}) which are resolved by Chandra.
These sources do not appear in the SIRIUS catalog due to crowding and/or saturation.  
The $K$-band magnitudes shown are estimated by splitting the $K = 5.84$ measurement from \citet{Chini98} into two equal halves.  
}

\tablenotetext{b}{This star is saturated in the SIRIUS data and does 
not appear in the catalog.  We use here position and photometry from 
the 2MASS catalog: 2MASS 18203586$-$1615431 with J=7.97, H=7.21 and 
K=6.82.}  

\tablenotetext{c}{This star is outside the field of view of the 
SIRIUS data.  We use here position and photometry from the 2MASS 
catalog: 2MASS 18200299$-$1602068 with J=8.68, H=8.09, and K=7.80.}  

\tablenotetext{d}{This star is outside the field of view of the 
SIRIUS data.  The available position is too inaccurate to reliably match to 2MASS.  }

\tablenotetext{e}{The X-ray emission may be variable.}

\tablenotetext{f}{The X-ray emission is definitely variable;  see              
Figure~\ref{fig:lightcurve} for light curve.} 

\end{deluxetable}                                           


\subsection{OB Star Luminosities \label{sec:OB_6618}}

\begin{figure}
\centering
\includegraphics[angle=0,width=0.5\textwidth]{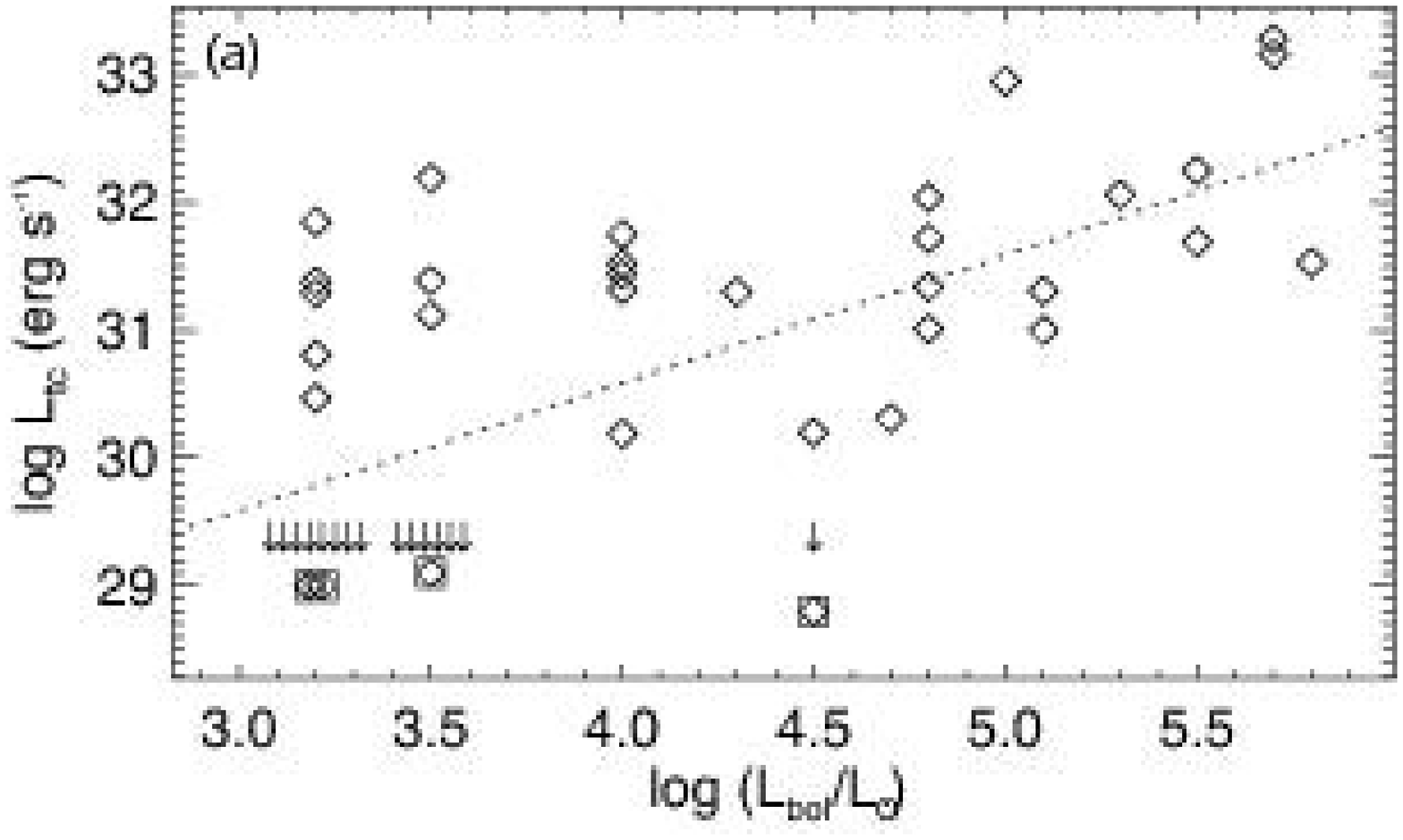} 
\includegraphics[angle=0,width=0.5\textwidth]{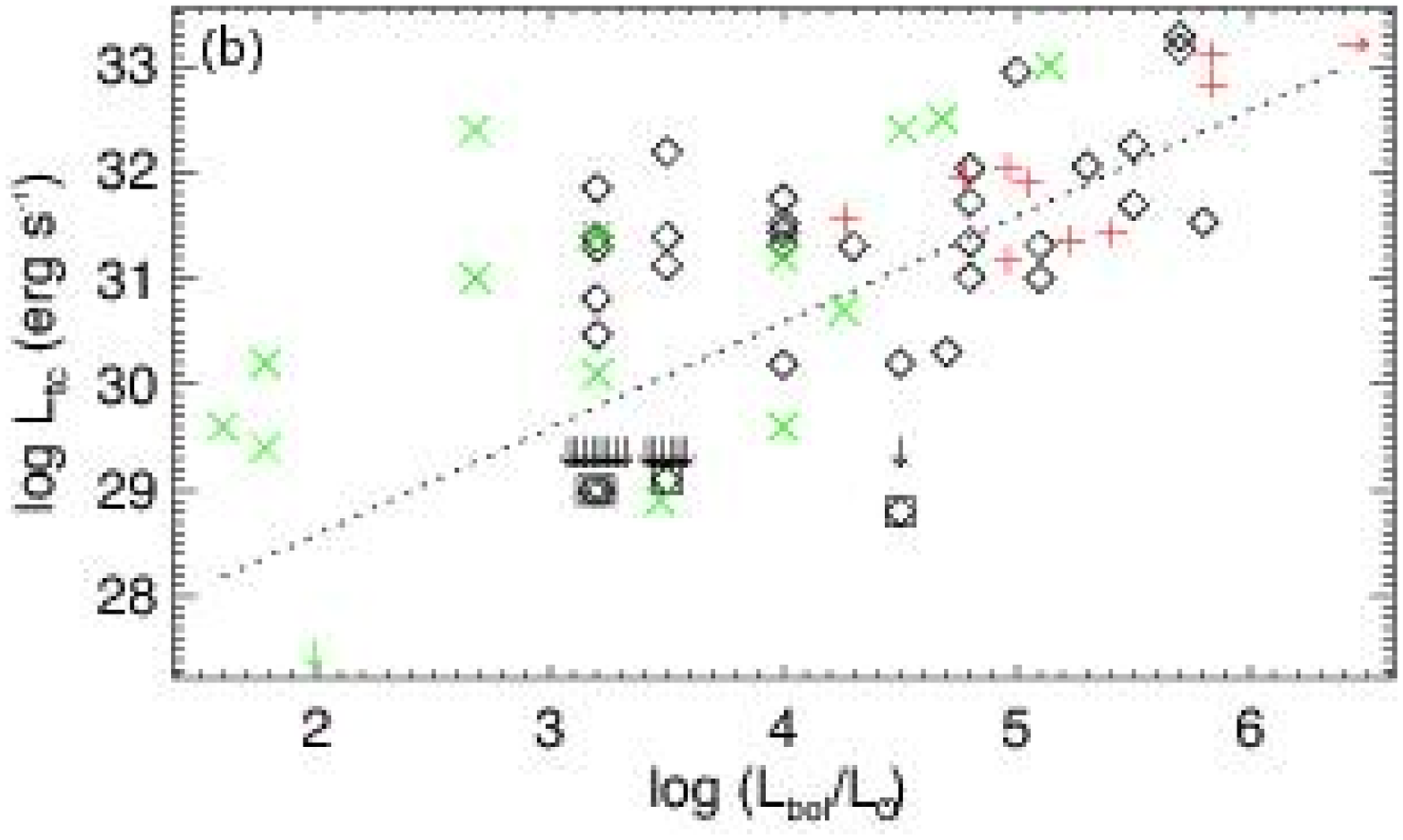} 
\caption{
Relationship between X-ray and bolometric luminosities for M17 OB stars in Table~\ref{OB.tab}.
Arrow tails represent estimated upper/lower limits; overlapping points are artificially displaced horizontally for readability.
Boxed points represent estimated values.
The dotted line represents the model $L_{t,c} = 1 \times 10^{-7} L_{bol}$.
{\bf (a)} Broad-band (0.5--8~keV) X-ray luminosity corrected for absorption vs. bolometric luminosity.
{\bf (b)} M17 OB stars (same symbols as in panel {\em a}) shown with Pismis 24 OB stars (red pluses and red arrow) from \citet{Wang06} and Orion Nebula Cluster OB stars (green $\times$'s and green arrow) from the COUP study \citep{Stelzer05}.
\label{Lx_Lbol.fig}} 
\end{figure}

The most fundamental question concerning OB X-ray emission for single stars is whether 
the X-ray production scales with the wind power, which in turn is 
expected to scale with the bolometric luminosity. 
Figure~\ref{Lx_Lbol.fig}{\it a} compares X-ray and bolometric luminosities for the cataloged M17 OB population.
The X-ray luminosity shown, $L_{t,c}$, is over the full (0.5--8~keV) band, corrected for absorption. 
We adopt $L_{bol}$ values here based on calibrations between spectral 
types and bolometric luminosities for main sequence OB stars \citep{Martins05,deJager87}. 
No attempt is made to use available photometry with absorption and bolometric corrections as the available measurements and reddening estimates are incomplete and unreliable. 


We used the ASURV survival analysis package \citep{Isobe86}\footnote{Available from the \anchor{http://astrostatistics.psu.edu/statcodes}{Center for Astrostatistics} at \url{http://astrostatistics.psu.edu/statcodes}}
to perform statistical tests for correlation between between $L_{t,c}$ and $L_{bol}$, with proper statistical treatment (generalized Kendall's tau correlation test) of the available upper limits.   
The correlation among M17 sources was very significant (null hypothesis probability $P<0.01\%$), supporting the long-standing 
$L_x \propto 10^{-7}~L_{bol}$ relationship \citep{Chlebowski89,Berghoefer97}.  
Most of the lower luminosity members are undetected; this group would 
be underrepresented in a plot that did not include X-ray upper limits 
based on a sample defined at other wavebands.  

For comparison, Figure~\ref{Lx_Lbol.fig}{\it b} shows the M17 OB stars in context with the OB stars in the Pismis 24 cluster in NGC~6357 \citep{Wang06} and in the Orion Nebula Cluster \citep{Stelzer05}, with $L_{bol}$ for these sources estimated from spectral types via the same method used for the M17 sources.  The red arrow represents a lower limit for $L_{bol}$ for the Wolf-Rayet + O7 binary WR93 in NGC~6357 and the green arrow represents an upper limit to the X-ray emission from a COUP B star.  
The $L_{t,c}$ value for the Orion source $\theta^1$~Ori~C was replaced by a more accurate value reported by \citet{Gagne05}.
This plot demonstrates that the M17 OB stars display a similar correlation between $L_{t,c}$ and $L_{bol}$, with similar scatter, to that seen in the other two populations.  As in the COUP sample, more scatter is seen in the M17 B stars than in the O stars; these B stars have emission comparable to lower mass pre-main sequence stars and sometimes that emission may actually come from unresolved low-mass companions rather than from the B stars themselves.  Only one of these M17 sources is variable: the B3 star \#350 ($\Leftrightarrow$ CEN~82) shows factor of $\sim$2 variations 
during the 12 hour exposure (Figure~\ref{fig:lightcurve}).  The 
behavior does not resemble the common fast-rise-slow-decay shape seen 
in low-mass stellar flares, so it is possible that it arises from large-scale 
shocks in a magnetically channeled wind \citep[e.g.][]{Gagne05}. 
Results shown in Figure~\ref{Lx_Lbol.fig} are consistent with those found for NGC~6231 by \citet{Sana06}.

\subsection{The central O4+O4 binary \label{sec:Kleinmann}}

{\em Chandra} resolves (Figure~\ref{central_img.fig}) the two 
luminous components of Kleinmann's Anonymous Star 
\citep{Kleinmann73a}, the binary O4+O4 system with separation 
1.8\arcsec\/ ($\sim$3000 AU projected) 
and total mass $M \sim 140$\Msol at the top of the stellar IMF of 
NGC~6618.  The spectral types for these stars were originally determined by $UBVRI$ photometry \citep{Chini80} using images where the two stars were not resolved, thus the identical spectral types is an assumption.  \citet{Hanson97} obtained a NIR spectrum for the unresolved pair and deduced an IR spectral type of O3--O4, again with no information on the individual components.  Given that these early O stars provide substantial ionization to power the M17 complex and lie near the center of the NGC~6618 cluster, independent spectral types for the two components via improved NIR and visual spectroscopy are clearly needed.

ACIS sources \#536 and \#543, the X-ray counterparts to the O4 stars, are the brightest and most luminous 
X-ray stars in the cluster with intrinsic total band (0.5-8~keV) 
luminosities $\log L_{t,c} \sim 33.2$ and 33.3~ergs~s$^{-1}$, respectively.  
We see no enhanced X-ray emission between the two O4 stars, implying that there is not a strong shock between their winds; this result is expected given their large separation.  The light curves show that the X-ray 
emission from each source is constant within $\sim$10\% during the 12 hour exposure.

Two-temperature thermal plasma models were needed to account for each O4 star's broad X-ray 
spectrum, and each star shows a strong hard component with emission 
extending beyond the 8~keV reflectivity limit of the {\em Chandra} 
mirrors (Figure~\ref{OB_spec.fig}{\it a,b}).  We recorded twice as many ACIS counts for source \#543 as for \#536
\footnote{Source \#543 may suffer slight photon pile-up with $\sim 0.3$ counts per
frame, but a careful analysis of spectra from annular regions in the
PSF wings of this source did not yield substantially different spectral
fit parameters; although the core might be slightly piled up, there are
not enough counts in the wings to yield a measurable change in the fit
parameters when core events are excluded.}.  
In both sources, the hard
component contributes substantially to the intrinsic flux.  For \#543,
96\% of the hard-band intrinsic flux comes from the $kT = 10$~keV plasma;
this component contributes 26\% of the soft-band and 52\% of the full-band
flux ($\log L_{t,c} \sim 33.0$~ergs~s$^{-1}$).  For \#536, 71\% of the
hard-band flux is due to the $kT = 13$~keV component, while just 10\%
of the soft-band and 25\% of the full-band flux ($\log L_{t,c} \sim
32.5$~ergs~s$^{-1}$) comes from this hard plasma.

\citet{Skinner02} provide a particularly cogent and helpful review of the
possible causes of hard X-ray emission in massive stars.  They describe an
{\em XMM-Newton} observation of the Wolf-Rayet star WR~110, which displays
the expected soft plasma ($kT = 0.55$~keV) seen in single massive stars
and an additional hard component with a poorly constrained temperature of
$kT > 3$~keV.  They assess several models, including magnetically-confined
wind shocks, and finally conclude that wind shocks with an unseen binary
companion are the most likely cause of the hard component.  

Another model mentioned by \citet{Skinner02} involves non-thermal hard X-ray
emission caused by inverse Compton scattering of
UV photons from the star off of relativistic charged particles created
in the wind shocks \citep{Chen91}.
Although the WR~110 data are too poor to constrain this model, we can
apply it to the M17 O4 stars.  Fitting the fainter O4 star (\#536) with a
power law instead of a hard thermal component (keeping $N_H$ and the soft
plasma component the same as in the original fit) yields an acceptable
fit, only slightly worse than the two-temperature thermal plasma fit, with a power
law photon index of $\Gamma = 1.3$.  Although this power law slope is
not well-constrained by these data, it is consistent with the slope of
1.5 predicted by \citet{Chen91}.  The full-band intrinsic luminosity
for this power law component is $\log L_{t,c} \sim 32.4$~ergs~s$^{-1}$,
three orders of magnitude brighter than the prediction of \citet{Chen91}.
However, \citet{Skinner02} note that this luminosity is strongly dependent
on the magnetic field strength of the star and its wind temperature and
mass loss rate.  Thus, inverse Compton scattering could be an alternative
explanation to colliding winds for source \#536, but extreme physical parameters would be required to generate such a high luminosity.

In contrast, fitting the brighter O4 star (\#543) with
this soft plasma + power law model yields an unacceptable fit; this
source has an Fe line that the power law model cannot reproduce (see
Figure~\ref{OB_spec.fig}{\it b}).  This Fe line is strong evidence that
the 10~keV plasma in source \#543 has a thermal origin, most likely caused
by colliding winds between the O4 star and an as yet undiscovered binary
companion.  Since the Fe line in source \#543 argues that its hard component
is thermal, it is reasonable to assume that the hard component seen in the
other O4 star (source \#536) is also thermal; it is probably just too
faint for its Fe line to be seen in this observation.  

This implies that
each O4 component of Kleinmann's Anonymous Star may itself be a massive
binary and that the system actually consists of at least 4 massive stars.  It is remarkable that these two X-ray sources show such similar and yet highly unusual spectra; thermal plasmas this hot are not seen even in Wolf-Rayet colliding-wind binaries or in $\eta$~Carinae.  Only the ``$\gamma$~Cas analogs'' \citep{Smith06,Rakowski06,Smith04} show such hard X-ray spectra, yet M17's O4 stars do not show the variable lightcurves typical of these objects.  Some complex process, likely involving both colliding winds in close binaries and unusual magnetic field properties, may be necessary to explain M17's O4 stars.
Again, visual and NIR spectroscopy of these two stars is highly warranted.

The NGC~6618 O4 stars and later OB stars (Figure \ref{OB_spec.fig}) thus give new examples of the extremely hot 
plasmas that can be generated in powerful O star winds.  The 
luminous, hard emission of these stars has implications for the 
study of Galactic massive star formation in the X-ray band.  Emitting 
$\sim$$10^{31.5}$~ergs~s$^{-1}$ above 5~keV, such stars can be detected in 
reasonable {\em Chandra} exposures anywhere in our half of the 
Galactic disk, even if they lie behind several spiral arms and are 
embedded in dense molecular clouds; they can even be studied in the Magellanic Clouds.  
Recently studied examples have been found in the MSFRs NGC~6231, NGC~3576, Wd~1, W3, W51A, and R136 in 30~Doradus \citep{Sana06,Townsley06c,Skinner06,Townsley05,Townsley06b}.

\subsection{Possible new OB stars \label{sec:new_OB}}

{\em Chandra} data have the potential for identifying new OB stars 
in MSFRs, either in the rich clusters 
illuminating \hii regions or deeply embedded in the surrounding molecular cloud. Recent {\em Chandra} studies of RCW~38 and NGC~6357 have identified 31 and 24 
candidate new OB stars, respectively \citep{Wolk06, Wang06}.  Such stars 
are bright in the NIR and already appear in IR catalogs, but 
cannot be easily distinguished from reddened field stars (e.g.\ red 
giants, which are generally X-ray faint) without X-ray study.  

\citet{Jiang02} find about 1000 OB candidates from IR photometry, but many of those are likely foreground or background stars.  X-ray emission should help to discriminate interlopers from M17 OB stars.  In Table~\ref{tbl:highmass}, we narrow these candidates to 143 X-ray selected sources that have masses $>2$\Msol based on the NIR color-magnitude diagram in Figure~\ref{fig:cmd}.  Twenty-nine of these are known OB stars from Tables~\ref{Nielbock_HM.tab} or \ref{OB.tab}, but the remaining 114 sources are candidate massive stars.  Two of these sources, \#51 and \#163 (marked in Figure~\ref{fig:cmd}), appear to be previously unremarked highly obscured O stars with $A_V \sim 40$.  
Spectroscopy of these OB candidates is warranted to confirm their spectral types. 

Two alternative indicators of candidate OB stars have been employed \citep[e.g.\ ][]{Wang06}: high X-ray luminosity, and X-ray detection coupled with bright $K$-band magnitude.
There are just four X-ray sources (described below) in our sample with $\log L_{t,c} > 32.0$~ergs~s$^{-1}$ that are not known O stars (Tables~\ref{Nielbock_HM.tab} or \ref{OB.tab}); none of them appears in our list of candidate protostars (Table~\ref{protostars.tab}).
\begin{description}
\item[ACIS \#187 ($\Leftrightarrow$ B335)]~~This source, located in the South Bar, appears in Table~\ref{tbl:highmass}, where it is estimated to be an early B star ($M \sim 11$\Msol).  No X-ray variability is seen.  Its X-ray luminosity and spectral characteristics are similar to the known B2 star CEN~85 ($\Leftrightarrow$ ACIS \#466), so it is plausibly a new high-mass M17 star.  

\item[ACIS \#230]~~This source is located near the top of the ACIS field, west of M17~North.  It shows no X-ray variability and its IR colors do not imply that it is massive.  Its X-ray spectral parameters are essentially identical to \#187 and \#466, however, suggesting that it too may be massive.   

\item[ACIS \#396]~~Ranking by detected counts, this source is the second brightest X-ray source in the field, located in the South Bar near the \hii region interface.  It is not listed as a bright NIR source, but IR photometry can be affected by nebular emission.  Its X-ray lightcurve is classified as ``definitely variable'' but this variability is not flare-like.  This hard X-ray source is equally well fit by a very hard thermal plasma ($kT > 10$~keV, Figure~\ref{OB_spec.fig}{\it f}) or a power law with photon index $\Gamma = 1.55$; both fits give the same absorption and flux.  Although its NIR colors do not indicate a massive star, this source's X-ray characteristics are similar to the most massive stars in M17.  Its flat X-ray spectrum could indicate that it is a bright AGN, but its location near the ionization front argues against that interpretation.

\item[ACIS \#655 ($\Leftrightarrow$ B148)]~~This source, located near the center of NGC~6618 (unfortunately in a chip gap), is quite faint for most of the  observation, but exhibits the strongest flare in the whole sample (see Figure~\ref{fig:lightcurve}).  This ``superflare'' behavior has been seen in low-mass stars in several star-forming regions \citep[e.g.,][]{Grosso04,Wang06,Getman06a,Favata05}.  From Table~\ref{tbl:highmass}, source \#655 is estimated to be an intermediate-mass star ($M \sim 2$\Msol), suggesting that the flare may be from an unresolved lower-mass companion.  It is unlikely that source \#655 is a new high-mass member of M17.  
\end{description}

The final indicator of OB candidacy that we consider---X-ray detection coupled with bright ($K<10$) $K$-band magnitude---produces 28 candidates. Seventeen of these are known massive stars and two are likely foreground stars: 
\#281 $\Leftrightarrow$ B293 $\Leftrightarrow$ CEN 7 (spectral type F8) and \#854 $\Leftrightarrow$ TYC~6265-1977-1 (spectral type G with large parallax).  
ACIS \#600 ($\Leftrightarrow$ B~163) is listed by \citet{Hanson97} as a candidate young stellar object.  The remaining sources are ACIS \#1, 9, 35, 402, 650, 698, 852, and 875, widely distributed across the ACIS field of view.  These are all listed in Table~\ref{tbl:highmass}.

\subsection{Protoplanetary Disks Around Higher-Mass M17 Stars 
\label{sec:disks}}

As discussed in \S \ref{sec:intro} and \citet{Feigelson06}, X-ray 
surveys of young stellar populations give samples that are largely 
unbiased with respect to the presence or absence of protoplanetary 
disks.  They thus provide an unusual opportunity to study disk 
evolution in coeval populations.  
Here we address this issue in a restricted fashion, considering only inner disks revealed through a color excess in $JHK$ photometry (\S \ref{sec:NIRprop}) arising from their heated inner edges; such disks are present mainly in actively accreting Class I and II young stellar systems.
Our analysis is insensitive to centrally-cleared, outer disks emitting only at longer wavelengths.

The stellar population we consider here is our X-ray-selected sample of intermediate- and high-mass stars (Table \ref{tbl:highmass} in \S \ref{sec:NIRprop}).
This X-ray sample is not complete even for the 1~Myr old NGC~6618 population.
Table \ref{tbl:highmass} is missing disk/envelope evolutionary classifications for some known massive stars (e.g.\ the O4+O4 binary) due to imperfections in the $JHK$ photometry (\S \ref{sec:NIRprop}).    
Half of the cataloged early B stars in the NGC~6618 cluster are undetected (\S \ref{sec:OB}) in X-rays, and a larger fraction of late-B stars will be missing.  
Our shallow X-ray data miss much of the massive population just now forming and thought to possess disks, e.g.\ many of the high-mass Class I sources studied by \citet{Nielbock01} and \citet{Hoffmeister06}, M17-SO1 \citep{Chini04b}, and CEN~92 \citep{Chini05}.  
Despite this incompleteness, our X-ray sample of higher mass stars in the coeval NGC~6618 population should be largely unbiased with respect to the presence of inner disks.

Our sample thus consists of the 138 stars in Table~\ref{tbl:highmass} with disk/envelope evolutionary classifications.  
A remarkably low fraction (12\%, or 16/138) are designated Class I or Class II, indicating $K$-band excesses consistent with inner disks.
A similar result 
emerged from the recent {\em Chandra} study of the rich Pismis~24 
cluster illuminating the NGC~6357 \hii region, where the X-ray selected NIR Class III:II 
ratio was $\sim$25:1 among intermediate-mass stars \citep{Wang06}.  
Considering that the NGC~6618 cluster is thought to be only $\sim$1 
Myr old (\S \ref{sec:intro}), we infer that the inner disks around its more massive 
stars evolved very rapidly.  
A similar conclusion was reached by \citet{Hillenbrand93} and \citet{Hernandez05} in their
studies of young stellar clusters with many disk-free
intermediate-mass stars.
It also suggests that studies of nearby
accreting Herbig Ae/Be disks with strong NIR emission \citep{Waters98} may 
characterize only a small fraction of the underlying population of 
coeval stars.  Disk-free stars are often missing from IR-selected catalogs of 
pre-main sequence populations.

\section{Summary \label{sec:summary}}

Extended X-ray emission had been detected from the M~17 \hii region by the
{\em Einstein}, {\em ROSAT}, and {\em ASCA} X-ray observatories since the 1980s, but the
finding had not been reported until the {\em Chandra X-ray Observatory}
revealed a wealth of morphological detail.  Our earlier study focused
on the outflowing 10~MK plasma attributed to shocked O star winds
(TFM03), while the present study concentrates on the stellar
population.  In a 40~ks ACIS-I exposure obtained in 2002, we find 886 unresolved
X-ray sources with a wide range of intrinsic X-ray luminosities ($L_{h,c} \simeq 10^{28}$ to almost $10^{33}$
erg~s$^{-1}$ in the hard $2-8$~keV band) and absorptions (column densities
$N_H<10^{20}$ to $10^{24}$~cm$^{-2}$).  Nearly 90\% of the sources have
counterparts in NIR (2MASS, SIRIUS) or MIR ({\em Spitzer} GLIMPSE) images; the
remaining sources are divided between background contaminants
and newly discovered members of the cloud population.  With {\em Chandra}, the
identification of X-ray sources with individual stars is unambiguous (median
offset 0.24\arcsec\/ from NIR sources) except for components of multiple
systems.     

The spatial distribution of X-ray stars suggests a more complex 
morphology of star formation than is generally discussed from 
optical- and IR-derived surveys.  In addition to the central NGC~6618 
cluster and three well-studied embedded groupings (M17-UC1, KWO, and 
M17-North), we find evidence for a new embedded cluster (designated 
M17-X) located 4\arcmin\/ (2 pc) north-northwest of NGC~6618 
(Figure~\ref{fig:stellar_density}{\it b}).  A $5\arcmin$-long arc of several 
dozen X-ray stars along the interface between the M17 \hii region and 
the M17-SW molecular cloud core is clearly traced, delineating 
recently triggered star formation and confirming the similar 
structure seen in NIR by \citet{Jiang02}.  
Substructures on scales of 0.1~pc are seen both 
within the central NGC~6618 cluster and in the embedded populations.  
These suggest that the populations are dynamical and young, and that 
equilibrium has not been reached.  
Our data support the argument of \citet{Jiang02} and others for widely 
distributed star formation that is not concentrated into clusters.  
This is seen in both the heavily and lightly obscured X-ray populations.  

We use our detailed knowledge of the Orion Nebula Cluster population 
and XLF to calibrate the observed XLF of the ACIS M17 field.  The 
inferred total population of heavily obscured stars is 
$5000-7000$ down to 0.1\Msol, outnumbering the lightly obscured 
($A_V \la 10$~mag) population of $2500-3500$ stars which are mostly concentrated in the 
central NGC~6618 cluster.  
These values are consistent with the $\sim$3600 
Class I-III young stars counted by \citet{Jiang02} in their NIR 
survey sensitive to $K < 19$~mag.  

Forty percent 
of the ACIS sources are heavily obscured, defined here by 
median energy $E_{median} > 2.5$~keV which is roughly equivalent to $A_V > 
10$~mag.  Concentrations of $10-20$ sources are seen around each of 
the three well-studied star formation subregions in the field: 
IRS~5/UC1, the KWO, and M17-North. Some high-$L_{IR}$, 
high-mass embedded sources are detected in X-rays (e.g.\ IRS~5S and Anon~1 
near M17-UC1, IRS~2, CEN~31, IRS~15), but others are not 
(the KWO, M17-SO1, CEN~92).  The situation for the most famous embedded 
objects is complicated: IRS~5N which ionized M17-UC1 is only 
tentatively detected with $<$5 photons; the KWO Herbig Be star is undetected but a nearby member of the KWO cluster is seen. 

A much greater number of heavily obscured stars are seen outside of 
the three known concentrations.  Many are located in the North Bar and 
South Bar where the \hii region ionization front is propagating into 
the cloud, but others appear in more distant molecular cloud cores or 
distributed more-or-less uniformly across the field.  
At our current completeness limit for heavily obscured stars (roughly estimated in \S \ref{sec:XLF} as $\log L_{h,c} 
\ga 30.9$~ergs~s$^{-1}$), only the tip of the embedded XLF is detected.
An upcoming deeper {\em Chandra} observation should identify hundreds more heavily obscured 
stars and elucidate any clustering patterns of the embedded stellar 
population.   

All of the cataloged O stars but only half of the cataloged early B stars (B0--B3) in the field are 
detected in this short 40 ks ACIS exposure.
These OB stars show intrinsic X-ray luminosities consistent with the long-standing $L_x \propto 10^{-7}~L_{bol}$ relationship, with scatter similar to that found for the NGC~6357, NGC~6231, and Orion Nebula Cluster populations.

Only the O4+O4 binary components, 
which are resolved at 1.8\arcsec\/ separation, are very X-ray luminous, 
emitting $\sim$$2 \times 10^{33}$ ergs~s$^{-1}$ in the total band, 
$20-30$\% of which appears in the penetrating hard band.  
Although there is no evidence for a shock between the winds from the two resolved components, each component itself exhibits remarkably hard thermal plasma emission even harder than that usually seen in colliding wind binaries.  This suggests that Kleinmann's Anonymous Star may be a system of at least 4 massive stars, two pairs of colliding-wind binaries both with extraordinary wind and/or magnetic properties.  

Many B stars have X-ray luminosities scattered in the $29 < 
\log L_t < 31$ ergs~s$^{-1}$ range, overlapping the low-mass pre-main 
sequence XLF.  Most of these stars are thus not readily distinguished from 
lower mass stars in the X-ray band and, not infrequently, the 
emission likely comes from a low-mass companion. 
Many likely new OB stars are found by the association of {\em Chandra} 
sources with NIR sources.

Our final result concerns the frequency of inner protoplanetary disks 
around the X-ray selected high- and intermediate-mass stars.  
X-ray selection is effective in establishing the underlying population of these stars in a disk-unbiased fashion.
We find that only $\sim$10--15\% of NGC~6618's OBA stars have disks with $K$-band excesses.

\acknowledgments

We appreciate several helpful comments from the anonymous referee.
Support for this work was provided to Gordon Garmire, the ACIS
Principal Investigator, by the National Aeronautics and Space
Administration (NASA) through NASA Contract NAS8-38252 and {\em
Chandra} Contract SV4-74018 issued by the {\em Chandra X-ray Observatory}
Center, which is operated by the Smithsonian Astrophysical Observatory
for and on behalf of NASA under Contract NAS8-03060.  
This publication makes use of data products from the Two
Micron All Sky Survey, which is a joint project of the University of
Massachusetts and the Infrared Processing and Analysis
Center/California Institute of Technology, funded by NASA and the
National Science Foundation.  
This work is based in part on observations made with the {\em Spitzer Space Telescope}, which is operated by the Jet Propulsion Laboratory, California Institute of Technology under a contract with NASA.
This publication makes use of data products from the {\em Midcourse Space Experiment}. 
This research made use of the SIMBAD
database and VizieR catalogue access tool, operated at CDS, Strasbourg,
France.  We would have been lost without the invaluable tools of NASA's
Astrophysics Data System, and without CIAO, XSPEC, and ds9.

{\it Facilities:} \facility{CXO (ACIS)}.

\appendix
\section{Estimation of Matching Reliability \label{sec:matching}}

Because the SIRIUS, 2MASS, and GLIMPSE catalogs have far more entries than our X-ray catalog, some of the associations between X-ray and IR sources reported in Table~\ref{tbl:counterparts} are expected be false positives, i.e.\ false associations found by chance positional coincidence.
To estimate the number of false positives expected when matching to a specific IR catalog, it is useful to view the X-ray catalog as a mixture of two populations: an ``associated population" of X-ray sources where a ``true'' counterpart {\bf does} exist in the IR catalog, and an ``isolated population'' of X-ray sources where {\bf no} counterpart exists.

We can easily simulate the behavior of the matching algorithm for the isolated population by applying an offset between the catalogs, running the algorithm, and tabulating the results. The offset must be large enough to ``break'' all the true associations, but excessive offsets should be avoided in order to preserve the relative spatial distributions of the two catalogs.  
We interpret the matches produced by this simulation as ``false positives.''
We refer to the X-ray sources which (correctly) did not produce matches as ``true negatives.''

With slightly more effort we can simulate the behavior of the matching algorithm for the associated population by creating a ``fake'' IR counterpart for each X-ray source, running the algorithm, and tabulating the results.  
Each fake IR position is randomly offset from the corresponding X-ray position by sampling the appropriate offset distribution, taken to be Gaussian with variance equal to the sum of that source's X-ray position variance and an IR position variance chosen randomly from the IR sources reported in Table~\ref{tbl:counterparts}.
We attempt to eliminate the actual associations from the IR catalog by removing the counterparts reported in Table~\ref{tbl:counterparts}.   
We also randomly offset the catalogs as described above because the ``background'' population in the IR catalog may be suppressed in the regions around the pruned sources (due to observational limitations). 
We interpret the matches produced by this simulation as falling into two categories: ``correct matches'' to the fake entry we created, and ``incorrect matches'' to some other source which lies closer by chance.
We refer to the X-ray sources which (incorrectly) did not produce matches as ``false negatives.''

By repeating the simulations many times, we estimate the probabilities of each of the five match types described above.  These probabilities, reported in Table~\ref{table:matchsims}, can be used to characterize Table~\ref{tbl:counterparts} only if we can estimate the true mixture ratio between the isolated and associated  populations, i.e.\ the number of X-ray sources having no counterparts (thus susceptible to falsely matching the background IR sources) vs. the number of X-ray sources having true counterparts (thus susceptible to only those background IR sources that lie closer than the true match).  We estimate the ``associated fraction'' for each catalog ($f_{A_{2MASS}}$, $f_{A_{SIRIUS}}$, $f_{A_{GLIMPSE}}$) by considering the number of X-ray sources for which no match was reported in Table~\ref{table:matchsims} (410 for 2MASS, 159 for SIRIUS, 662 for GLIMPSE).  Those ``negative'' matches are a mixture of the true negative and false negative match types in the simulations (columns 4 and 5 in Table~\ref{table:matchsims}):
\[ \begin{array}{rclcl}
 410 & = & 7.6 \: f_{A_{2MASS}}   & + & 865.5 \: (1-f_{A_{2MASS}})  \\
 159 & = & 7.6 \: f_{A_{SIRIUS}}  & + & 739.2 \: (1-f_{A_{SIRIUS}}) \\
 662 & = & 8.8 \: f_{A_{GLIMPSE}} & + & 861.6 \: (1-f_{A_{GLIMPSE}}) 
\end{array} \]
\begin{deluxetable}{lcccccc}
\tablecaption{Simulations of matching algorithm 
\label{table:matchsims}} 
\tablehead{
                  & \multicolumn{3}{c}{Associated Population} && \multicolumn{2}{c}{Isolated Population} \\
				   \cline{2-4}                                              \cline{6-7}
\colhead{Catalog} & \colhead{Correct Matches} & \colhead{Incorrect Matches} & \colhead{False Neg.} && \colhead{True Neg.} & \colhead{False Pos.} \\
\colhead{(1)} & \colhead{(2)} & \colhead{(3)} & \colhead{(4)} && \colhead{(5)} & \colhead{(6)}  
}
\startdata
2MASS   & 874.2 & \phn4.2 &  7.6 && 865.6 & \phn20.4 \\
SIRIUS  & 849.9 &    28.5 &  7.6 && 739.2 &    146.8 \\
GLIMPSE & 872.7 & \phn4.6 &  8.8 && 861.6 & \phn24.4 \\
\enddata
\tablecomments{{\bf Columns 2--4}: Expected number of population (1) match types.
{\bf Columns 5--6}: Expected number of population (2) match types.}
\end{deluxetable}

After solving for these associated fractions
($f_{A_{2MASS}}=0.53$, $f_{A_{SIRIUS}}=0.79$, $f_{A_{GLIMPSE}}=0.23$),
we can characterize the expected flaws in Table~\ref{tbl:counterparts}.
For example, the expected number of reported matches which are wrong
is estimated as a weighted sum of ``incorrect matches'' from the associated population plus ``false positives'' from the isolated population:
\[ \begin{array}{lclcrr}
 \phn4.2 \: f_{A_{2MASS}}   & + & \phn20.4 \: (1-f_{A_{2MASS}})   & = & 12 & (2.5\%) \\
    28.5 \: f_{A_{SIRIUS}}  & + &    146.8 \: (1-f_{A_{SIRIUS}})  & = & 53 & (7\%)   \\
 \phn4.6 \: f_{A_{GLIMPSE}} & + & \phn24.4 \: (1-f_{A_{GLIMPSE}}) & = & 20 & (9\%) 
\end{array} \]

Note that these are averages across the catalog; the reliability of individual matches is complex since it depends on both the position uncertainties (e.g.\ larger uncertainties lead to larger matching ``footprints'' which have larger susceptibility to spurious matches) and on the local density of IR sources (e.g.\ crowded regions are more likely to generate spurious matches).

\end{document}